\documentclass[twocolumn,tighten,times,twocolappendix]{aastex63}

\usepackage{amsmath} %%-->for use of eqref
\usepackage{amssymb}
\usepackage{afterpage}
\extrafloats{100}

\received{\today}
\revised{\today}
\accepted{\today}
\newcommand{\galprop}{\textsc{GalProp}}
\newcommand{\helmod}{\textsc{HelMod}}
\newcommand{\fermilat}{\emph{Fermi}-LAT}

\newcommand{\hi}{H\,{\sc i}}
\newcommand{\hii}{H\,{\sc ii}}
\newcommand{\htwo}{H$_2$}
\newcommand{\gray}{$\gamma$-ray}

\newcommand{\planck}{{\it Planck}}

\shorttitle{Local Interstellar Spectra of Cosmic Ray Nuclei $Z\le 28$}
\shortauthors{Boschini et al.}

\begin{document}

\title{
Inference of the Local Interstellar Spectra of Cosmic Ray Nuclei $Z\le 28$ with the \galprop{}--\helmod{} Framework
}

\author[0000-0002-6401-0457]{M.~J.~Boschini}
\affiliation{INFN, Milano-Bicocca, Milano, Italy}
\affiliation{CINECA, Segrate, Milano, Italy}

\author[0000-0002-7669-0859]{S.~{Della~Torre}}
\affiliation{INFN, Milano-Bicocca, Milano, Italy}

\author[0000-0003-3884-0905]{M.~Gervasi}
\affiliation{INFN, Milano-Bicocca, Milano, Italy}
\affiliation{Physics Department, University of Milano-Bicocca, Milano, Italy}

\author[0000-0003-1942-8587]{D.~Grandi}
\affiliation{INFN, Milano-Bicocca, Milano, Italy}
\affiliation{Physics Department, University of Milano-Bicocca, Milano, Italy}

\author[0000-0003-1458-7036]{G.~J\'{o}hannesson} 
\affiliation{Science Institute, University of Iceland, Dunhaga 3, IS-107 Reykjavik, Iceland}
\affiliation{NORDITA,  Roslagstullsbacken 23, 106 91 Stockholm, Sweden}

\author[0000-0002-2168-9447]{G.~{La~Vacca}}
\affiliation{INFN, Milano-Bicocca, Milano, Italy}
\affiliation{Physics Department, University of Milano-Bicocca, Milano, Italy}

\author[0000-0002-3729-7608]{N.~Masi}
\affiliation{INFN, Bologna, Italy}
\affiliation{Physics Department, University of Bologna, Bologna, Italy}

\author[0000-0001-6141-458X]{I.~V.~Moskalenko} 
\affiliation{Hansen Experimental Physics Laboratory, Stanford University, Stanford, CA 94305}
\affiliation{Kavli Institute for Particle Astrophysics and Cosmology, Stanford University, Stanford, CA 94305}

\author{S.~Pensotti}
\affiliation{INFN, Milano-Bicocca, Milano, Italy}
\affiliation{Physics Department, University of Milano-Bicocca, Milano, Italy}

\author[0000-0002-2621-4440]{T.~A.~Porter} 
\affiliation{Hansen Experimental Physics Laboratory, Stanford University, Stanford, CA 94305}
\affiliation{Kavli Institute for Particle Astrophysics and Cosmology, Stanford University, Stanford, CA 94305}

\author{L.~Quadrani}
\affiliation{INFN, Bologna, Italy}
\affiliation{Physics Department, University of Bologna, Bologna, Italy}

\author[0000-0002-1990-4283]{P.~G.~Rancoita}
\affiliation{INFN, Milano-Bicocca, Milano, Italy}

\author[0000-0002-7378-6353]{D.~Rozza}
\affiliation{INFN, Milano-Bicocca, Milano, Italy}
\affiliation{Physics Department, University of Milano-Bicocca, Milano, Italy}

\author[0000-0002-9344-6305]{M.~Tacconi}
\affiliation{INFN, Milano-Bicocca, Milano, Italy}
\affiliation{Physics Department, University of Milano-Bicocca, Milano, Italy}
%\affil{institute}
%\email{\myemail}

%% Notice that each of these authors has alternate affiliations, which
%% are identified by the \altaffilmark after each name.  Specify alternate
%% affiliation information with \altaffiltext, with one command per each
%% affiliation.

%% Mark off your abstract in the ``abstract'' environment. In the manuscript
%% style, abstract will output a Received/Accepted line after the
%% title and affiliation information. No date will appear since the author
%% does not have this information. The dates will be filled in by the
%% editorial office after submission.

\begin{abstract}

Composition and spectra of Galactic cosmic rays (CRs) are vital for studies of high-energy processes in a variety of environments and on different scales, for interpretation of \gray{} and microwave observations, disentangling possible signatures of new phenomena, and for understanding of our local Galactic neighborhood. Since its launch, AMS-02 has delivered outstanding quality measurements of the spectra of $\bar{p}$, $e^{\pm}$, and nuclei: $_1$H--\,$_8$O, $_{10}$Ne, $_{12}$Mg, $_{14}$Si. These measurements resulted in a number of breakthroughs, however, spectra of heavier nuclei and especially low-abundance nuclei are not expected until later in the mission. Meanwhile, a comparison of published AMS-02 results with earlier data from HEAO-3-C2 indicate that HEAO-3-C2 data may be affected by undocumented systematic errors. Utilizing such data to compensate for the lack of AMS-02 measurements could result in significant errors. In this paper we show that {\it a fraction} of HEAO-3-C2 data {\it match} available AMS-02 measurements quite well and can be used together with Voyager 1 and ACE-CRIS data to make \emph{predictions} for the local interstellar spectra (LIS) of nuclei that are not yet released by AMS-02. We are also updating our already published LIS to provide a complete set from $_{1}$H--\,$_{28}$Ni in the energy range from 1 MeV nucleon$^{-1}$ to $\sim$100--500 TeV nucleon$^{-1}$ thus covering 8--9 orders of magnitude in energy. Our calculations employ the \galprop{}--\helmod{} framework that is proved to be a reliable tool in deriving the LIS of CR $\bar{p}$, $e^{-}$, and nuclei $_1$H--\,$_8$O.
\end{abstract}

%% Keywords should appear after the \end{abstract} command. The uncommented
%% example has been keyed in ApJ style. See the instructions to authors
%% for the journal to which you are submitting your paper to determine
%% what keyword punctuation is appropriate.

\keywords{cosmic rays --- diffusion --- elementary particles --- interplanetary medium --- ISM: general --- Sun: heliosphere}

\section{Introduction} \label{Intro}
%%%%%%%%%%%%%%%%%%%%%%%%%%%%%%%%%%%%%%%%%%%%%%%%%%%
%%%%%%%%%%%%%%%%%%%%%%%%%%%%%%%%%%%%%%%%%%%%%%%%%%%

The last decade brought impressive advances in astrophysics of CRs and \gray{} astronomy. Launches of new missions that are employing forefront detector technologies were followed by a series of remarkable discoveries\footnote{The 1st Workshop on the Next Generation of Astro\-Particle Experiments in Space (NextGAPES-2019), \url http://www.sinp.msu.ru/contrib/NextGAPES/} even in the energy range that is deemed as well-studied. Many of those missions have the discovery of dark matter (DM) as one of their primary goals \citep{2011ARA&A..49..155P,2019arXiv191208821H}. 

Among those missions are the Payload for Antimatter Matter Exploration and Light-nuclei Astrophysics \citep[PAMELA,][]{2007APh....27..296P,2014PhR...544..323A}, the {\it Fermi} Large Area Telescope \citep[\fermilat,][]{2009ApJ...697.1071A}, the Alpha Magnetic Spectrometer -- 02 \citep[AMS-02,][]{2013PhRvL.110n1102A}, NUCLEON experiment \citep{2019AdSpR..64.2546G, 2019AdSpR..64.2559G}, CALorimetric Electron Telescope -- \citep[CALET,][]{2019PhRvL.122r1102A, 2019AdSpR..64.2531T, 2020PAN....82..766M}, DArk Matter Particle Explorer mission -- \citep[DAMPE,][]{2017APh....95....6C,2017Natur.552...63D,eaax3793}, and Cosmic-Ray Energetics and Mass investigation -- \citep[ISS-CREAM,][]{2014AdSpR..53.1451S}. Outstanding results have been also delivered by more mature missions, such as the Cosmic Ray Isotope Spectrometer onboard of the Advanced Composition Explorer \citep[ACE-CRIS,][]{2018ApJ...865...69I, 2016Sci...352..677B} operating at the L1 Lagrange point for more than two decades, and by the grandparents of the current instrumentation -- Voyager~1,~2 spacecraft \citep{1977SSRv...21..355S}, which were built with the technology of 1970s and launched in 1977 at the dawn of the space era. The latter are providing unique data on the elemental spectra and composition at the {\it interstellar reaches} of the solar system \citep{2013Sci...341..150S, 2016ApJ...831...18C, 2019NatAs...3.1013S} currently at 149 AU and 124 AU from the sun, correspondingly. The modern technologies employed by many of these missions have enabled measurements with unmatched precision, which allow for searches of subtle signatures of new phenomena and DM in CR and \gray{} data, e.g., the claimed precision of AMS-02 data is 1-3\%. 

Indirect observations of CR acceleration sites and CR propagation in the Galaxy and beyond are made by \gray{} telescopes covering the range from MeV to multi-TeV. Among these instruments are the International Gamma-Ray Astrophysics Laboratory \citep[INTEGRAL,][]{2008ApJ...679.1315B, 2010ApJ...720.1772B, 2011ApJ...739...29B}, \fermilat{} \citep{2012ApJ...750....3A}, the High-Altitude Water Cherenkov \gray{} observatory \citep[HAWC,][]{2017ApJ...843...40A,2019JCAP...07..022A}, the Large High Altitude Air Shower Observatory \citep[LHAASO,][]{2019arXiv190502773B} currently under construction, and three atmospheric Cherenkov telescopes: the High Energy Stereoscopic System \citep[H.E.S.S.,][]{2009ARA&A..47..523H, 2018A&A...612A...1H}, Major Atmospheric Gamma-ray Imaging Cherenkov Telescopes \citep[MAGIC,][]{2016APh....72...76A}, and the Very Energetic Radiation Imaging Telescope Array System \citep[VERITAS,][]{2006APh....25..391H,2019AdSpR..64.2578H}. Construction of the next generation Cherenkov Telescope Array (CTA) has begun in 2017 \citep{2013APh....43....3A,cta}. High-resolution data in the microwave domain are provided by the {\it Wilkinson} Microwave Anisotropy Probe \citep[WMAP,][]{2003ApJ...583....1B,2013ApJS..208...20B} and {\it Planck} mission \citep{2010A&A...520A...1T,2016A&A...594A..25P}.

Recent technological advances in astrophysics of CRs and \gray\ astronomy and the reached level of precision implicate that {\it we are on the verge of major discoveries.} Meanwhile, disentangling faint signals of new physics from the conventional astrophysical processes requires a high degree of sophistication, including a proper description of all variety of interactions of CR species in the interstellar medium (ISM), a description of the detailed properties of the ISM itself, and understanding of the propagation of energetic particles in the Galaxy and in the heliosphere. Such a description has to be self-consistent, i.e.\ account for the inter-relationships between CR species and their associated photon emissions (radio- to \gray{s}), and compare well with available data. Understanding the conventional astrophysical backgrounds and having most up-to-date knowledge of physics of the ISM -- these are the keys to extract weak signals of new physics. {\it If the signatures of DM were too obvious we would have discovered them long ago.}

The direct precise measurements of spectra of CR species in the wide energy range form a basis for propagation models, for interpretation of \gray{} and microwave observations, and for disentangling possible hints of new phenomena. Composition and spectra of CR species are vital for studies of galactic nucleosynthesis and high-energy processes in variety of environments, from CR sources to properties of the ISM and the Milky Way galaxy as the whole, and they are equally important for understanding of our local Galactic environment. Since the beginning of its operation, AMS-02 has delivered outstanding quality measurements of the spectra of CR protons, $\bar{p}$, $e^{\pm}$, and nuclei from $_2$He--\,$_8$O, $_{10}$Ne, $_{12}$Mg, $_{14}$Si \citep{2014PhRvL.113v1102A, 2015PhRvL.114q1103A, 2015PhRvL.115u1101A, 2016PhRvL.117i1103A, 2016PhRvL.117w1102A, 2017PhRvL.119y1101A, 2018PhRvL.120b1101A, 2018PhRvL.121e1103A, 2019PhRvL.122d1102A, 2019PhRvL.122j1101A, PhysRevLett.124.211102}. These measurements already resulted in a number of breakthroughs. However, the spectra of heavier nuclei and especially low-abundance nuclei, such as $_{9}$F, $_{11}$Na, $_{13}$Al, $_{15}$P--\,$_{28}$Ni, are expected to be available only later in the mission. Meanwhile, a comparison of published AMS-02 results with data from earlier experiments, such as HEAO-3-C2 \citep{1990A&A...233...96E}, indicate that earlier data may be affected by significant undocumented systematic errors \citep[see][]{2018ApJ...858...61B, 2020ApJ...889..167B}. Therefore, we found ourselves in an awkward position: we are keen to use new data to move forward, but cannot do it because the new data are still incomplete. At the same time, using the old data to make up for the lack of AMS-02 measurements could lead to significant errors.  

In this paper we are using the existing AMS-02 data to test HEAO-3-C2 results \citep{1990A&A...233...96E}. We show that the LIS built with available AMS-02 measurements match well {\it some part of the HEAO-3-C2 data} when modulated appropriately to the solar activity observed during the HEAO-3-C2 flight. This part of the HEAO-3-C2 data  can be used together with Voyager 1 \citep{2016ApJ...831...18C, 2019NatAs...3.1013S} and ACE-CRIS data to make \emph{predictions} for the LIS of nuclei that are not yet released by the AMS-02 collaboration. We are also updating our already published LIS to provide a complete set of LIS from $_{1}$H--\,$_{28}$Ni in the kinetic energy range from 1 MeV nucleon$^{-1}$ to $\sim$100--500 TeV nucleon$^{-1}$ thus covering 8--9 orders of magnitude in energy. Our calculations employ the \galprop{}--\helmod{} framework that is proved to be a reliable tool in deriving the LIS of CR species: $\bar{p}$, $e^{-}$, and nuclei from $_1$H to $_8$O.

\section{CR transport in the Galaxy and the heliosphere} \label{sec2}
%%%%%%%%%%%%%%%%%%%%%%%%%%%%%%%%%%%%%%%%%%%%%%%%%%%
%%%%%%%%%%%%%%%%%%%%%%%%%%%%%%%%%%%%%%%%%%%%%%%%%%%

Here we provide short descriptions of the two dedicated codes that are used in the present work and that complement each other: \galprop{}\footnote{Available from http://galprop.stanford.edu \label{galprop-site}} -- for description of the interstellar propagation, and \helmod{}\footnote{http://www.helmod.org/ \label{helmod-footnote}} -- for description of the heliospheric transport.  More details can be found in the referenced papers.

\subsection{\galprop{} Framework for Galactic CR Propagation and diffuse emissions}\label{galprop}
%%%%%%%%%%%%%%%%%%%%%%%%%%%%%%%%%%%%%%%%%%%%%%%%%%%
%%%%%%%%%%%%%%%%%%%%%%%%%%%%%%%%%%%%%%%%%%%%%%%%%%%
%
Our main research tool is the state-of-the-art fully numerical \galprop{} code that describes propagation of Galactic CRs and production of the associated diffuse emissions (radio, X-rays, \gray{s}). It has about 23 years of development behind it \citep{1998ApJ...493..694M, 1998ApJ...509..212S}. Over these years, the \galprop{} code has proven to be invaluable tool in sophisticated analyses in many areas of astrophysics including numerous searches for DM signatures \citep[e.g.,][]{2012ApJ...750....3A, 2015ApJ...799...86A, 2016ApJ...819...44A, 2020ApJS..247...33A, 2016A&A...594A..10P, 2016A&A...594A..25P, 2012ApJ...752...68V, 2016ApJ...831...18C, 2017JCAP...12..040A, 2019PhRvD..99j3026C, 2017PhRvD..95j3005K, 2019ApJ...880...95K}.

The \galprop{} code uses information from astronomy, particle, and nuclear physics to predict CRs, \gray{s}, synchrotron emission and its polarization in a self-consistent manner. The key concept underlying the \galprop{} code is that various kinds of data, e.g., direct CR measurements, $\bar{p}$, $e^\pm$, \gray{s}, synchrotron radiation, and so forth, are all related to the same Galaxy and hence have to be modeled self-consistently \citep{1998A&A...338L..75M}. It provides the modeling framework unifying results of many individual experiments in physics and astronomy spanning in energy coverage, types of instrumentation, and the nature of detected species. The goal for the \galprop{}-based models is to be as realistic as possible and to make use of available information with a minimum of simplifying assumptions \citep{2007ARNPS..57..285S}. The range of physical validity of the \galprop{} code extends from sub-keV---PeV energies for particles and from $10^{-6}$ eV---PeV for photons.  

The \galprop{} code solves a system of about 90 time-dependent transport equations (partial differential equations in 3D or 4D: spatial variables plus energy) with a given source distribution and boundary conditions for all CR species: $^1_1$H--\,$^{64}_{28}$Ni, $\bar{p}$, $e^\pm$ \citep{1998ApJ...509..212S, 2007ARNPS..57..285S, 2009arXiv0907.0559S}. This includes convection, distributed reacceleration, energy losses, nuclear fragmentation, radioactive decay, and production of secondary particles and isotopes. The numerical solution is based on a Crank-Nicholson implicit second-order scheme \citep{1992nrfa.book.....P}. The spatial boundary conditions assume free particle escape. For a given halo size the diffusion coefficient, as a function of rigidity and propagation parameters, is determined from secondary-to-primary nuclei ratios, typically $_{5}$B/$_{6}$C, [$_{21}$Sc+$_{22}$Ti+$_{23}$V]/$_{26}$Fe, and/or $\bar{p}/p$. If reacceleration is included, the momentum-space diffusion coefficient $D_{pp}$ is related to the spatial coefficient $D_{xx} = \beta D_0 R^\delta$ \citep{1994ApJ...431..705S}, where $\beta=v/c$ is the particle velocity, $R$ is the magnetic rigidity, $\delta = 1/3$ for a Kolmogorov spectrum of interstellar turbulence \citep{1941DoSSR..30..301K}, or $\delta = 1/2$ for an Iroshnikov-Kraichnan cascade \citep{1964SvA.....7..566I, 1965PhFl....8.1385K}, or can also be arbitrary. Also arbitrary can be the spatial dependence of the diffusion coefficient, but for calculations in this paper it is assumed uniform throughout the Galaxy. For more details and fitted parameter values see Section~\ref{calcs} and Table~\ref{tbl-prop}.

Distributions of CR sources and gas components can be customized at the user discretion. Their configurations are defined with an XML file (using the galstruct library distributed with the galtoolslib package), where each entry is a single component that can be added to the aggregate distribution. Possible components include axi-symmetric disk, spiral arms, various central bulges, and other structures. Each basic component can be further split up and fine-tuned with different radial profiles, allowing for a very flexible description of a galaxy.  

In this work we are using a standard pulsar distribution for all CR species:
\begin{equation}  \label{eq:0}
S(r)\propto \left(\frac{r}{R_\odot}\right)^\alpha  \exp\left(-\beta\, \frac{r-R_\odot}{R_\odot}\right), 
\end{equation}
where $r$ (kpc) is the distance from the Galactic center, and the parameter values $\alpha=1.9, \beta=5.0, R_\odot=8.5$ kpc correspond to the C-model from \citet{2006MNRAS.372..777L}. Meanwhile, the results are not sensitive to the used CR source distribution, and other parameterizations, such as the SNR distribution \citep{1998ApJ...504..761C} or another pulsar distribution \citep{2004A&A...422..545Y}, work as well. The total injected CR power does not vary at most more than a few percent over these similar 2D, or more detailed 3D, CR source distributions \citep{2010ApJ...722L..58S, 2017ApJ...846...67P}. The injection spectra of CR species are parametrized by the rigidity-dependent function:
\begin{equation}  \label{eq:1}
q(R) \propto (R/R_0)^{-\gamma_0}\prod_{i=0}^2\bigg[1 + (R/R_i)^\frac{\gamma_i - \gamma_{i+1}}{s_i}\bigg]^{s_i},
\end{equation}
where $\gamma_{i =0,1,2,3}$ are the spectral indices, $R_{i = 0,1,2}$ are the break rigidities, $s_i$ are the smoothing parameters ($s_i$ is negative/positive for $|\gamma_i |\lessgtr |\gamma_{i+1} |$).

The \galprop{} code computes a complete network of primary, secondary, and tertiary isotope production starting from input CR source abundances. Since the decay branching ratios and half-lifes of fully stripped and hydrogen-like ions may differ, \galprop{} includes the processes of K-electron capture, electron pick-up from neutral ISM gas and formation of hydrogen-like ions as well as the inverse process of electron stripping \citep{1973RvMP...45..273P, 1978PhDT........12W, 1979PhDT........67C}. Meanwhile, the fully stripped and hydrogen-like ions are treated as separate species. Also included are knock-on electrons \citep{1966PhRv..150.1088A, 2003ApJ...594..709B} that may significantly contribute to hard X-ray---soft \gray{} diffuse emission through inverse Compton scattering and Bremsstrahlung \citep{2008ApJ...682..400P}. 

The nuclear reaction network is built using the {\it sixty four} volumes of Nuclear Data Sheets \citep[see][for Cumulated Index to $A$-Chains for $A=1-64$ nuclei]{2018NDS...151D...3.}. Included are multistage chains of $p$, $n$, $d$, $t$, $^3$He, $\alpha$, $\beta^\pm$-decays, and electron K-capture, and, in many cases, more complicated reactions. This accounts for up to 4 stages of 3 decay branchings ($\mathcal{B}$) each in any of the decay channels, i.e.\ up to $3^4=81$ total daughter nuclei in the final state for \emph{each fragment} produced in spallation of the initial target nucleus plus unlimited number of $p$, $n$, and $\beta^\pm$-decays. For example, for a neutron-deficient fragment $^{20}_{12}$Mg we have a 2-stage decay chain with 3 daughter nuclei in the final state:
\begin{align}
^{20}_{12}{\rm Mg} \to {} ^{19}_{10}{\rm Ne}\, (\mathcal{B}=0.03)&\xrightarrow{\beta^+} {}^{19}_{\phn 9}{\rm F} \nonumber\\
{^\searrow} ^{20}_{11}{\rm Na}\, (\mathcal{B}=0.97) &\xrightarrow{\beta^+} {}^{20}_{10}{\rm Ne}\, (\mathcal{B}=0.80) \nonumber\\
&{^\searrow} ^{16}_{\phn 8}{\rm O}\, (\mathcal{B}=0.20).\nonumber
\end{align}
Another example is a neutron-rich fragment $^{14}_{\phn 4}$Be, where we have a 2-stage decay chain with 4 daughter nuclei in the final state:
\begin{align}
^{14}_{\phn 4}{\rm Be} \to {} ^{14}_{\phn 5}{\rm B}\, (\mathcal{B}=0.14) & \xrightarrow{\beta^-} {}^{14}_{\phn 6}{\rm C} \xrightarrow{\beta^-} {}^{14}_{\phn 7}{\rm N} \nonumber\\
{^\searrow} ^{13}_{\phn 5}{\rm B}\, (\mathcal{B}=0.81) & \xrightarrow{\beta^-} {}^{13}_{\phn 6}{\rm C} \nonumber\\
{^\searrow} ^{12}_{\phn 5}{\rm B} (\mathcal{B}=0.05) & \xrightarrow{\beta^-} {}^{12}_{\phn 6}{\rm C}\, (\mathcal{B}=0.97) \nonumber\\
&{^\searrow} 2\times\, ^{4}_{2}{\rm He}\, (\mathcal{B}=0.06). \nonumber
\end{align}
Here we have 2 $\alpha$-particles with branching $\mathcal{B}=0.03$, and thus the branching of this chain is $\mathcal{B}=0.06$. The individual branchings in the consequent decay chains are multiplied to obtain the yields of the final products.   

The routines for the isotopic production cross sections are built using a systematic approach tuned to all available data extracted from Los Alamos (LANL) and Experimental Nuclear Reaction Data (EXFOR) databases, as well as from an extensive literature search. To account for different measurement techniques that were introduced in experimental nuclear physics over decades of research since 1950s, the distinction was made between the {\it individual, direct, decaying, charge-changing, cumulative, differential, total, and isobaric cross sections, or reactions with metastable final states, with the target that could be a particular isotope, a natural sample with mixed isotopic composition, or a chemical compound.} Often, experimental cross sections for the same reaction published by various groups were found to differ by a significant factor. A (tough) decision on which set to be used was based on examination of the descriptions of particular experimental setups in the original papers. 

One example is a hard-to-find 26-page document PRVCAN-58-074812-24, which is a supplement to a paper by \citet{1998PhRvC..58.3539W} detailing the secondary beams quality used for measurements of individual charge changing and isotopic production cross sections along with the cross section values themselves measured with 1.52 g cm$^2$ thick hydrogen target at SATURNE. For many secondary beams the quality is marked as $>$90\%, which means the percentage of the desired isotope in the beam. This may be acceptable given the accuracy of CR measurements at that time. However, many other secondary beams have much smaller fraction of the isotope in question, such as $^{22}_{11}$Na (72\%), $^{30}_{15}$P (57\%),  $^{33}_{16}$S (61\%), $^{34}_{17}$Cl (65\%), $^{52}_{24}$Cr (81\%), $^{54}_{26}$Fe (60\%), and these are just a few examples. At the same time the typical accuracy of the measured cross sections is claimed to be 3\%-5\% (labeled B), 5\%-8\% (C) or 8\%-12\% (D), which perhaps represent only statistical errors. The true beam energy varies from 496 MeV nucleon$^{-1}$ to 577 MeV nucleon$^{-1}$, not being 600 MeV nucleon$^{-1}$ for all beams as claimed. The total error for such measurements used in the fits and in \galprop{} routines was increased appropriately reaching up to 50\% in some cases. Meanwhile, even such quality measurements are often the only measurement available for a particular reaction, as many astrophysically important reactions were not measured at all. 

The isotopic production cross sections were ranked by their contributions to the production of a particular isotope \citep[e.g., see][]{2013ICRC...33..803M}. The most effort was devoted to the main contributing channels. The approach to the description of each channel depended on the accuracy and availability of experimental data. If the cross section data were detailed enough, they were approximated with fitted functional dependences or provided as a table for interpolation. If only a few or no data points were available, such cross sections were approximated using the results of the Los Alamos nuclear codes \citep{2001ICRC....5.1836M, 2003ApJ...586.1050M, 2003ICRC....4.1969M, 2004AdSpR..34.1288M, 2005AIPC..769.1612M}, such as a version of the Cascade-Exciton Model \citep[CEM2k,][]{2004AdSpR..34.1288M} and the ALICE code with the Hybrid Monte Carlo Simulation model \citep[HMS-ALICE,][]{1996PhRvC..54.1341B, 1998PhRvC..57..233B}. In general, parameterizations of all isotopic production cross sections are provided from a few MeV/nucleon to several GeV/nucleon, above which it is assumed a constant.

In the case of a minor contribution channel, the best of the available semi-empirical formulae by Webber et al. \citep[WNEWTR code with modifications made in 2003,][]{2003ApJS..144..153W} or parametric formulae by Silberberg and Tsao \citep[YIELDX code,][]{1998ApJ...501..911S,1998ApJ...501..920T} normalized to the data when exists was used. Each of the 1000s channels was tested to ensure the best description of the available data. \emph{A very limited database} of the measured cross section points is supplied with \galprop{} routines to renormalize the output of WNEWTR and YIELDX codes. The data points to include into this database were selected for the stated validity range of the semi-empirical formulae \citep[typically $>$150 MeV nucleon$^{-1}$, ][]{2003ApJS..144..153W}, while \emph{the data points outside of this validity range were excluded from the auxiliary files.} 

The total (inelastic) fragmentation cross sections for $pA$- and $AA$-reactions are calculated using CRN6 code by \citet{BarPol1994}, or using optional parameterizations by \citet{1983ApJS...51..271L} or by \citet[][with corrections provided by the authors]{1996PhRvC..54.1329W} and $A$-scaling dependencies.

Though the overall process was very laborious and often impossible to automate, it produced probably the most accurate package (nuc\_package.cc and auxiliary files) for massive calculations of the production nuclear cross sections so far. Since it is the core part of \galprop{}, it was used in numerous studies where the \galprop{} code was employed. It was also used in many studies of the accuracy of the isotopic production cross sections employed in astrophysical applications \citep[e.g.,][]{2015arXiv151009212T,2018PhRvC..98c4611G, 2019PhRvD..99j3023E}, and in other Galactic propagation codes, such as, e.g., Diffusion of cosmic RAys in galaxy modelizatiON code \citep[DRAGON,][]{2008JCAP...10..018E, 2016JCAP...04E.001E}. A more recent attempt to characterize the uncertainties in the calculation of the isotopic production cross sections was made in the framework of the ongoing ISOtopic PROduction Cross Sections (ISOPROCS) project \citep{2011ICRC....6..283M, 2013ICRC...33..803M}.

Production of secondary particles in \galprop{} is calculated taking into account $pp$-, $pA$-, $Ap$-, and $AA$-reactions. Calculations of $\bar{p}$ production and propagation are detailed in \citet{2002ApJ...565..280M,2003ApJ...586.1050M},  \citet{2015ApJ...803...54K}, and \citet{2019CoPhC.24506846K}, while inelastically scattered (tertiary) $\bar{p}$ and (secondary) $p$ are treated as separate species due to the catastrophic energy losses. Production of neutral mesons ($\pi^0$, $K^0$, $\bar{K}^0$, etc.), and secondary $e^\pm$ is calculated using the formalism by \citet{1986A&A...157..223D, 1986ApJ...307...47D} as described in \citep{1998ApJ...493..694M} or recent parameterizations \citep{2006ApJ...647..692K, 2012PhRvD..86d3004K, 2014ApJ...789..136K, 2019CoPhC.24506846K}. 

Production of \gray{s} is calculated using the propagated CR distributions, including primary $e^-$, secondary $e^\pm$, and knock-on $e^-$, as well as inelastically scattered (secondary) protons \citep{2004ApJ...613..962S, 2008ApJ...682..400P}. The inverse Compton scattering is treated using the formalism for an anisotropic background photon distribution \citep{2000ApJ...528..357M} with full Galactic interstellar radiation field on the 2D or 3D grid \citep{2006ApJ...640L.155M, 2006ApJ...648L..29P}. Electron bremsstrahlung cross section is calculated as described in \citet{2000ApJ...537..763S}. Gas-related \gray{} intensities ($\pi^0$-decay, bremsstrahlung) are computed from the emissivities using the column densities of \htwo{} + \hi{} (+ \hii{}, ionized hydrogen) gas for Galactocentric annuli based on 2.6-mm carbon monoxide CO (a tracer of molecular hydrogen \htwo{}) and 21-cm \hi{} (atomic hydrogen) survey data. The synchrotron emission\footnote{\galprop{} calculations of the foreground synchrotron emission were used by the \planck{} Collaboration \citep{2016A&A...594A..10P,2016A&A...594A..25P} to study anisotropies in Cosmic Microwave Background (CMB) with many important implications for the DM studies.} and its polarization are computed \citep{2013MNRAS.436.2127O} using published models of the Galactic magnetic field for regular, random, and striated components \citep{2008A&A...477..573S, 2010RAA....10.1287S, 2011ApJ...738..192P, 2012ApJ...761L..11J}. The line-of-sight integration of the corresponding emissivities with the distributions of gas, interstellar radiation and magnetic fields yields \gray{} and synchrotron sky maps. Spectra of CR species and the \gray{} and synchrotron sky maps are output in standard astronomical formats.

Similarly to ordinary CR species and their diffuse emissions, {\it \galprop{} has well-developed options to propagate particles produced in {\it exotic sources and processes}}, such as annihilation or decay of DM particles, and calculate the associated emissions (DM \gray{} and synchrotron skymaps). It can be used alone or run in conjunction with dedicated packages, such as DarkSUSY; the appropriate interface is also provided. 

Recent updates and developments of the \galprop{} code are detailed in \citet{2017ApJ...846...67P, 2019ApJ...887..250P}, and \citet{2018ApJ...856...45J, 2019ApJ...879...91J}. All details on \galprop{} including the description of all involved processes and reactions can be found in dedicated publications \citep{1997A&A...325..401M,1998ApJ...493..694M, 2000ApJ...528..357M,1998ApJ...509..212S,  2000ApJ...537..763S, 2004ApJ...613..962S, 2007ARNPS..57..285S, 2010ApJ...722L..58S,2011A&A...534A..54S, 2002ApJ...565..280M, 2003ApJ...586.1050M, PoS(ICRC2017)279, 2006ApJ...642..902P, 2011CoPhC.182.1156V, 2012ApJ...752...68V, 2013MNRAS.436.2127O, 2018PhRvC..98c4611G}.

\subsection{\helmod{} Model for heliospheric transport}\label{Sect::Helmod}
%%%%%%%%%%%%%%%%%%%%%%%%%%%%%%%%%%%%%%%%%%%%%%%%%%%
%%%%%%%%%%%%%%%%%%%%%%%%%%%%%%%%%%%%%%%%%%%%%%%%%%%

Before reaching the Earth orbit, CR particles pass through the interplanetary medium, called \textit{heliosphere}. Though many processes are similar to the Galactic propagation, heliosphere has its own specifics. Solar activity changes on weekly and monthly scales, the solar wind and magnetic field varies with time and position in space, and therefore requires a dedicated modeling to understand all factors involved \citep[see discussion in][]{2017ApJ...840..115B}.  The heliospheric propagation of CR species leads to the suppression of the particle flux below $\la$50 GV; this phenomenon is called the solar modulation and the strength of the suppression depends on the solar activity, particle charge sign, polarity of the solar magnetic field and other conditions. In this work, the particle transport within the heliosphere, is treated by means of \helmod{} model \citep[and reference therein]{2019HelMod}. The \helmod{} model, now version 4.0 (\helmod{}-4), numerically solves the \citet{1965P&SS...13....9P} transport equation\footnote{
Parker's equation describes the particle transport through the heliosphere as long as particle gyration radii are not too large with respect to the size of irregularities of the interplanetary magnetic field. For larger or much larger radii, i.e., for particle rigidities above several tens of GV, a simple diffusive approximation becomes invalid, and finally particles propagate in the ballistic regime \citep[e.g., see][]{2017PhRvD..95b3007M}. On the other hand, above several tens of GV the heliospheric modulation becomes negligible implying that we see the LIS. In turn, the LIS is calculated using GalProp that employs the diffusive approximation on a much larger Galactic scale. 
},
\begin{align}
\label{EQ::FPE}
 \frac{\partial U}{\partial t}= &\frac{\partial}{\partial x_i} \left( K^S_{ij}\frac{\partial \mathrm{U} }{\partial x_j}\right)\\
&+\frac{1}{3}\frac{\partial V_{ \mathrm{sw},i} }{\partial x_i} \frac{\partial }{\partial T}\left(\alpha_{\mathrm{rel} }T\mathrm{U} \right)
- \frac{\partial}{\partial x_i} [ (V_{ \mathrm{sw},i}+v_{d,i})\mathrm{U}],\nonumber
\end{align}
using a Monte Carlo approach involving stochastic differential equations \citep[see a discussion in, e.g.,][and references therein]{Bobik2011ApJ,BobikEtAl2016}. Here $U$ is the number density of CR species per unit of kinetic energy $T$, $t$ is the time, $V_{ \mathrm{sw},i}$ is the solar wind velocity along the axis $x_i$, $K^S_{ij}$ is the symmetric part of the diffusion tensor, $v_{d,i}$ is the particle magnetic drift velocity (related to the antisymmetric part of the diffusion tensor), and finally $\alpha_{\mathrm{rel} }=\frac{T+2m_r c^2}{T+m_r c^2} $, with $m_r$ the particle rest mass in units of GeV/nucleon. Parker's transport equation describes: (i) the \textit{diffusion} of CR species due to magnetic irregularities, (ii) the so-called \textit{adiabatic-energy changes} associated with expansions and compressions of cosmic radiation, (iii) an \textit{effective convection} resulting from the convection with \textit{solar wind} (SW, with velocity $\vec{V}_{{\rm sw}}$), and (iv) the drift effects related to the \textit{drift velocity} ($\vec{v}_d$).

The particle transport within the heliosphere is computed from the outer boundary (i.e.\ the heliopause) down to Earth orbit. In the latest version the actual dimensions of the heliosphere and its boundaries were taken into account based on Voyager~1 measurements \citep{2019HelMod}. The heliopause (HP) represents the extreme limit beyond which solar modulation does not affect the CR flux. Thus, the CR spectra measured by Voyager 1 outside HP are the truly pristine LIS of CR species\footnote{
The spectra measured by Voyager probes beyond the HP can be regarded as truly interstellar spectra considering that (i) Voyager 2 is now in the interstellar space \citep{2019NatAs...3.1013S} confirming the data from its sister spacecraft Voyager 1 while being $>$170 AU apart, and (ii) using the data from the IBEX spacecraft, \citet{2012Sci...336.1291M} found a ``bow wave'' of enhanced density {\it instead of a bow shock,} and a broadened H wall ahead of the heliosphere. 
}. 
Using the Parker's model of the heliosphere \citep{1961ApJ...134...20P, 1963idp..book.....P} in combination with Voyager~1 observations, we were able to estimate the time dependence of the positions of the termination shock (TS, $R_{\rm TS}$) and the HP ($R_{\rm HP}$) as \citep{2019HelMod}:
\begin{align}
R_{\rm TS} &= R_{\rm obs} \left( \frac{\rho_{\rm obs} u_{\rm obs}^2}{P_{\rm ISM}}
\right)^{\frac{1}{2}} \left[ \frac{\gamma+3}{2(\gamma+1)}
\right]^{\frac{1}{2}}, \\
\frac{R_{\rm HP}}{R'_{\rm TS}}&=1.58 \pm 0.05,
\label{RTS.incompr}
\end{align}
where $\rho_{\rm obs}$ and $u_{\rm obs}$ are respectively plasma density and plasma velocity measured \emph{in situ} at distance $R_{\rm obs}$, $P_{\rm ISM}$ is the stagnation pressure discussed in section 4 in \citet{2019HelMod}, and $\gamma=5/3$. $R'_{\rm TS}$ is defined as the TS position at the time when it was left by the SW stream that is currently reaching the HP \citep[for more details see][]{2019HelMod}; this typically takes about 4 years, but depends on the SW speed. Therefore, the actual dimensions of the heliosphere used in \helmod{}-4 evolve with time. The predicted TS distances are in good agreement with those observed: for Voyager 1 (Voyager 2) the detected TS position\footnote{The predicted TS is also compatible with the {\it putative near-TS crossing} by Voyager 1 on 1 August 2002 at heliocentric distance slightly larger than 85 AU \citep{2003Natur.426...45K}. In fact, since the spacecraft speed is slower than that of solar wind, the probe might experience several TS crossings.} is 93.8 AU on 16 December 2004 (83.6 AU on 30 August 2007) and the predicted is 91.8 AU (86.3 AU), i.e.\ within 3 AU. Regarding the HP, based on the $R_{\rm HP}$ observed by Voyager 1, the predicted $R_{\rm HP}$ at the time of the Voyager 2 crossing was 120.7 AU on 5 November 2018 while in reality it is 119 AU \citep[see also,][]{2019NatAs...3..997K, 2019GeoRL..46.7911D}. 

In the present code, particular attention is paid to the quality of description of the high solar activity periods, which is evaluated though a comparison of \helmod{} calculations and the CR proton data by AMS-02 \citep{2018PhRvL.pHe}, and to transitions from/to solar minima. This was achieved through introduction of a drift suppression factor and particle diffusion parameters which depend of the level of solar disturbances \citep[see a discussion in][]{2019HelMod}.    

\section{HEAO-3-C2 data and CR transport}\label{calcs} 
%%%%%%%%%%%%%%%%%%%%%%%%%%%%%%%%%%%%%%%%%%%%%%%%%%%
%%%%%%%%%%%%%%%%%%%%%%%%%%%%%%%%%%%%%%%%%%%%%%%%%%%

Observations of new features in the CR proton and He spectra in the energy range that is deemed well-studied by PAMELA \citep{2011Sci...332...69A} and their confirmation by \fermilat{} \citep{2014PhRvL.112o1103A} emphasize the importance of high statistics and accuracy in CR studies. Precise measurements by AMS-02 experiment \citep{2015PhRvL.114q1103A, 2015PhRvL.115u1101A} provided important details about the observed hardening and extended the accurate measurements of proton and He spectra into the TV rigidity range (see also: ATIC-2 -- \citealt{2009BRASP..73..564P}, and CREAM -- \citealt{2010ApJ...714L..89A}). Besides the hardening observed at the same rigidity ($\sim$370 GV) and the same spectral index change for both species, the He spectrum appears to be flatter than the spectrum of protons in the whole range. Consequently, the observed p/He ratio is smooth and monotonically decreasing with rigidity.  

Understanding the nature of these features requires accurate measurements of other CR species, which was accomplished by AMS-02. The detailed spectra of heavier species $_3$Li--\,$_8$O \citep{2017PhRvL.119y1101A, 2018PhRvL.120b1101A, 2018PhRvL.121e1103A}, and $_{10}$Ne, $_{12}$Mg, $_{14}$Si \citep{PhysRevLett.124.211102}, exhibit the hardening (breaks) similar to that observed in the spectra of CR protons and He, where the break rigidity $\sim$370 GV is about the same for all species. Observations show similarity between the spectra of mostly primary (p, He, C, O, Ne, Mg, Si) and secondary (Li, Be, B) nuclei, while the spectral slopes of these groups of nuclei are different. Nitrogen is about half-primary--half-secondary and behaves as being in between the other two groups.

Meanwhile, getting into the higher $Z$ species is important as they provide complementary information about properties of CR sources and ISM in our local Galactic environment.  Especially important are the spectra of the iron group nuclei, as they have large fragmentation cross sections, and, therefore, their fragmentation time scale at relatively low energies is shorter than the timescale of their escape from the Galaxy. Such nuclei can reach us only if accelerated in nearby sources. However, it may take a while to get such data from AMS-02 which is currently collecting statistics. In this energy range available are data of earlier missions, such as HEAO-3-C2 \citep{1990A&A...233...96E}, ATIC-2 \citep{2009BRASP..73..564P}, CREAM \citep{2008APh....30..133A, 2009ApJ...707..593A}, and NUCLEON \citep{2019AdSpR..64.2546G, 2019AdSpR..64.2559G}, while at low energies Voyager~1 \citep{2016ApJ...831...18C} and ACE-CRIS can be used. These data do not provide a continuous coverage comparable in quality to the AMS-02 data, but can be used to deduce the LIS of $_9$F--\,$_{28}$Ni nuclei --- with some caveats discussed below.

A comparison of the available AMS-02 data with HEAO-3-C2 measurements using our derived LIS modulated to the corresponding level of solar activity with the previous version of \helmod{} (\helmod{}-3) gives an important insight into the accuracy of the calibration technique used in 1970s. In particular, the precise measurements of CR nuclei $_4$Be--\,$_8$O by AMS-02 experiment \citep{2017PhRvL.119y1101A, 2018PhRvL.120b1101A, 2018PhRvL.121e1103A} indicate that there are clear discrepancies with HEAO-3-C2 data at low and high energies \citep[see Figure 7 in][]{2018ApJ...858...61B}, while in the middle range between 2.65 and 10.6 GeV nucleon$^{-1}$ the agreement is fair. Calculations repeated with \galprop{}--\helmod{}-4 framework yield the LIS consistent with our previous results within 1--2\% and confirm our previous findings. The likely reasons of these systematic discrepancies in HEAO-3-C2 data at low and high energies are discussed in detail in Section~\ref{heao3} in the Appendix. For details and differences between \helmod{}-3 and \helmod{}-4, see a description in \citet{2019AdSpR..64.2459B} and at the dedicated Web-site\textsuperscript{\ref{helmod-footnote}}.  

In our following analysis we are using the middle range of HEAO-3-C2 data 2.65--10.6 GeV nucleon$^{-1}$, we call it the ``plateau'' because it mimics a plateau in the spectral residual plots, while the data outside this range are discarded. Here we provide the details of the calibration procedure we used for the HEAO-3-C2  \citep{1990A&A...233...96E} data using AMS-02 measurements of the CR $_4$Be--\,$_8$O, $_{10}$Ne, $_{12}$Mg, and $_{14}$Si spectra (Section~\ref{calibration}). Calculations of the interstellar and heliospheric transport are described in Section~\ref{prop}. 

\subsection{Calibration of HEAO-3-C2 spectra with AMS-02 data}\label{calibration}
%%%%%%%%%%%%%%%%%%%%%%%%%%%%%%%%%%%%%%%%%%%%%%%%%%%

As already mentioned, the HEAO-3-C2 energy range is significantly affected by the heliospheric modulation. Therefore, a proper evaluation of the systematic errors of the HEAO-3-C2 data can be done through their comparison with LIS, tuned to the AMS-02 data, that are modulated appropriately to the solar activity observed during the HEAO-3-C2 flight as it was done in \citet{2018ApJ...858...61B, 2020ApJ...889..167B}. The LIS spectra tuned to AMS-02 data clearly overlap with the middle part of the HEAO-3-C2 data, exhibiting a flat region at a few GeV nucleon$^{-1}$ in the relative difference plots. This proper range from 2.65 GeV nucleon$^{-1}$ to 10.6 GeV nucleon$^{-1}$ is referred in the present work to as the ``plateau.'' A numerical procedure elaborated with the HEAO-3-C2 dataset proves that in the wider energy region outside the ``plateau'' there is a systematic discrepancy between the dataset and calculated spectra that is similar in the spectra of different elements and that is likely a calibration artifact.

\begin{figure*}[ptb!]
	\centering
	\includegraphics[width=0.98\textwidth]{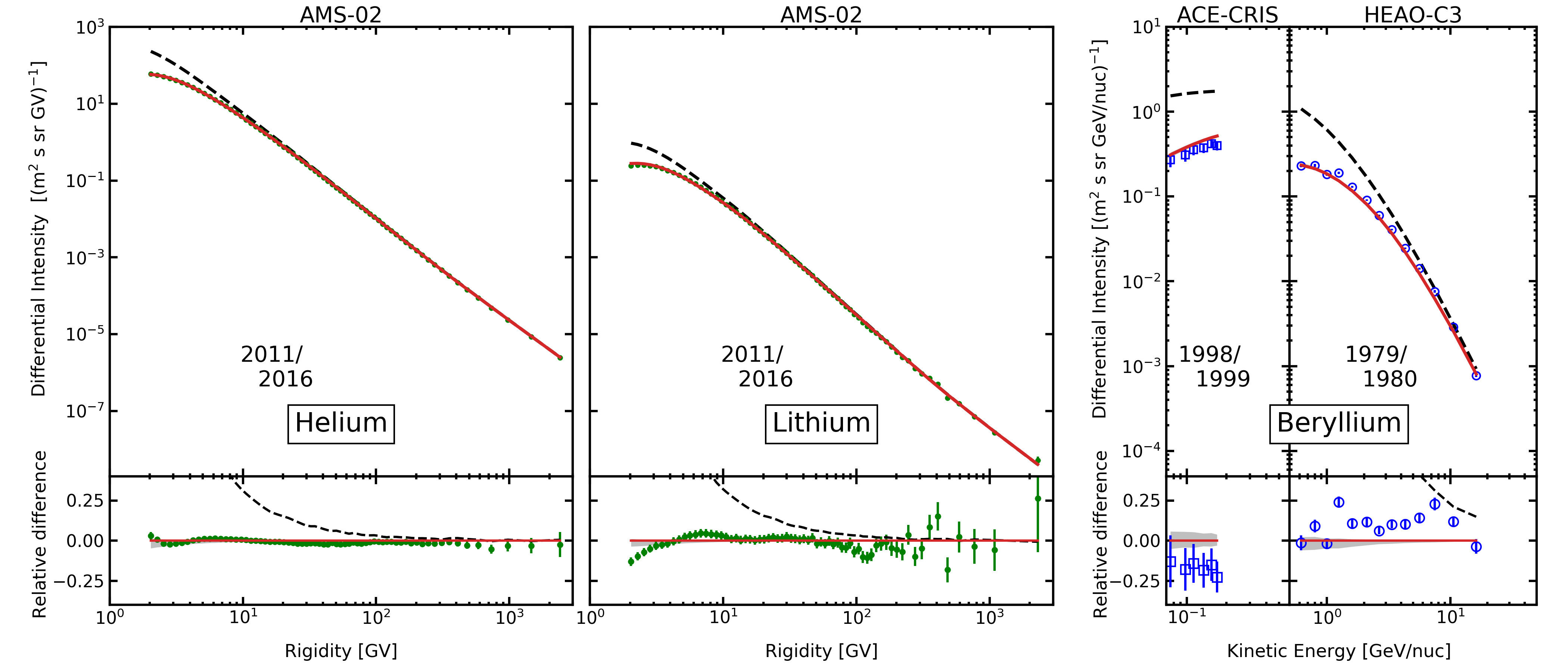}
	\includegraphics[width=0.98\textwidth]{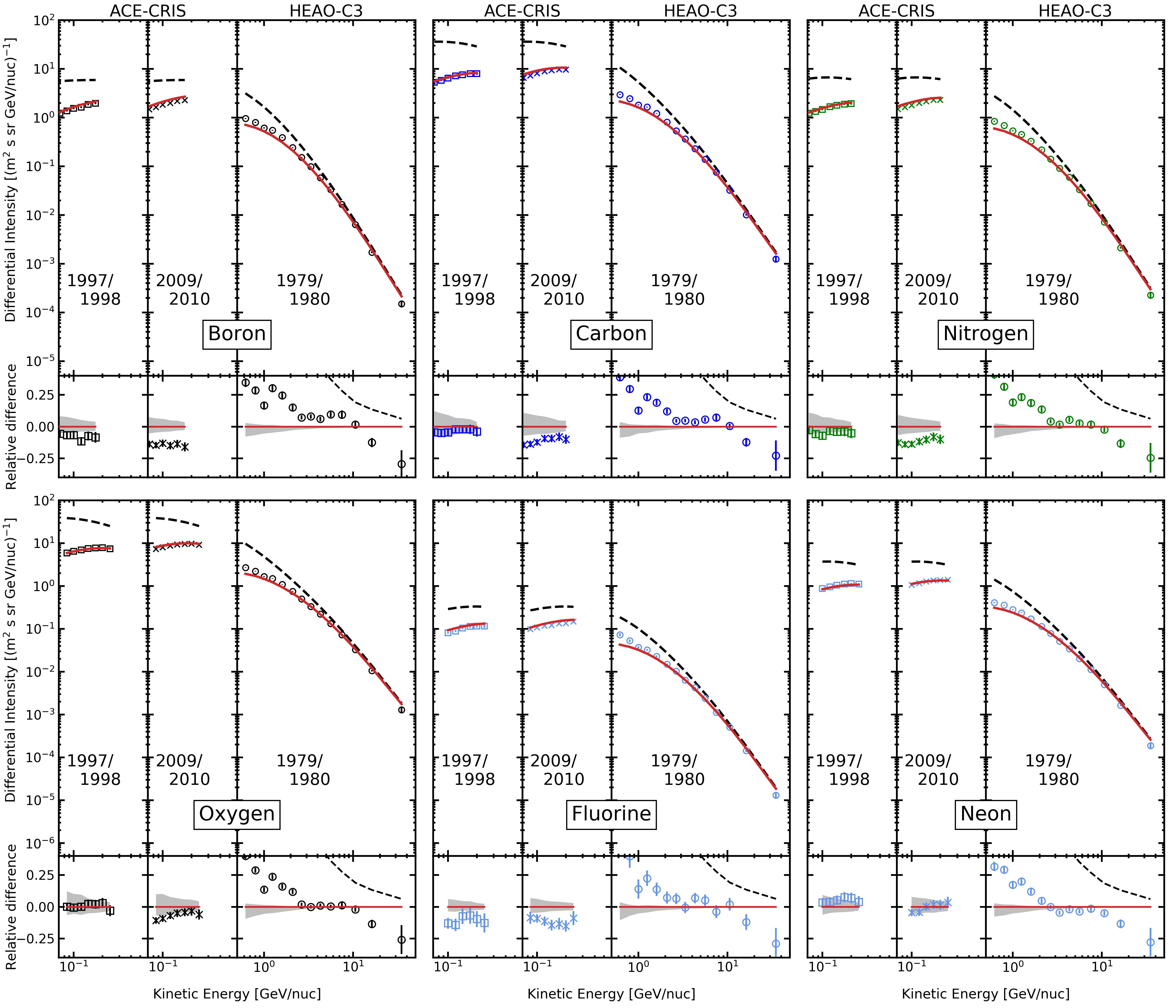}
	\caption{
	Calculated elemental spectra: $_2$He--\,$_{10}$Ne. Black dashed lines show the calculated LIS spectra, the red solid lines are modulated to the levels that correspond to the periods of data taking. Data for $Z\ge4$: ACE-CRIS and HEAO-3-C2 \citep{1990A&A...233...96E}. AMS-02 data for $_2$He and $_3$Li are compared to the {\it I}-scenario calculations, see Section~\ref{prop} for details. Bottom panels in each plot show the relative difference between the calculations and a corresponding data set.
	}
	\label{fig:He-Ne}
\end{figure*}

\begin{figure*}[ptb!]
	\centering
	\includegraphics[width=0.98\textwidth]{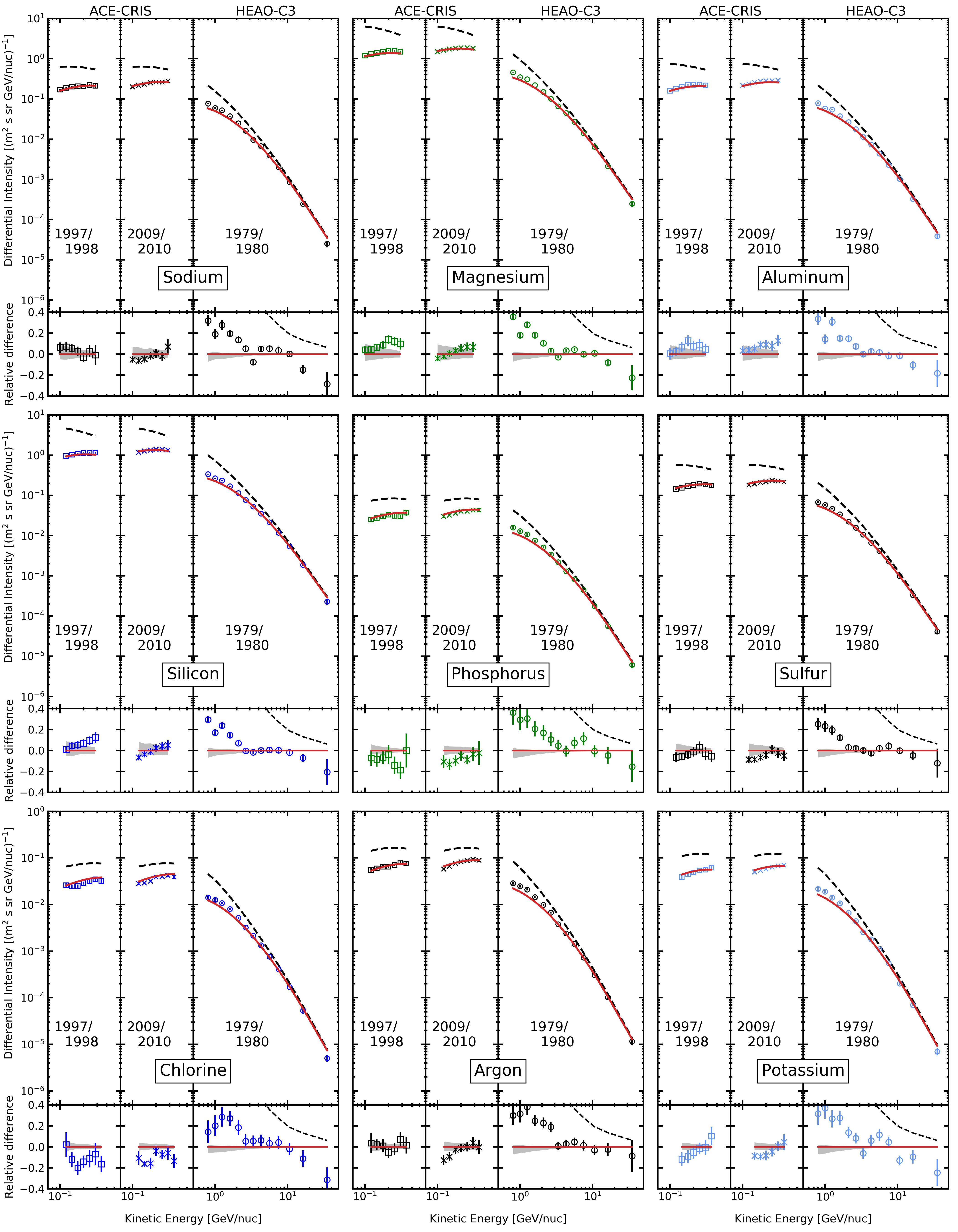}
	\caption{Calculated elemental spectra: $_{11}$Na--\,$_{19}$K. The line coding and data as in Figure \ref{fig:He-Ne}.
	}
	\label{fig:Na-K}
\end{figure*}

\begin{figure*}[tb!]
	\centering
	\includegraphics[width=0.33\textwidth]{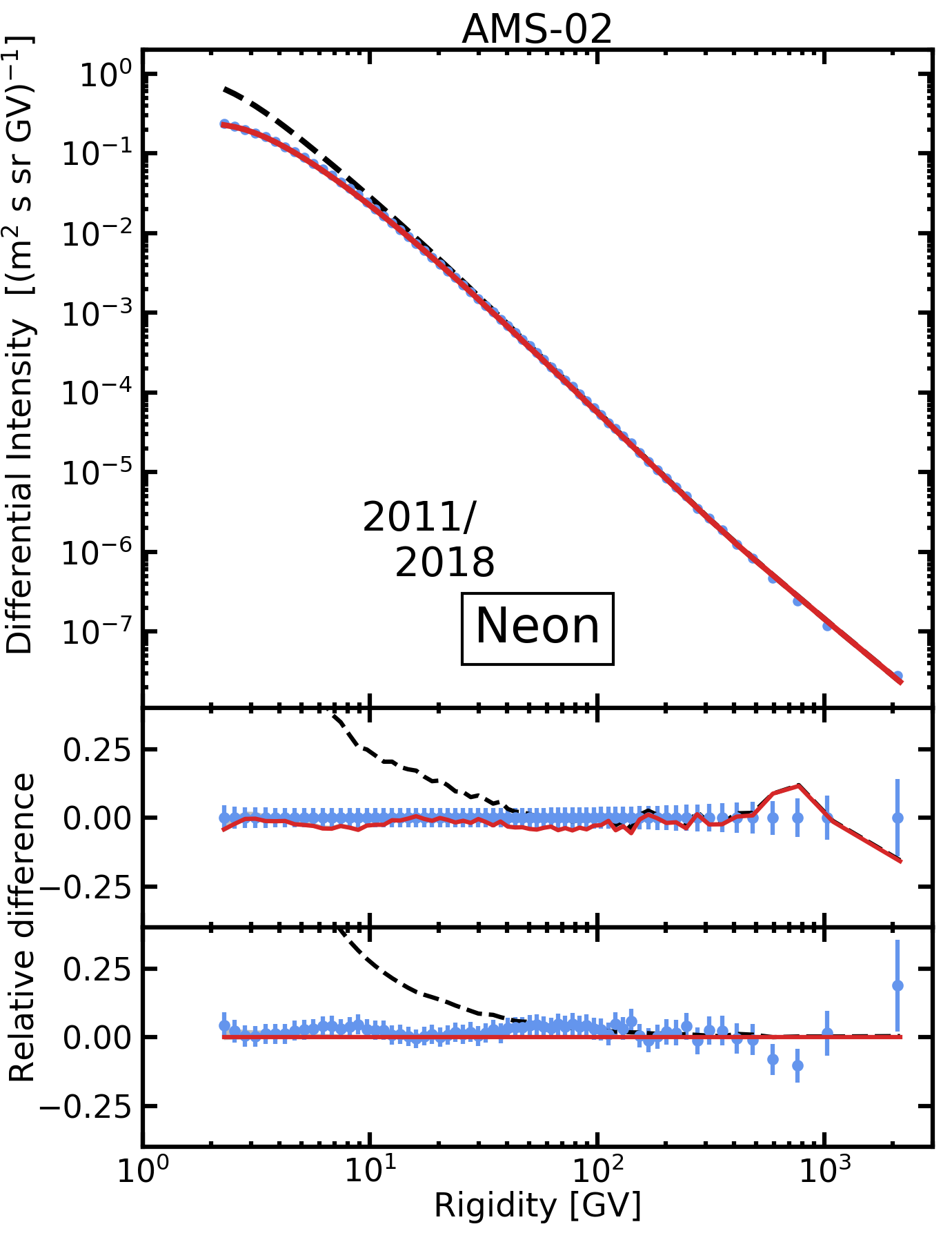}
	\includegraphics[width=0.33\textwidth]{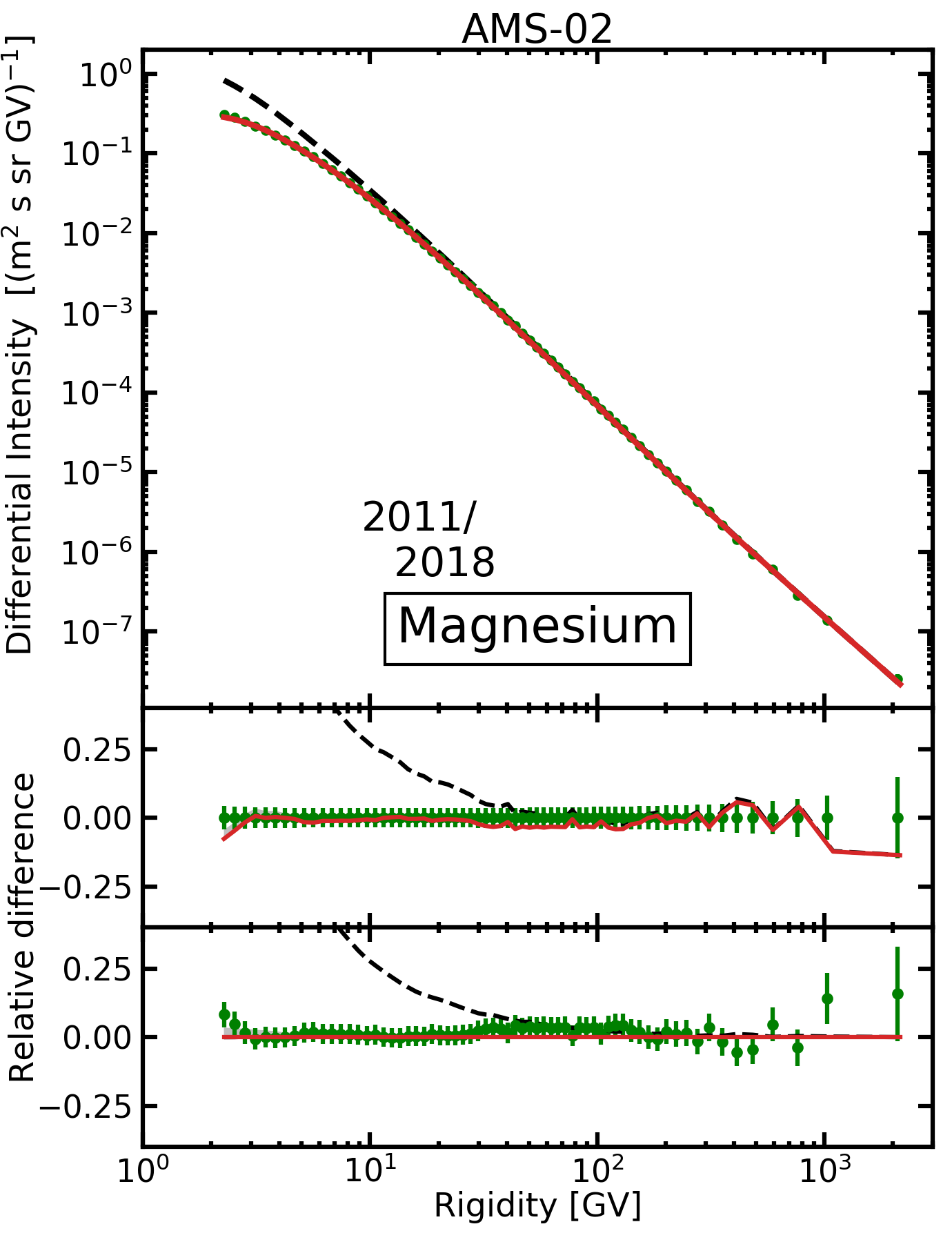}
	\includegraphics[width=0.33\textwidth]{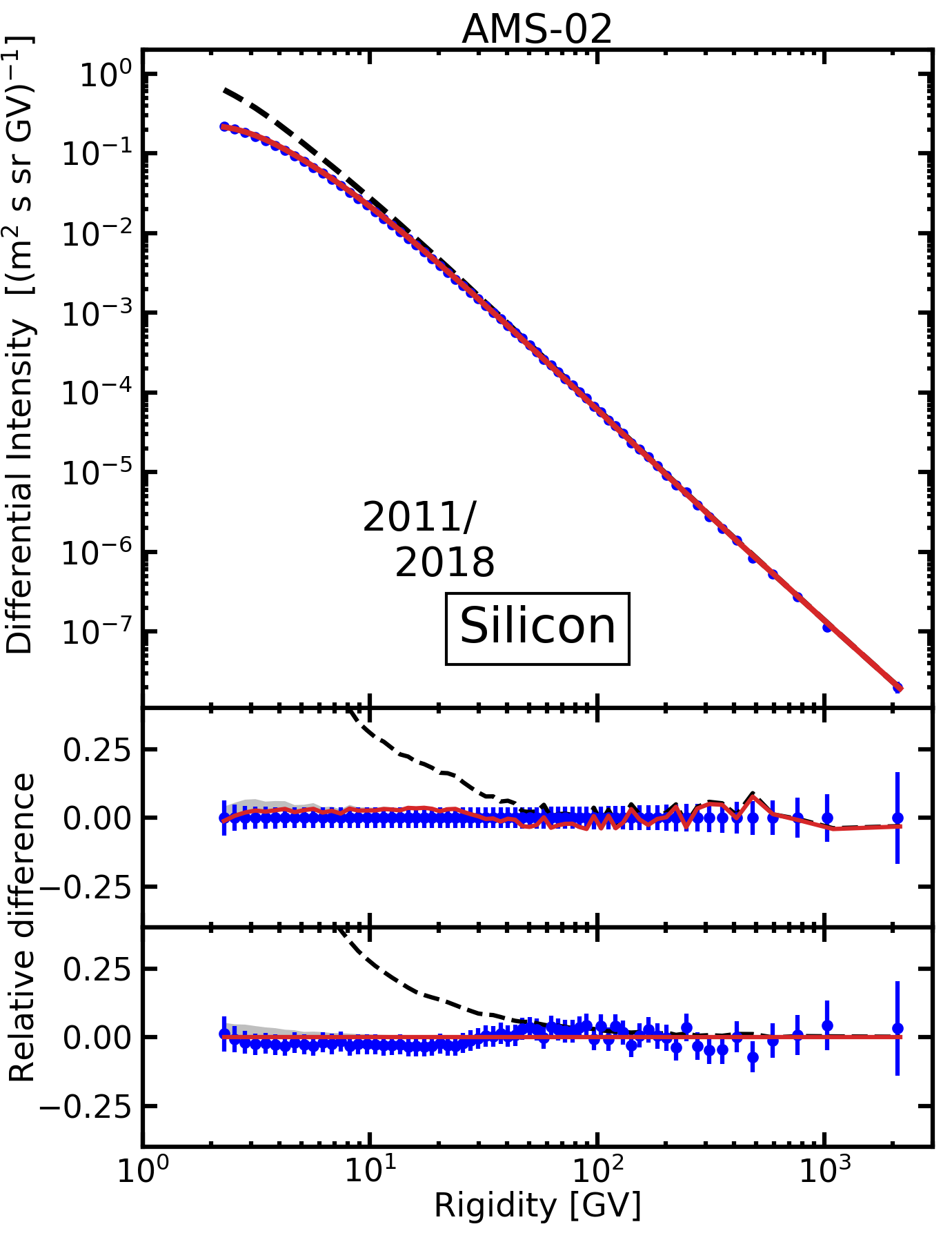}
	\caption{The \galprop{}--\helmod{} LIS and modulated spectra are compared with AMS-02 data for $_{10}$Ne, $_{12}$Mg, and $_{14}$Si \citep{PhysRevLett.124.211102}. Shown are the calculations made in the {\it I}-scenario. The lower panels show the relative difference in two ways: AMS-02-data-centered as in our previous papers, and in a more traditional model-predictions-centered view. Line coding as in Figure~\ref{fig:He-Ne}. 
	}
	\label{fig:Ne-Si}
\end{figure*}

The HEAO-3-C2 spectra of $_4$Be through $_{10}$Ne are shown in Figure~\ref{fig:He-Ne}, where we also show the spectra of $_2$He and $_3$Li for completeness. The HEAO-3-C2 spectra of $_{11}$Na through $_{19}$K are shown in Figure~\ref{fig:Na-K}. Figure \ref{fig:Ne-Si} shows our LIS and modulated spectra for $_{10}$Ne, $_{12}$Mg, and $_{14}$Si derived using the latest AMS-02 data \citep{PhysRevLett.124.211102}. Shown are the calculations made in the {\it I}-scenario, the {\it P}-scenario calculations look similarly. 

Only $Z\ge4$ nuclei (Be and heavier) were measured by HEAO-3-C2 and the time interval corresponds to October 1979 through June 1980 \citep{1990A&A...233...96E}. One can see the flat region from 2.65 GeV nucleon$^{-1}$ through 10.6 GeV nucleon$^{-1}$ in the relative difference plots of the spectra of B, C, N, O, Ne, Mg, and Si nuclei, which corresponds to the aerogel block counter (up to 5.60 GeV nucleon$^{-1}$) and the first two energy bins of the aerogel sand counter (for a detailed description see Section~\ref{heao3} in the Appendix). In this energy interval the spectra of N, O, Ne, Mg, and Si nuclei measured by HEAO-3-C2 and our modulated LIS agree very well. Meanwhile, the absolute normalizations of the spectra of B and C nuclei are slightly off by $\la$5--8\%, while the spectrum of Be demonstrates worst agreement still being within $\la$15\% of our calculations. The latter is not surprising since Be is the lightest nucleus measured by HEAO-3-C2, and given that ``The time-of-flight system is performing well for $Z>6$. For lowest charges the resolution is rather poor...'' -- \citet{1982Ap&SS..84....3B}, see discussion in Section \ref{counters}. 

Overall, the differences in normalizations between the appropriately modulated AMS-02 and HEAO-3-C2 ``plateau'' data are quite small, e.g., the average discrepancy for C, N, O, Ne, Mg, Si nuclei is below 1\%. The reference element used in the HEAO-3-C2 data analysis is O and the spectra of the other nuclei were derived from their relative abundances with respect to the reference element \citep[Table 2 in][]{1990A&A...233...96E}. One can see that our calculations for O agree well with HEAO-3-C2 data and, therefore, in our further analysis we rely on the HEAO-3-C2 data points in the energy range from 2.65 GeV nucleon$^{-1}$ through 10.6 GeV nucleon$^{-1}$ the ``plateau" region with the data points outside of this range being discarded from further analysis. The HEAO-3-C2 data are used at their nominal values, i.e.\ no renormalization is applied.

\subsection{Source abundances and CR transport}\label{prop} %\section{Normalization procedure with HEAO-3-C2}
%%%%%%%%%%%%%%%%%%%%%%%%%%%%%%%%%%%%%%%%%%%%%%%%%%%

\begin{figure*}[htb!]
	\centering
	\includegraphics[width=0.33\textwidth]{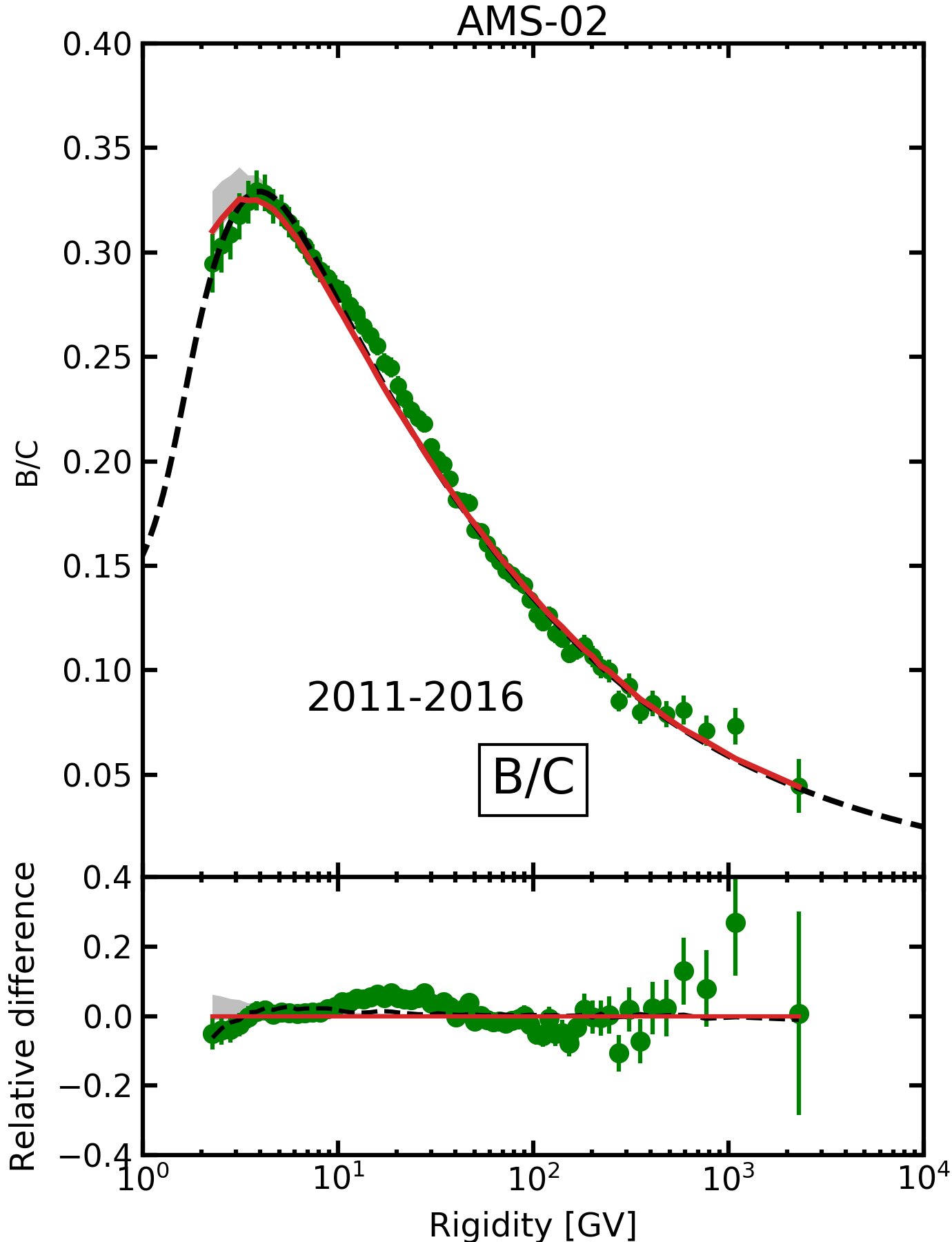}
	\includegraphics[width=0.33\textwidth]{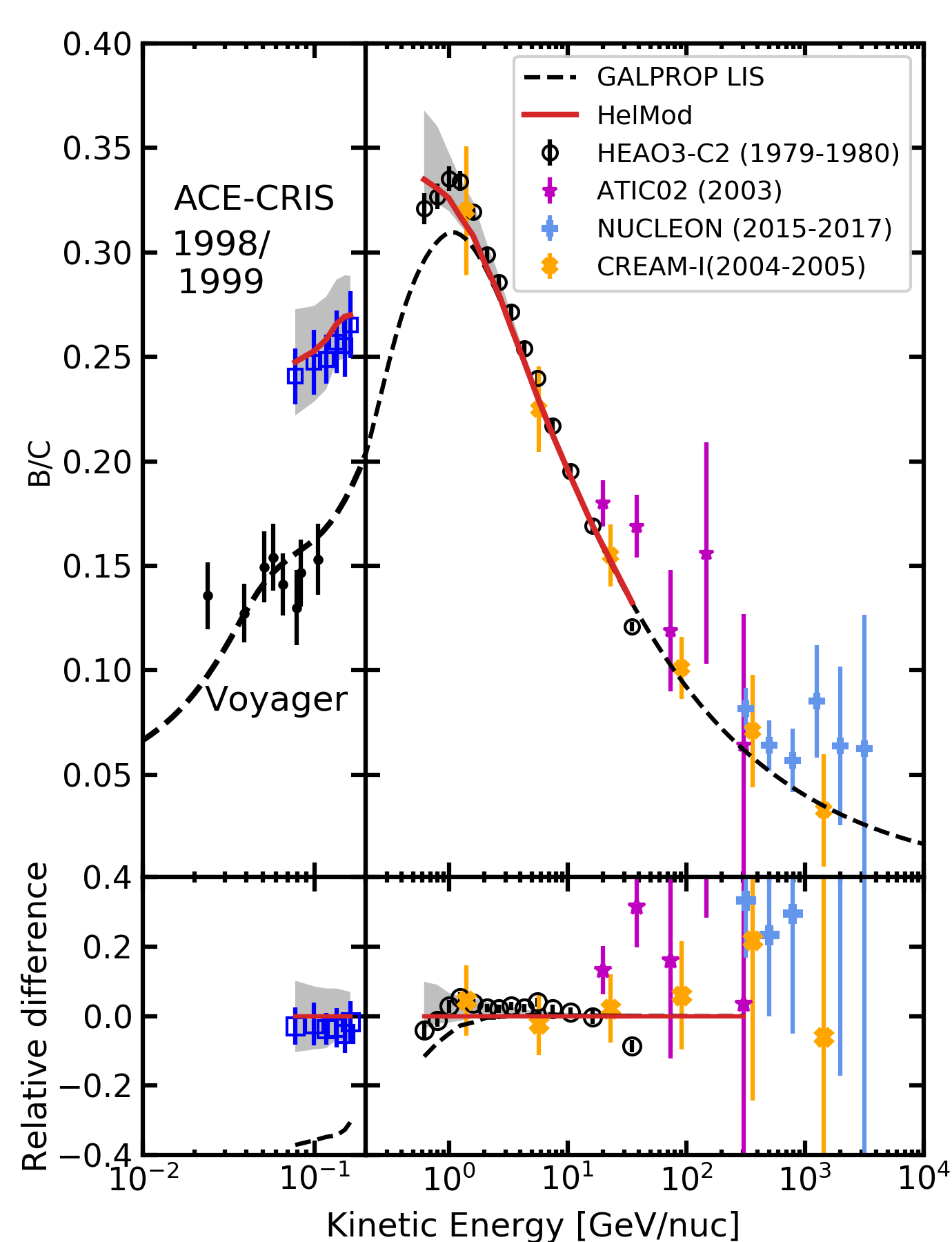}
	\includegraphics[width=0.33\textwidth]{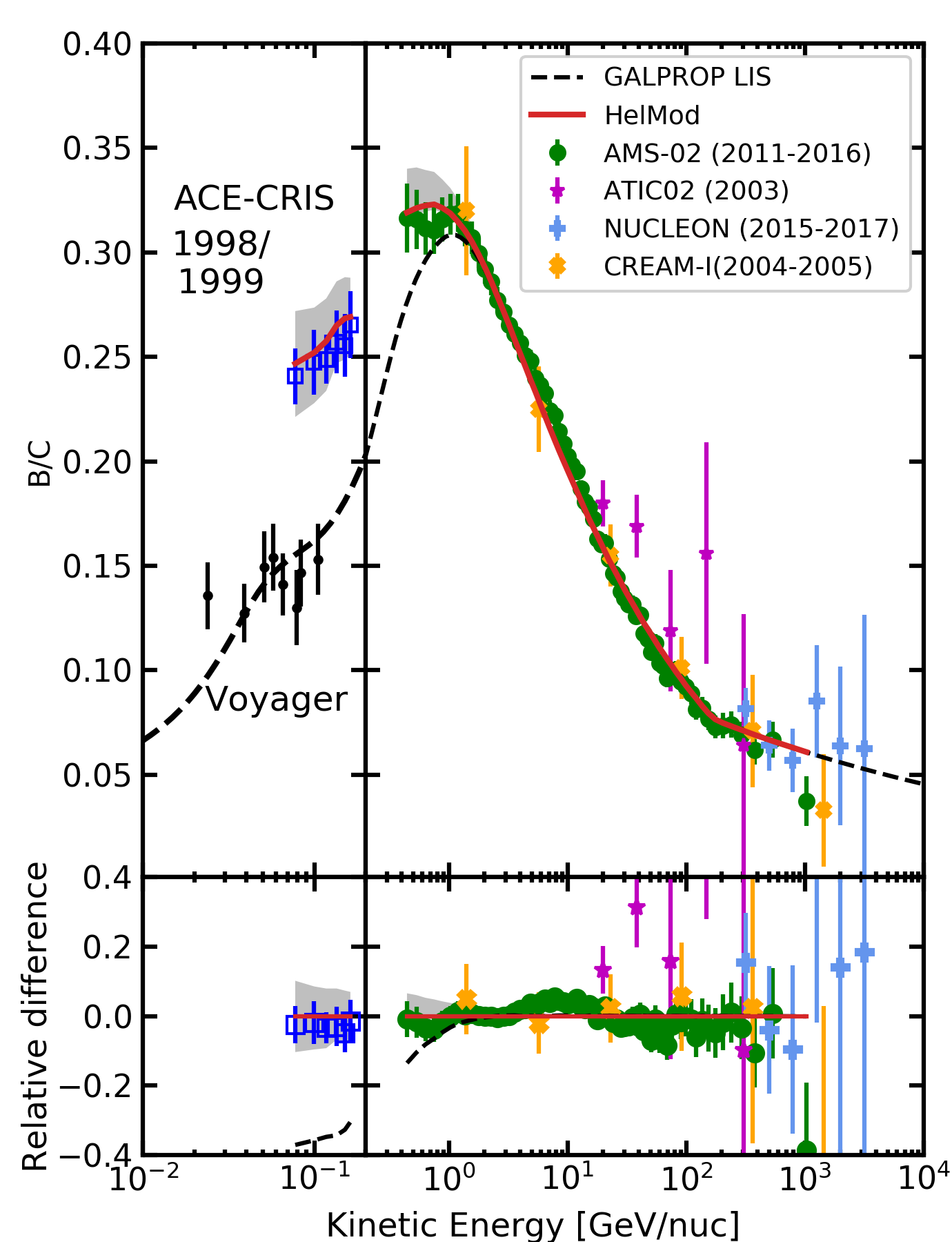}
	\caption{
	The B/C ratio calculated for the {\it I}-scenario is compared to AMS-02 \citep[left,][]{2018PhRvL.120b1101A}  and HEAO-3-C2 \citep[middle,][]{1990A&A...233...96E} data. The right plot shows a comparison of the B/C ratio calculated for the {\it P}-scenario with available measurements, where HEAO-3-C2 data are not shown for clarity. In the middle and right plots shown are the measurements by Voyager~1 \citep{2016ApJ...831...18C}, ACE-CRIS (1998--1999), ATIC-2 \citep{2009BRASP..73..564P}, CREAM \citep{2008APh....30..133A, 2009ApJ...707..593A}, and NUCLEON \citep{2019AdSpR..64.2559G} measurements. Note that the units in the left plot is rigidity, while the middle and right plots are shown vs.\ $E_{\rm kin}$ per nucleon. Line coding as in Figure~\ref{fig:He-Ne}.
	}
	\label{fig:BC}
\end{figure*}

\begin{deluxetable}{rlcc}[tb!]
	\def\arraystretch{0.9}
	\tablewidth{0mm}
	\tablecaption{Best-fit propagation parameters for {\it I}- and {\it P}-scenarios\label{tbl-prop}}
	\tablehead{
		\colhead{Parameter}& \multicolumn{1}{l}{Units}& \colhead{Best Value}& \colhead{Error} 
	}
	\startdata
	$z_h$ & kpc &4.0 &0.6\\
	$D_0 (R= 4\ {\rm GV})$ & cm$^{2}$ s$^{-1}$  & $4.3\times10^{28}$ &0.7\\
	$\delta$\tablenotemark{a} & &0.415 &0.025\\
	$V_{\rm Alf}$ & km s$^{-1}$ &30 &3\\
	$dV_{\rm conv}/dz$ & km s$^{-1}$ kpc$^{-1}$ & 9.8 &0.8
%	$\eta$ && 0.91 
	\enddata
	\tablenotetext{a}{The {\it P}-scenario assumes a break in the diffusion coefficient with index $\delta_1=\delta$ below the break and index $\delta_2=0.15\pm 0.03$ above the break at $R=370\pm 25$ GV, see text for details.}
\end{deluxetable}

Diffusion of CRs in the Galaxy is well-described by the transport equations \citep{1990acr..book.....B}, but the exact values of the propagation parameters depend on the assumed propagation model and selected CR datasets. In this work we are using the same propagation model with distributed reacceleration and convection that was used in our previous analyses \citep[for more details see][]{2017ApJ...840..115B,  2018ApJ...854...94B, 2018ApJ...858...61B, 2020ApJ...889..167B}. 

The values of propagation parameters along with their confidence limits are derived from the best available CR data using the Markov Chain Monte Carlo (MCMC) routine. Five main propagation parameters, that affect the overall shape of CR spectra, were left free in the scan using \galprop{} running in the 2D mode: the Galactic halo half-width $z_h$, the normalization of the diffusion coefficient $D_0$ at the reference rigidity $R=4$ GV and the index of its rigidity dependence $\delta$, the Alfv\'en velocity $V_{\rm Alf}$, and the gradient of the convection velocity $dV_{\rm conv}/dz$ ($V_{\rm conv}=0$ in the plane, $z=0$). Their best-fit values tuned to the AMS-02 data are listed in Table~\ref{tbl-prop} and are the same as obtained in \citet{2020ApJ...889..167B} within the quoted error bars. The radial size of the Galaxy does not significantly affect the values of propagation parameters and was set to 20 kpc. Besides, we introduced a factor $\beta^\eta$ in the diffusion coefficient, where $\beta=v/c$, and $\eta$ was left free. The best fit value of $\eta=0.70$ improves the agreement at low energies, and slightly affects the choice of injection indices ${\gamma}_0$ an ${\gamma}_1$ shown in Table~\ref{tbl-inject}.

Table~\ref{tbl-prop} shows propagation parameters for two scenarios. The ``propagation'' {\it P}-scenario assumes that the observed break at $\sim$370 GV is the result of a change in the spectrum of interstellar turbulence that translates into a break in the index of the diffusion coefficient. The ``injection'' {\it I}-scenario proposes that the break is due to the presence of populations of CR sources injecting particles with softer and harder spectra. The {\it P}-scenario predicts that the break should be observed in spectra of all CR species at about the same rigidity since the interstellar turbulence acts similarly on all particles. The predicted change in the spectral index of secondary species (difference between the spectral indices below and above the break) would then be twice the value of the break observed in the spectra of primary species. The recent AMS-02 measurements of the secondary species Li, Be, and B \citep{2018PhRvL.120b1101A} prefer the {\it P}-scenario. For more details see \citet{2012ApJ...752...68V}, \citet{2020ApJ...889..167B}, and references therein.

The corresponding B/C ratio is shown in Figure~\ref{fig:BC} for both scenarios, which are identical below the break, with the {\it P}-scenario predicting the flatter ratio above the break. Separately shown is a comparison with AMS-02 \citep{2018PhRvL.120b1101A} and HEAO-3-C2 \citep{1990A&A...233...96E}, where the left and middle plots are shown for the {\it I}-scenario, while the right plot shows a comparison for the {\it P}-scenario. The calculated B/C ratio compares well with all available measurements: Voyager~1 \citep{2016ApJ...831...18C}, ACE-CRIS, AMS-02 \citep{2018PhRvL.120b1101A}, ATIC-2 \citep{2009BRASP..73..564P}, CREAM \citep{2008APh....30..133A, 2009ApJ...707..593A}, and NUCLEON \citep{2019AdSpR..64.2559G}. For the most part HEAO-3-C2 \citep{1990A&A...233...96E} data agree well with the modulated LIS ratio, but shows a noticeable discrepancy below 2 GeV nucleon$^{-1}$. The latter supports our conclusion about the unaccounted variations in the rigidity cutoff along the orbit of the instrument that was used for the event selection at low energies (see Sections \ref{cutoff}, \ref{summaryHEAO}).  

To derive the LIS of CR species we use an optimization procedure that was employed in our previous analyses \citep{2017ApJ...840..115B, 2018ApJ...854...94B, 2018ApJ...858...61B, 2020ApJ...889..167B}, but with some modifications. Abundances and injection parameters for nuclei from $_1$H--\,$_8$O were fixed along with the parameters of heliospheric transport as detailed in our previous analysis \citep{2020ApJ...889..167B}. In the current study, these parameters are only slightly changed while CR source abundances for nuclei $_9$F--\,$_{28}$Ni are tuned to the data with the MCMC interface to \galprop{} v.56 \citep{MasiDM2016} through sampling their abundance space. The MCMC scan uses new AMS-02 data for $_{10}$Ne, $_{12}$Mg, $_{14}$Si nuclei \citep{PhysRevLett.124.211102} and for the rest uses the subset of HEAO-2-C2 data between 2.65 GeV nucleon$^{-1}$ and 10.6 GeV nucleon$^{-1}$, the ``plateau'' region as described in Section \ref{calibration}. 

At this step, the same parametrization was assigned to the injection spectra of $_{9}$F, $_{11}$Na, $_{13}$Al, and $_{15}$P--\,$_{28}$Ni nuclei: three break rigidities $R_{0,1,2} = 1.0, 7.0, 350$ GV, and four spectral indices $\gamma_{0,1,2,3} = 1.5, 1.98, 2.48, 2.14$. The break rigidities $R_{0,1,2}$ and two spectral indices $\gamma_{1,2}$ were chosen to be close to the values found in our recent C, N, O analysis \citep{2018ApJ...858...61B, 2020ApJ...889..167B}. Indices ${\gamma}_0$ and ${\gamma}_3$ are used for the energy ranges outside of the HEAO-3-C2 ``plateau'' region: to describe the spectral bending at Voyager 1 energies \citep{2018ApJ...858...61B, 2020ApJ...889..167B} and the well-known now spectral hardening revealed by AMS-02 \citep{2017PhRvL.119y1101A, 2018PhRvL.120b1101A, 2018PhRvL.121e1103A} correspondingly. The fitting procedure is not sensitive to the position of the high rigidity break $R_{2}$, so the choice of the {\it I}-scenario or the {\it P}-scenario with a break in the diffusion coefficient \citep[see][for details]{2020ApJ...889..167B} is a matter of preference as we restrict our analysis to the region below the break. The {\it P}-scenario also implies that $\gamma_3=\gamma_2$.

Consistency with data is provided through manual fine tuning of the source abundances and parameters of the injection spectra for each element from $_{28}$Ni down to $_{6}$C. The default  \galprop{} abundances that were originally tuned to ACE-CRIS data \citep{2001SSRv...99...15W} are now adjusted to match the HEAO-3-C2 ``plateau'' data \citep{1990A&A...233...96E}. Meanwhile, relative isotopic abundances for each element are kept consistent with ACE-CRIS. 

\begin{figure}[tb!]
	\centering
	\includegraphics[width=0.47\textwidth]{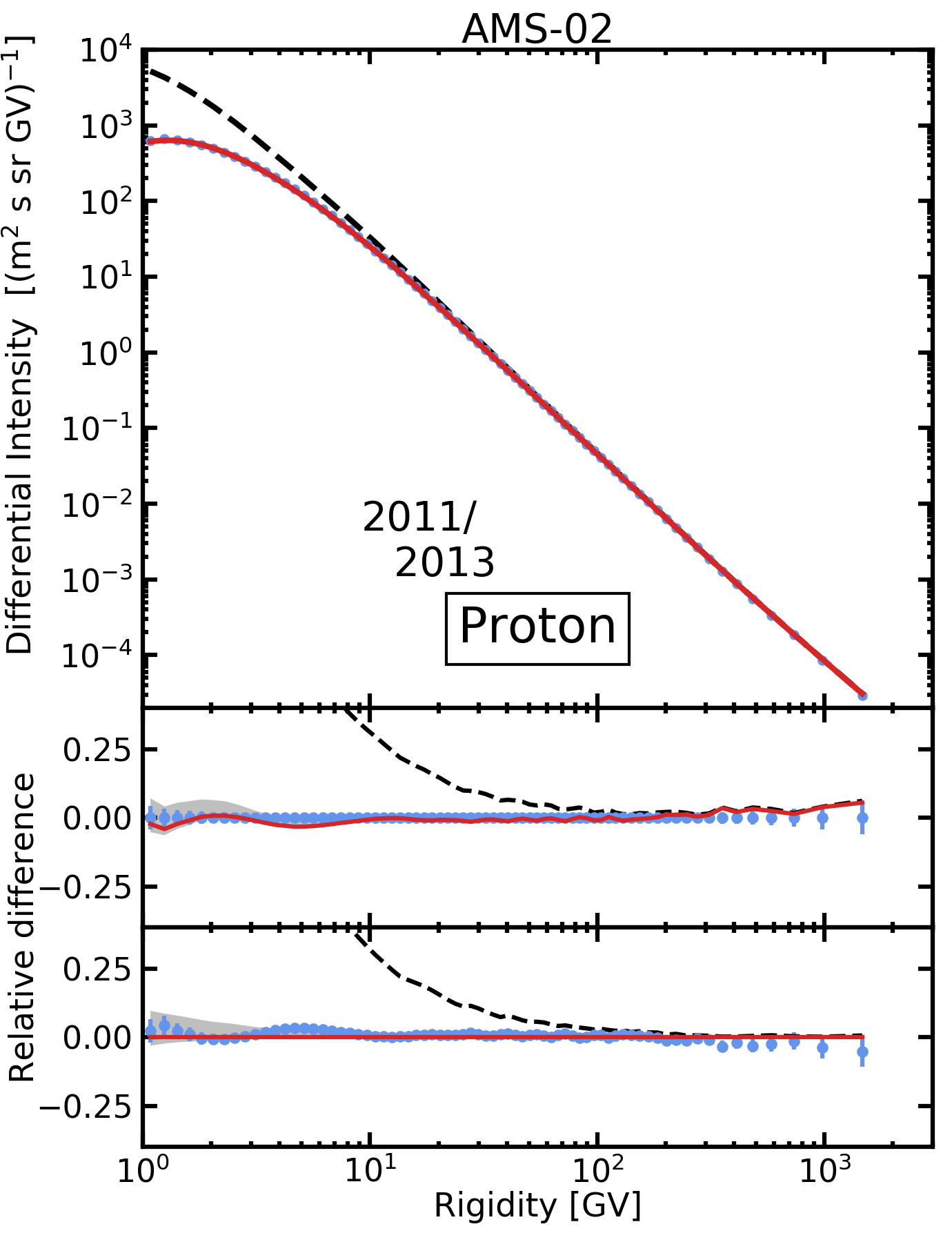}
	\caption{The \galprop{}--\helmod{} LIS (black dashed line) for CR protons calculated in the {\it I}-scenario and modulated spectrum (red solid line) are compared to the AMS-02 data \citep{2015PhRvL.114q1103A}. The lower panels show the relative difference in two ways: AMS-02-data-centered as in our previous papers, and in a more traditional model-predictions-centered view. 
	}
	\label{fig:p}
\end{figure}

At the final step, the ACE-CRIS and Voyager~1 \citep{2016ApJ...831...18C} data were added to the analysis to refine the LIS at low energies. The spectral indices $\gamma_{0,1,2}$ and the break rigidities $R_{0,1}$ were tuned to ACE-CRIS data taken in 1997--1998 (active sun) and 2009--2010 (quiet sun) and to Voyager~1 local interstellar data, which resulted only in minor modifications of the default parametrization set at the first step excluding $\gamma_{0}$ that required larger modifications. Our benchmark CR proton spectrum\footnote{The slightly modified proton LIS does not impact the overall agreement with Voyager~1 and 2 data shown in Figure 6 of \citet{2019AdSpR..64.2459B}.} is shown in Figure~\ref{fig:p} compared to AMS-02 data \citep{2015PhRvL.114q1103A}. Here we again went down in $Z$ from $_{28}$Ni to $_{6}$C. In particular, adjustment of the break rigidity $R_0$ improves the agreement with ACE-CRIS data for $_{10}$Ne--\,$_{14}$Si, $_{20}$Ca, and $_{24}$Cr--\,$_{28}$Ni nuclei. The best fit values for injection parameters of all species are listed in Table~\ref{tbl-inject}, for their definitions see Eq.~(\ref{eq:1}). 

The injection spectra of protons and helium are extended up to 200 TV into the region where their CR spectra exhibit another break (softening) at $\sim$10--20 TeV nucleon$^{-1}$, see Figure~\ref{fig:TeV_H_He}. It adds another break $R_3$ and spectral index $\gamma_4$ above this break, which are valid only in the {\it I}-scenario. CR source abundances are provided in Table~\ref{tbl-Ab}.

\section{Results and discussion}\label{results}
%%%%%%%%%%%%%%%%%%%%%%%%%%%%%%%%%%%%%%%%%%%%%%%%%%%
%%%%%%%%%%%%%%%%%%%%%%%%%%%%%%%%%%%%%%%%%%%%%%%%%%%

Figure~\ref{fig:abund1} shows a comparison of the derived elemental abundances in CR sources and propagated abundances at 100 MeV nucleon$^{-1}$ and 10 GeV nucleon$^{-1}$ normalized to Si = 100. The source abundances were derived using the AMS-02 data for $_1$H--\,$_8$O, $_{10}$Ne, $_{12}$Mg, and $_{14}$Si and HEAO-3-C2 data at 10 GeV nucleon$^{-1}$ for the rest. One can see that the {\it relative} source abundances of mostly primary elements, C, O, Ne, Mg, Si, S, Fe, and Ni remain unchanged after the propagation. The purely secondary elements are Be, B, and V. There are also partially primary/secondary species with a wide range of primary contributions, ranging from about half-and-half for N, Na, Al, Ca, Cr, and Mn nuclei to very low-abundant species that are still present in the sources, such as Li, F, P, Cl, Ar, K, Sc, and Ti. One exception is Co, which looks like mostly primary in this Figure, but it may be connected with somewhat underestimated contributions from fragmentations of heavier ($Z\ge29$) nuclei, which were not included in the analysis. One can notice that propagated abundances at 100 MeV nucleon$^{-1}$ and 10 GeV nucleon$^{-1}$ do not always match. This is an indication of a difference in the spectral shape between mostly primary elements and those where the secondary contribution is essential, the effect of concurrent fragmentation, production of secondaries, and ionization energy losses as low energies. 
%%%%%%%%%%%% Fluorine %%%%%%%%%%%%%%%%%%%%%%%%%%%
Our analysis also hints at a presence of \emph{primary} fluorine $_{9}$F whose source abundance was set to non-zero, see a discussion in Supplementary Section~\ref{fluorine}.
%%%%%%%%%%%%%%%%%%%%%%%%%%%%%%%%%%%%%%%%%%%%

\begin{deluxetable}{@{}r@{}l@{}l@{\hspace{5pt}}l@{\hspace{5pt}}l@{\hspace{5pt}}l@{}l@{}}[tb!] 
        \tabletypesize{\footnotesize}
        \def\arraystretch{1.1}
        \tablecolumns{7}
        \tablewidth{0mm}
        \tablecaption{Injection spectra of CR species for {\it I}- and {\it P}-scenarios \label{tbl-inject}}
        \tablehead{
                \multicolumn{2}{l}{Nuc-}&
                \multicolumn{5}{c}{Spectral parameters}\\
                \cline{3-7}
                \multicolumn{2}{l}{leus}&
                \multicolumn{1}{l}{\!\!$\gamma_0 {}^{R_0 {\rm (GV)}} s_0$} &
                \multicolumn{1}{l}{\!\!$\gamma_1 {}^{R_1 {\rm (GV)}} s_1$} &
                \multicolumn{1}{l}{\!\!\!$\gamma_2 {}^{R_2 {\rm (GV)}}  s_2$} &
                \multicolumn{1}{l}{\!\!\!\!$\gamma_3 {}^{R_3 {\rm (TV)}}\! s_3$} & 
                \multicolumn{1}{l}{\!\!\!$\gamma_4$}
                }
	\startdata
	$_{1}$ & H & 
	$2.24\, {}^{0.95}\, 0.29$ & 
	$1.70\, {}^{6.97}\, 0.22$ & 
	$2.44\, {}^{400}\, 0.09$ &
	$2.19\, {}^{16}\, 0.09$ &
	2.37\\
	$_{2}$ & He &
	$2.05\, {}^{1.00}\, 0.26$ & 
	$1.76\, {}^{7.49}\, 0.33$ & 
	$2.41\, {}^{340}\, 0.13$ &
	$2.12\, {}^{30}\, 0.10$ &
	2.37\\
	$^{7}_3$ & Li\tablenotemark{a}\! & 
	\nodata\, ${}^{\ \ \ }$ \nodata & 
	$1.10\, {}^{12.0}\, 0.16$ & 
	$2.72\, {}^{355}\, 0.13$ &
	$1.90\, {}^{\ \ }$ \nodata &
	\nodata\\
	$_{6}$ & C & 
	$1.00\, {}^{1.10}\, 0.19$ & 
	$1.98\, {}^{6.54}\, 0.31$ & 
	$2.43\, {}^{348}\, 0.17$ &
	$2.12\, {}^{\ \ }$ \nodata &
	\nodata\\
	$^{14}_{\phn7}$ & N & 
	$1.00\, {}^{1.30}\, 0.17$ & 
	$1.96\, {}^{7.00}\, 0.20$ & 
	$2.46\, {}^{300}\, 0.17$ &
	$1.90\, {}^{\ \ }$ \nodata &
	\nodata\\
	$_{8}$ & O & 
	$0.95\, {}^{0.90}\, 0.18$ & 
	$1.99\, {}^{7.50}\, 0.30$ & 
	$2.46\, {}^{365}\, 0.17$ &
	$2.13\, {}^{\ \ }$ \nodata &
	\nodata\\
	$_{9}$ & F &
	$0.20\, {}^{1.50}\, 0.19$ & 
	$1.97\, {}^{7.00}\, 0.20$ & 
	$2.48\, {}^{355}\, 0.17$ &
	$2.14\, {}^{\ \ }$ \nodata &
	\nodata\\
	$_{10}$ & Ne &
	$0.60\, {}^{1.15}\, 0.17$ & 
	$1.92\, {}^{9.42}\, 0.26$ & 
	$2.44\, {}^{355}\, 0.17$ &
	$1.97\, {}^{\ \ }$ \nodata &
	\nodata\\
	$_{11}$ & Na &
	$0.50\, {}^{0.75}\, 0.17$ & 
	$1.98\, {}^{7.00}\, 0.21$ & 
	$2.49\, {}^{355}\, 0.17$ &
	$2.14\, {}^{\ \ }$ \nodata &
	\nodata\\
	$_{12}$ & Mg &
	$0.20\, {}^{0.85}\, 0.12$ & 
	$1.99\, {}^{7.00}\, 0.23$ & 
	$2.48\, {}^{355}\, 0.17$ &
	$2.15\, {}^{\ \ }$ \nodata &
	\nodata\\
	$_{13}$ & Al &
	$0.20\, {}^{0.60}\, 0.17$ & 
	$2.04\, {}^{7.00}\, 0.20$ & 
	$2.48\, {}^{355}\, 0.17$ &
	$2.14\, {}^{\ \ }$ \nodata &
	\nodata\\
	$_{14}$ & Si &
	$0.20\, {}^{0.85}\, 0.17$ & 
	$1.97\, {}^{7.00}\, 0.26$ & 
	$2.47\, {}^{355}\, 0.17$ &
	$2.19\, {}^{\ \ }$ \nodata &
	\nodata\\
	$_{15}$ & P &
	$0.25\, {}^{1.60}\, 0.19$ & 
	$1.95\, {}^{7.00}\, 0.20$ & 
	$2.48\, {}^{355}\, 0.17$ &
	$2.14\, {}^{\ \ }$ \nodata &
	\nodata\\
	$_{16}$ & S &
	$0.80\, {}^{1.30}\, 0.17$ & 
	$1.96\, {}^{7.00}\, 0.20$ & 
	$2.49\, {}^{355}\, 0.17$ &
	$2.14\, {}^{\ \ }$ \nodata &
	\nodata\\
	$_{17}$ & Cl &
	$1.10\, {}^{1.50}\, 0.17$ & 
	$1.98\, {}^{7.20}\, 0.20$ & 
	$2.53\, {}^{355}\, 0.17$ &
	$2.14\, {}^{\ \ }$ \nodata &
	\nodata\\
	$_{18}$ & Ar &
	$0.20\, {}^{1.30}\, 0.17$ & 
	$1.96\, {}^{7.00}\, 0.20$ & 
	$2.46\, {}^{355}\, 0.17$ &
	$2.09\, {}^{\ \ }$ \nodata &
	\nodata\\
	$_{19}$ & K &
	$0.20\, {}^{1.40}\, 0.15$ & 
	$1.96\, {}^{7.00}\, 0.20$ & 
	$2.53\, {}^{355}\, 0.17$ &
	$2.14\, {}^{\ \ }$ \nodata &
	\nodata\\
	$_{20}$ & Ca &
	$0.30\, {}^{1.00}\, 0.11$ & 
	$2.07\, {}^{7.00}\, 0.20$ & 
	$2.48\, {}^{355}\, 0.17$ &
	$2.14\, {}^{\ \ }$ \nodata &
	\nodata\\
	$_{21}$ & Sc &
	$0.20\, {}^{1.40}\, 0.17$ & 
	$1.97\, {}^{7.00}\, 0.22$ & 
	$2.53\, {}^{355}\, 0.17$ &
	$2.14\, {}^{\ \ }$ \nodata &
	\nodata\\
	$_{22}$ & Ti &
	$1.50\, {}^{0.90}\, 0.17$ & 
	$1.98\, {}^{7.00}\, 0.22$ & 
	$2.57\, {}^{355}\, 0.17$ &
	$2.14\, {}^{\ \ }$ \nodata &
	\nodata\\
	$_{23}$ & V &
	$1.10\, {}^{0.80}\, 0.17$ & 
	$1.98\, {}^{7.00}\, 0.22$ & 
	$2.53\, {}^{355}\, 0.17$ &
	$2.14\, {}^{\ \ }$ \nodata &
	\nodata\\
	$_{24}$ & Cr &
	$1.70\, {}^{0.65}\, 0.17$ & 
	$1.99\, {}^{7.00}\, 0.20$ & 
	$2.48\, {}^{355}\, 0.17$ &
	$2.14\, {}^{\ \ }$ \nodata &
	\nodata\\
	$_{25}$ & Mn &
	$0.20\, {}^{0.85}\, 0.10$ & 
	$2.08\, {}^{7.00}\, 0.20$ & 
	$2.48\, {}^{355}\, 0.17$ &
	$2.14\, {}^{\ \ }$ \nodata &
	\nodata\\
	$_{26}$ & Fe &
	$0.27\, {}^{1.04}\, 0.18$ & 
	$1.99\, {}^{7.00}\, 0.20$ & 
	$2.51\, {}^{355}\, 0.17$ &
	$2.19\, {}^{\ \ }$ \nodata &
	\nodata\\
	$_{27}$ & Co &
	$0.80\, {}^{0.70}\, 0.15$ & 
	$1.98\, {}^{7.00}\, 0.20$ & 
	$2.49\, {}^{355}\, 0.17$ &
	$2.14\, {}^{\ \ }$ \nodata &
	\nodata\\
	$_{28}$ & Ni &
	$1.50\, {}^{0.65}\, 0.17$ & 
	$1.98\, {}^{7.00}\, 0.20$ & 
	$2.48\, {}^{355}\, 0.17$ &
	$2.14\, {}^{\ \ }$ \nodata &
	\nodata\\
	\enddata
  \tablenotetext{a}{Primary lithium.}
  \tablecomments{For the {\it P}-scenario $\gamma_3=\gamma_2$. For parameter definitions see Eq.~(\ref{eq:1}). The fit errors: $\gamma_{0,1} \pm0.06$, $\gamma_{2,3} \pm0.04$, $R_0 \pm0.5$~GV, $R_1\pm1$ GV, and $R_2\pm15$ GV. For protons and He ({\it I}-scenario only): $R^p_3\pm5$ TV, $\gamma^p_{4}\pm 0.08$, $R^{\rm He}_3\pm15$ TV, $\gamma^{\rm He}_{4}\pm 0.1$.}
\end{deluxetable}

\begin{deluxetable}{l@{\hspace{1ex}}C@{\hspace{3ex}}l@{\hspace{1ex}}C@{\hspace{3ex}}l@{\hspace{1ex}}C}[htb!]
\tabletypesize{\footnotesize}
\def\arraystretch{1.1}
\tablecolumns{6}
\tablewidth{0mm}
\tablecaption{Source Abundances of CR species\label{tbl-Ab}}
\tablehead{
\multicolumn{1}{l}{Nuc-} &     
\multicolumn{1}{l}{Source} &
\multicolumn{1}{l}{Nuc-} &     
\multicolumn{1}{l}{Source} &
\multicolumn{1}{l}{Nuc-} &     
\multicolumn{1}{l}{Source}\\
\multicolumn{1}{l}{leus}  &     
\multicolumn{1}{l}{Abundance} &
\multicolumn{1}{l}{leus}  &     
\multicolumn{1}{l}{Abundance} &
\multicolumn{1}{l}{leus}  &     
\multicolumn{1}{l}{Abundance}
}
\startdata
$^{\phn 1}_{\phn 1}$H & 8.77\!\times\!10^{5} &
 \phm{-}$^{27}_{13}$Al & 51.1 &
 \phm{---}$^{48}$Ti & <\!10^{-4} \\ %6.05e-07 \\
\phm{--}$^{\phn 2}$H & 35 &
 \phm{-}$^{28}_{14}$Si & 580 &
 \phm{---}$^{49}$Ti & <\!10^{-4} \\%5.85e-09 \\
$^{\phn 3}_{\phn 2}$He& <\!10^{-4} &
 \phm{---}$^{29}$Si & 35 &
 \phm{---}$^{50}$Ti & <\!10^{-4} \\ %6.08e-07 \\
\phm{--}$^{\phn 4}$He& 7.74\!\times\!10^{4} &
\phm{---}$^{30}$Si & 24.7 &
\phm{-}$^{50}_{23}$V & <\!10^{-4} \\ %1.82e-05 \\
$^{\phn 6}_{\phn 3}$Li & <\!10^{-4} & %1.00e-06 &
\phm{-}$^{31}_{15}$P & 5.7 &
\phm{---}$^{51}$V & <\!10^{-4} \\ %5.99e-09 \\
\phm{--}$^{\phn 7}$Li & 52 &
\phm{-}$^{32}_{16}$S & 82.1 &
\phm{-}$^{50}_{24}$Cr & 4 \\
$^{\phn 7}_{\phn 4}$Be& 0 &
\phm{---}$^{33}$S & 0.306 &
\phm{---}$^{51}$Cr & 0 \\
\phm{--}$^{\phn 9}$Be & <\!10^{-4} & %2.65e-05 &
\phm{---}$^{34}$S & 3.42 & 
\phm{---}$^{52}$Cr & 11.1 \\
\phm{--}$^{10}$Be & <\!10^{-4} & %5.30e-06 &
\phm{---}$^{36}$S & 4.28\!\times\!10^{-4} &
\phm{---}$^{53}$Cr & 3.01\!\times\!10^{-3} \\
$^{10}_{\phn 5}$B & 1.80\!\times\!10^{-4} &
\phm{--}$^{35}_{17}$Cl & 2.5 &
\phm{---}$^{54}$Cr & 0.5 \\
\phm{--}$^{11}$B & 7.42\!\times\!10^{-4} &
\phm{---}$^{37}$Cl & 1.17\!\times\!10^{-3} &
\phm{-}$^{53}_{25}$Mn & 12.6 \\
$^{12}_{\phn 6}$C & 2720 &
\phm{-}$^{36}_{18}$Ar & 11.4 &
\phm{---}$^{55}$Mn & 2.9 \\
\phm{--}$^{13}$C & <\!10^{-4} & %5.27e-07 &
\phm{---}$^{38}$Ar & 0.74 &
\phm{-}$^{54}_{26}$Fe & 30.1 \\
$^{14}_{\phn 7}$N & 207 &
\phm{---}$^{40}$Ar & 1.74\!\times\!10^{-3} &
\phm{---}$^{55}$Fe & 0 \\
\phm{--}$^{15}$N & <\!10^{-4} & %5.96e-05 &
\phm{-}$^{39}_{19}$K & 1.39 &
\phm{---}$^{56}$Fe & 515 \\
$^{16}_{\phn 8}$O & 3510 &
\phm{---}$^{40}$K & 2.80 &
\phm{---}$^{57}$Fe & 17.7 \\ 
\phm{--}$^{17}$O & <\!10^{-4} & %6.71e-07 &
\phm{---}$^{41}$K & 3.34\!\times\!10^{-4} &
\phm{---}$^{58}$Fe & 5.34 \\
\phm{--}$^{18}$O & 1.29 &
\phm{-}$^{40}_{20}$Ca & 36.1 &
\phm{-}$^{59}_{27}$Co & 1.40 \\
$^{19}_{\phn 9}$F & 0.95 &
\phm{---}$^{41}$Ca & 1.97 &
\phm{-}$^{58}_{28}$Ni & 22.3 \\
$^{20}_{10}$Ne & 338 &
\phm{---}$^{42}$Ca & <\!10^{-4} &  %1.13e-06 \\
\phm{---}$^{59}$Ni & 0 \\
\phm{--}$^{21}$Ne & 3.56\!\times\!10^{-3} &
\phm{---}$^{43}$Ca & <\!10^{-4} & %2.12e-06 \\
\phm{---}$^{60}$Ni & 8.99 \\
\phm{--}$^{22}$Ne & 107 &
\phm{---}$^{44}$Ca & <\!10^{-4} & %9.93e-05 \\
\phm{---}$^{61}$Ni & 0.599 \\
$^{23}_{11}$Na & 24.1 &
\phm{---}$^{48}$Ca & 0.11 &
\phm{---}$^{62}$Ni & 1.43 \\
$^{24}_{12}$Mg & 490 &
\phm{-}$^{45}_{21}$Sc & 1.46 &
\phm{---}$^{64}$Ni & 0.304 \\
\phm{--}$^{25}$Mg & 70 &
\phm{-}$^{46}_{22}$Ti & 4.9 &
\nodata & \nodata \\
\phm{--}$^{26}$Mg & 90 &
\phm{---}$^{47}$Ti & <\!10^{-4} & %8.95e-06 \\
\nodata & \nodata 
\enddata
\tablecomments{For all non-zero relative source abundances below $10^{-4}$ we provide an upper limit. {\it Propagated} relative isotopic abundances for each element are tuned to the ACE-CRIS data.}
\end{deluxetable}

\begin{figure}[tb!]
	\centering
	\includegraphics[width=0.47\textwidth]{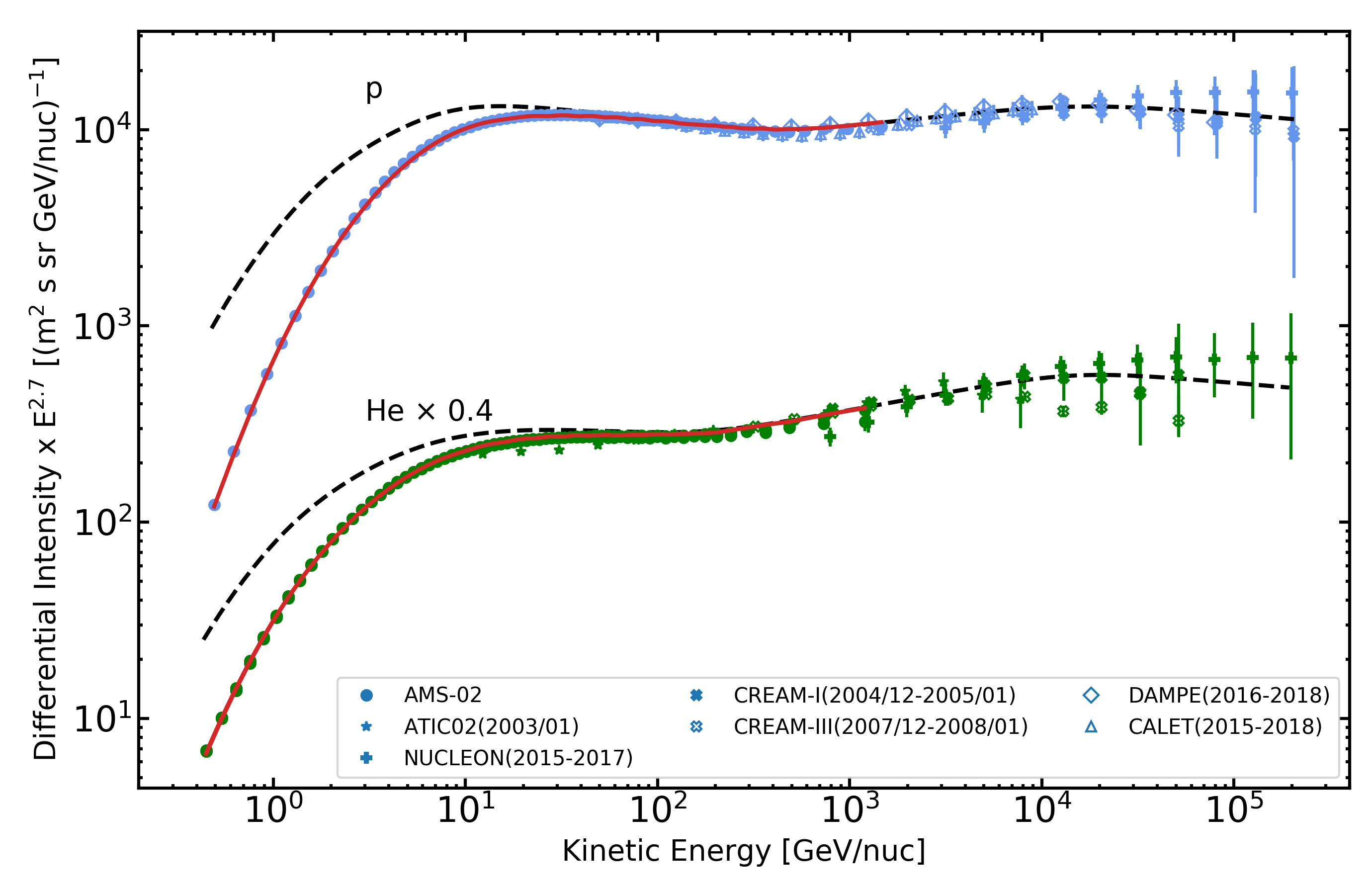}
	\includegraphics[width=0.47\textwidth]{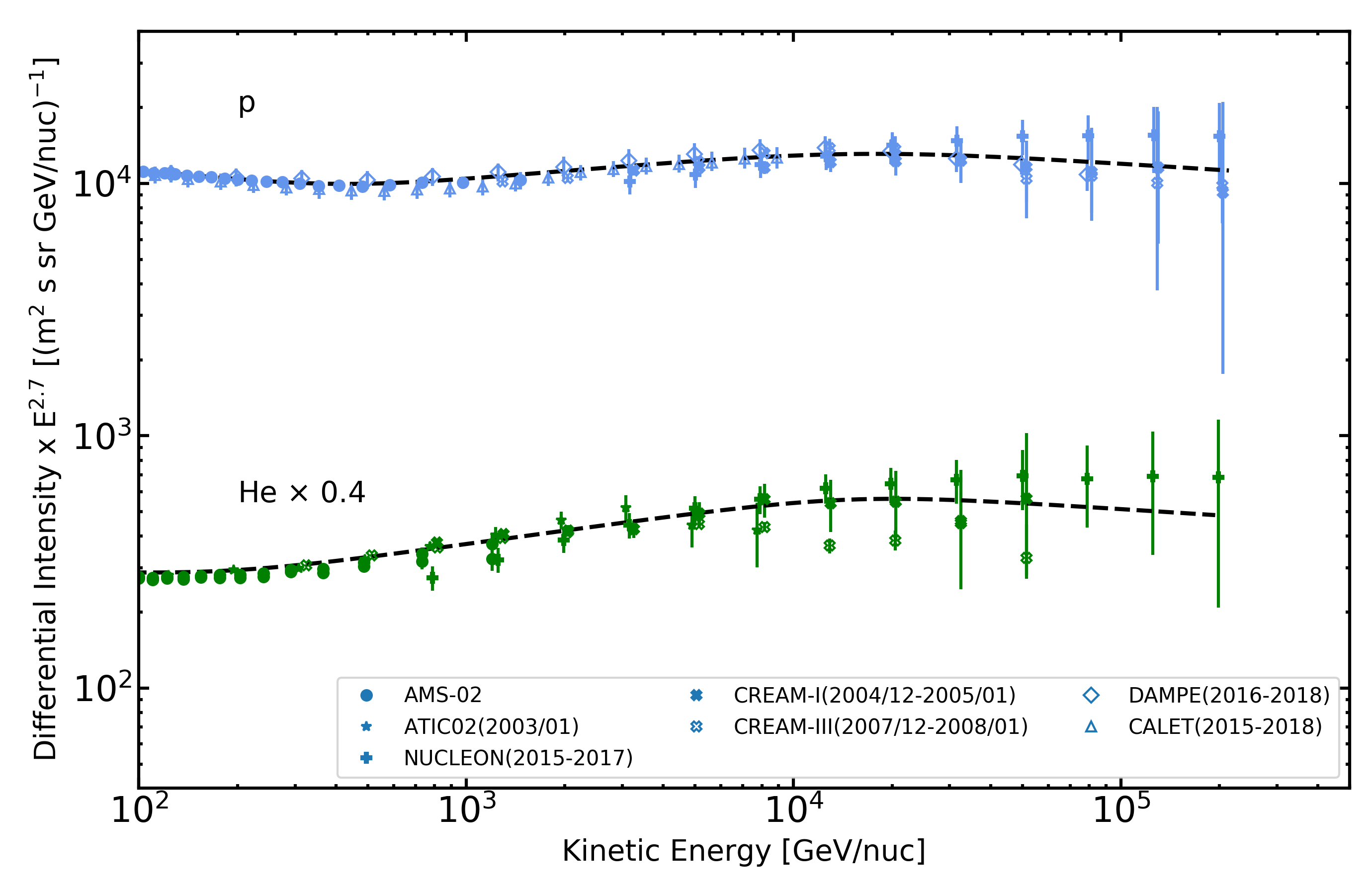}
	\caption{Our model calculations of the proton and He spectra for {\it I}-scenario and a comparison with available data: HEAO-3-C2 \citep{1990A&A...233...96E}, ATIC-2 \citep{2009BRASP..73..564P}, CREAM \citep{2008APh....30..133A}, NUCLEON \citep{2019AdSpR..64.2546G, 2019AdSpR..64.2559G}, AMS-02 \citep{2015PhRvL.114q1103A,2015PhRvL.115u1101A}, CALET \citep{2019PhRvL.122r1102A}, and DAMPE \citep{eaax3793}. The expanded high energy parts of the spectra $>$100 GeV nucleon$^{-1}$ are shown in the bottom. The line coding as in Figure \ref{fig:He-Ne}.}
	\label{fig:TeV_H_He}
\end{figure}

\begin{figure*}[tb!]
	\centering
	\includegraphics[width=0.8\textwidth]{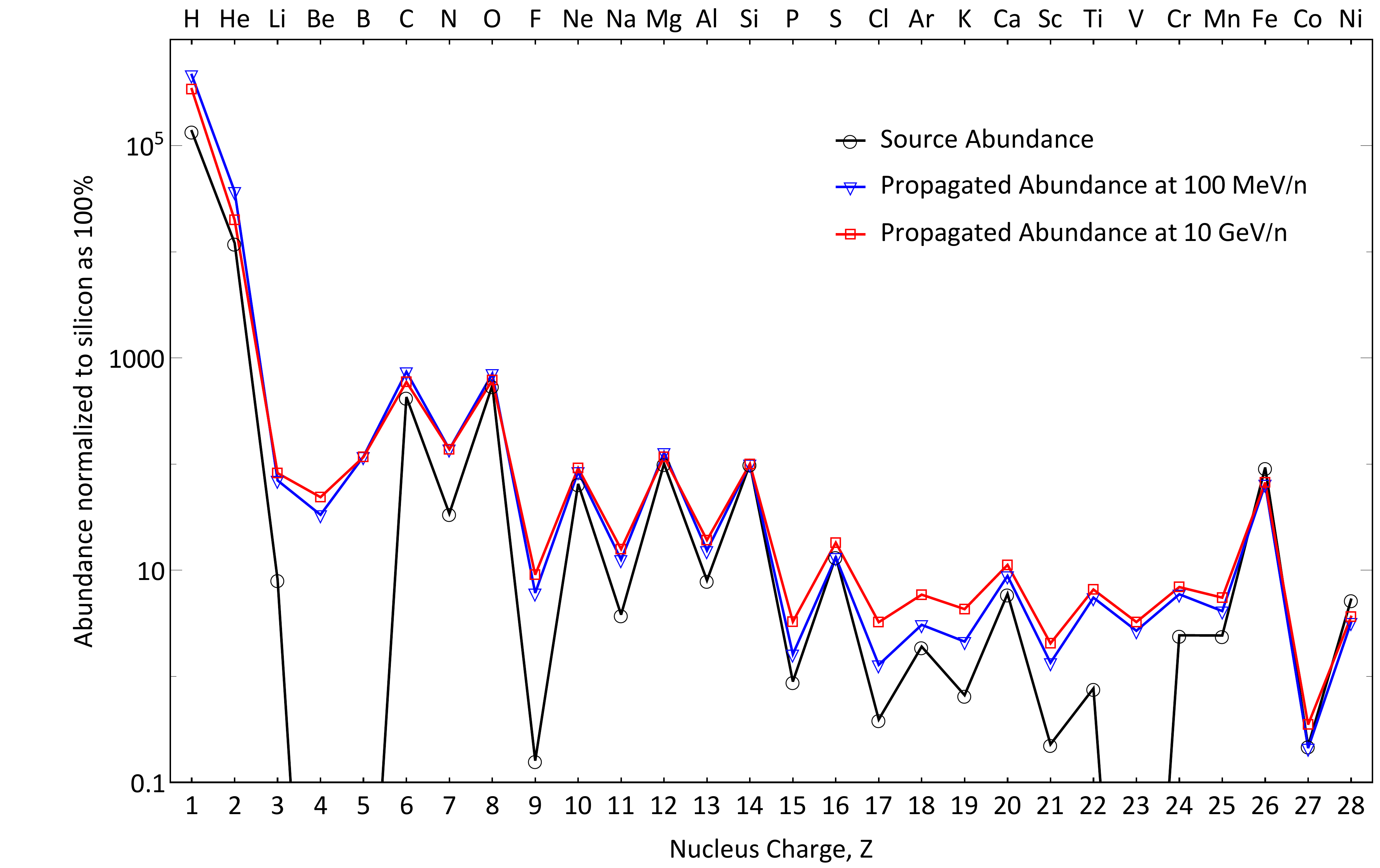} 
	\caption{\galprop{} source and LIS abundances for elements $Z = 1 - 28$ at 100 MeV nucleon$^{-1}$ and 10 GeV nucleon$^{-1}$ w.r.t.\ silicon (Si = 100) in our model. The abundances are tuned to Voyager 1 data \citep{2016ApJ...831...18C} and the HEAO-3-C2 ``plateau'' region \citep{1990A&A...233...96E}.
	}
	\label{fig:abund1}
\end{figure*}

\begin{figure*}[tb!]
	\centering
	\includegraphics[width=0.8\textwidth]{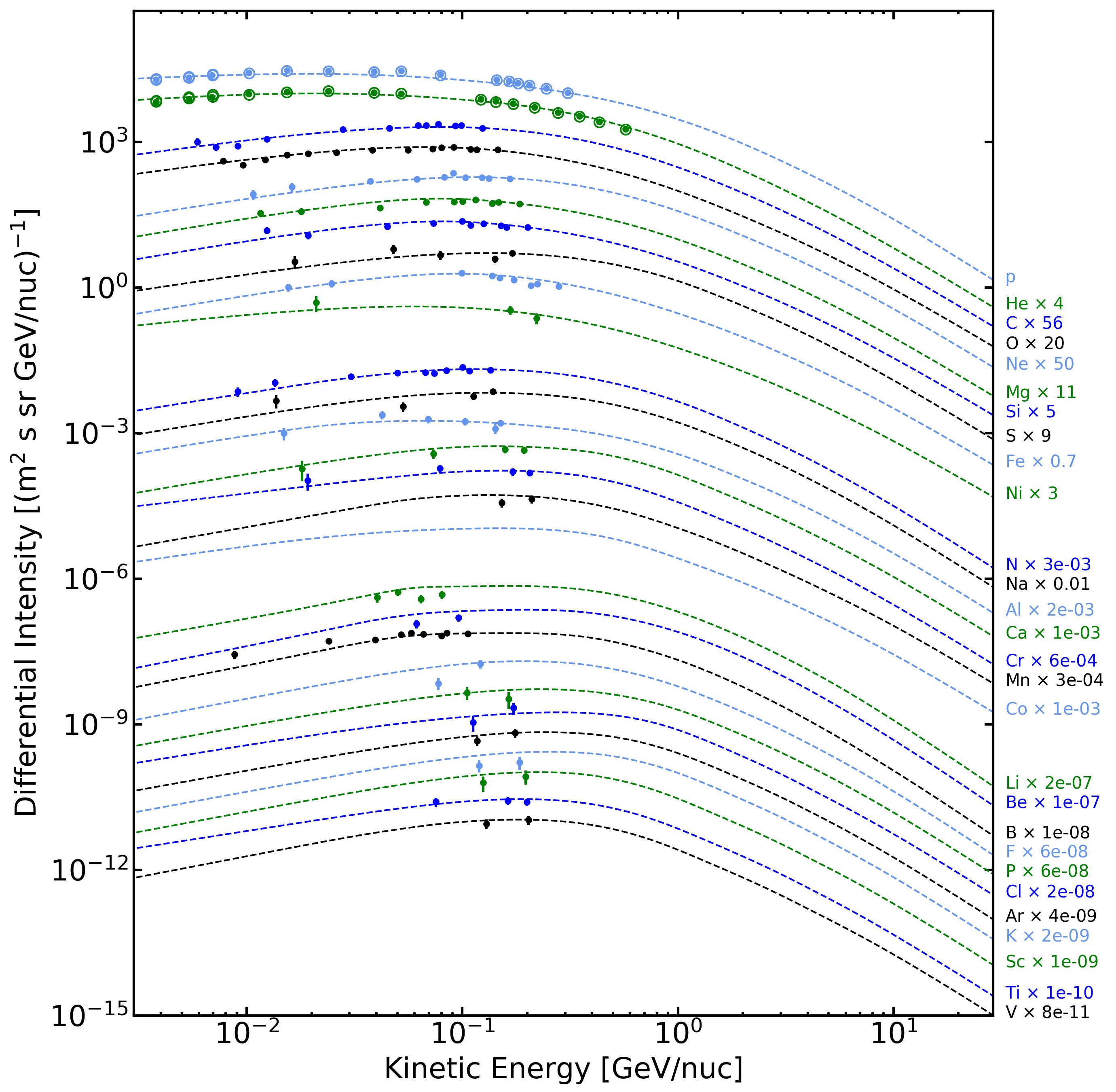}
	\caption{The \galprop{} LIS for all CR species (dashed lines) are compared to the Voyager 1 data \citep[filled circles,][]{2016ApJ...831...18C}. We also show updated Voyager 1 data for H and He (open circles) taken from September 1, 2012 to November 13, 2019, as posted on the NASA page\textsuperscript{\ref{v1}}. The elements are sorted by approximate amount of primary contribution: first group is mostly primary, second -- with significant primary contribution, and third -- mostly secondary. 
	}
	\label{fig:Voy}
\end{figure*}

\begin{figure*}[tb!]
	\centering
	\includegraphics[width=0.8\textwidth]{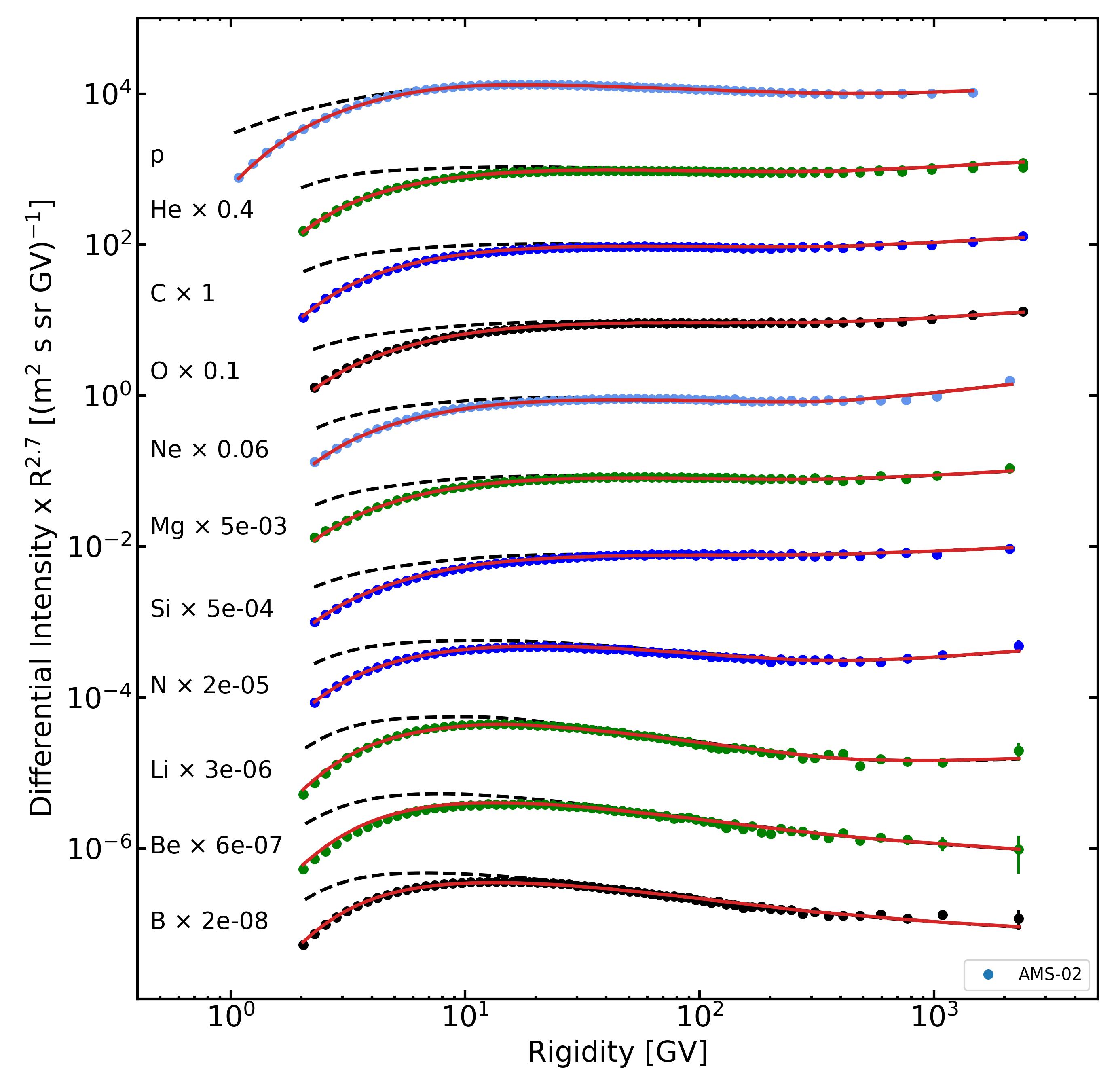}
	\caption{A summary plot of our model calculations for {\it I}-scenario and available AMS-02 data. The color of data points for each element is the same as in Figure~\ref{fig:Voy}. Line coding as in Figure~\ref{fig:He-Ne}.
	}
	\label{fig:TeVAMS}
\end{figure*}

Figure \ref{fig:Voy} shows a comparison of the derived LIS for all elements with Voyager 1 data \citep{2016ApJ...831...18C}. The open circles show updated Voyager 1 data for H and He taken from September 1, 2012 to November 13, 2019, as posted on the NASA page\footnote{https://voyager.gsfc.nasa.gov/spectra.html \label{v1}}---the remarkable agreement between the two data sets is another evidence that we see the local interstellar spectra.  The elements are sorted into three groups by approximate amount of primary contribution as discussed in Section~\ref{prop}, while within each group they are ordered by $Z$. One can see a clear difference in the spectral shapes above $\sim$1 GeV nucleon$^{-1}$ between the groups: the spectra become steeper and steeper as we move from the group of primaries to significantly secondary and to fully secondary groups. Meanwhile, the spectral slopes within each group are pretty similar.

\begin{figure*}[!bth]
	\centering
	\includegraphics[width=0.49\textwidth]{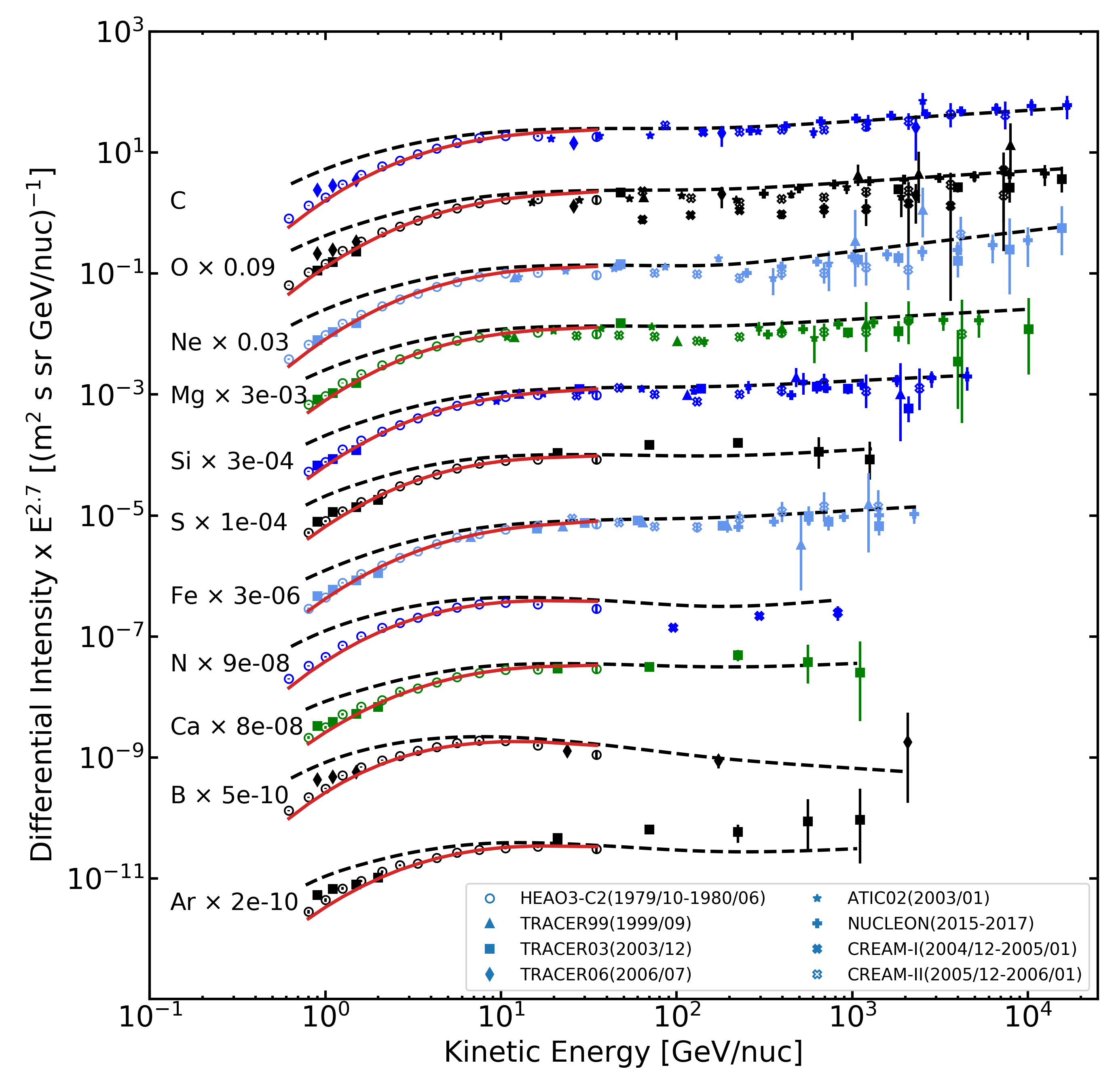}
	\includegraphics[width=0.49\textwidth]{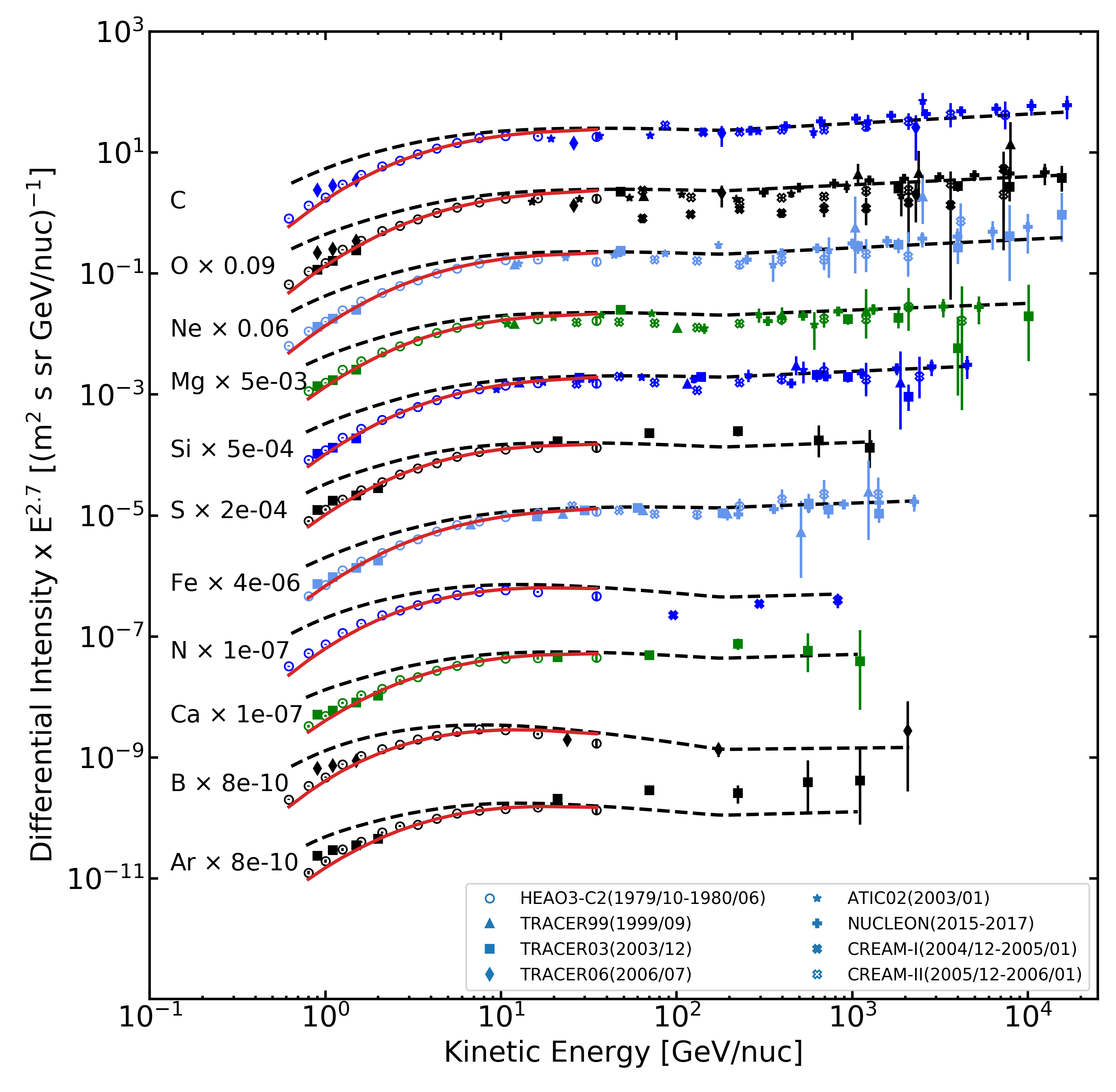}
	\caption{Our model calculations for {\it I}-scenario (left) and {\it P}-scenario (right) are compared to HEAO-3-C2 \citep{1990A&A...233...96E}, TRACER \citep{2004ApJ...607..333G, 2008ApJ...678..262A, 2011ApJ...742...14O}, ATIC-2 \citep{2009BRASP..73..564P}, CREAM \citep{2008APh....30..133A}, and NUCLEON \citep{2019AdSpR..64.2546G, 2019AdSpR..64.2559G} data. AMS-02 data are not shown for clarity. Black dashed lines show the calculated LIS spectra, and the red solid lines are the spectra modulated to the level that corresponds to the period of the HEAO-3-C2 mission. The color of data points for each element is the same as in Figure~\ref{fig:Voy}. 
	}
	\label{fig:TeV}
\end{figure*}

A summary plot of all available nuclei species measured by AMS-02 along with our model calculations is shown in Figure~\ref{fig:TeVAMS}. The achieved agreement is impressive and allows a detailed comparison with the HEAO-3-C2 data for those elements where both AMS-02 and HEAO-3-C2 data are available. The extension of our LIS into TeV energies and a comparison with HEAO-3-C2 \citep{1990A&A...233...96E}, TRACER \citep{2004ApJ...607..333G, 2008ApJ...678..262A, 2011ApJ...742...14O}, ATIC-2 \citep{2009BRASP..73..564P}, CREAM \citep{2008APh....30..133A, 2009ApJ...707..593A}, and NUCLEON \citep{2019AdSpR..64.2546G, 2019AdSpR..64.2559G} data is illustrated in Figure~\ref{fig:TeV} for the most abundant species for which such data are available: B, C, N, O, Ne, Mg, Si, S, Ar, Ca, and Fe. AMS-02 data are not shown in this plot for clarity. Black dashed lines show the calculated LIS spectra, and the red solid lines are the spectra modulated to the level that corresponds to the period of the HEAO-3-C2 mission. We show calculations for both scenarios: the {\it I}-scenario is on the l.h.s., and the {\it P}-scenario is on the r.h.s.

As we discussed in Section~\ref{calibration}, one can see some discrepancies between our modulated spectra (shown with red solid lines) and HEAO-3-C2 data at low and high energies, while the middle range ``plateau'' was used for the fits. The extension of the ``plateau'' fits to higher energies agrees well with data taken by other instruments, e.g., for C, O, Ne, Mg, Si, and Fe the agreement is very good, where the highest energy points reaching $\sim$200 TeV nucleon$^{-1}$ are coming from CREAM, TRACER, and NUCLEON experiments. This agreement is especially important as it provides a long lever arm to test our derived LIS in the kinetic energy range from 1 MeV nucleon$^{-1}$ to $\sim$100--500 TeV nucleon$^{-1}$ covering 8--9 orders of magnitude in energy.

\begin{figure*}[!ptb]
	\centering
	\includegraphics[width=0.98\textwidth]{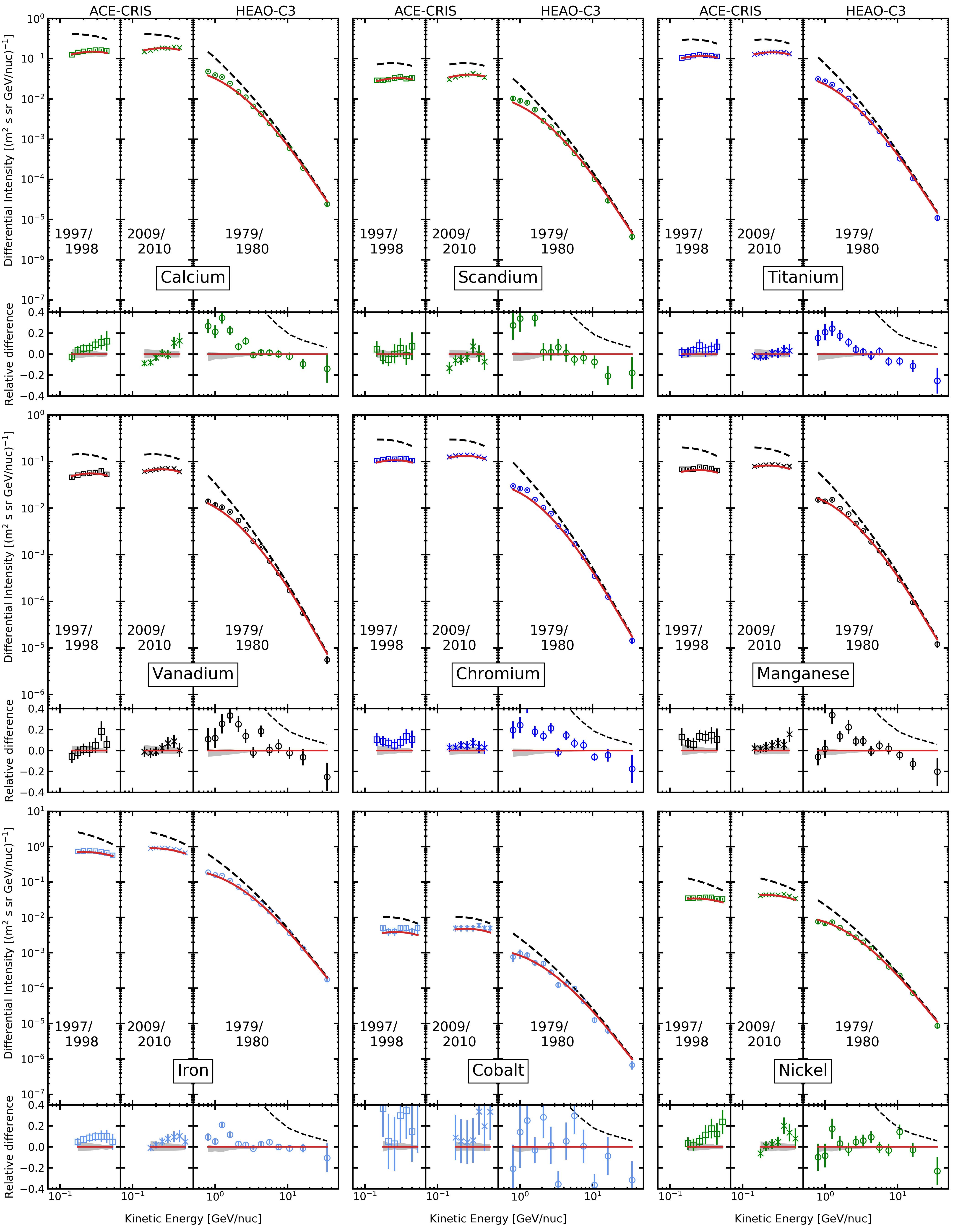}
	\caption{Calculated elemental spectra: $_{20}$Ca--\,$_{28}$Ni. The line coding and data as in Figure \ref{fig:He-Ne}.
	}
	\label{fig:Ca-Ni}
\end{figure*}

A detailed comparison of our modulated LIS for elements $_9$F to $_{28}$Ni with ACE-CRIS and HEAO-3-C2 ``plateau'' data is illustrated in a series of plots shown in Figures~\ref{fig:He-Ne}, \ref{fig:Na-K}, \ref{fig:Ca-Ni}. The modulation is calculated separately for each instrument appropriately to the data taking periods. The noticeable differences in the HEAO-3-C2 ``plateau'' region appear only in cases of rare species, such as P and K, where the scatter of the data points and the error bars are large indicating that the collected statistics is insufficient. 

The ACE-CRIS data are shown for the periods of active 1997--1998 and quiet 2009--2010 sun. In the case of abundant species, the agreement with ACE-CRIS data is also good within 5\%--10\% and is consistent with our calculations within the model uncertainties. Again, less abundant species, such as F, P, Cl, demonstrate worse agreement, but still within 20\%. This is mostly related to the period of active sun when the heliospheric magnetic field is highly turbulent and therefore is not surprising. The significant scattering of the data points in the spectra and residual plots of Na, P, Cl is hinting at the systematics. One can also see some discrepancies with Ni measurements due to the scattering of the data points, but large error bars make our calculations consistent within 2$\sigma$. 

\begin{figure*}[!hp] 
	\centering
	\includegraphics[width=0.33\textwidth]{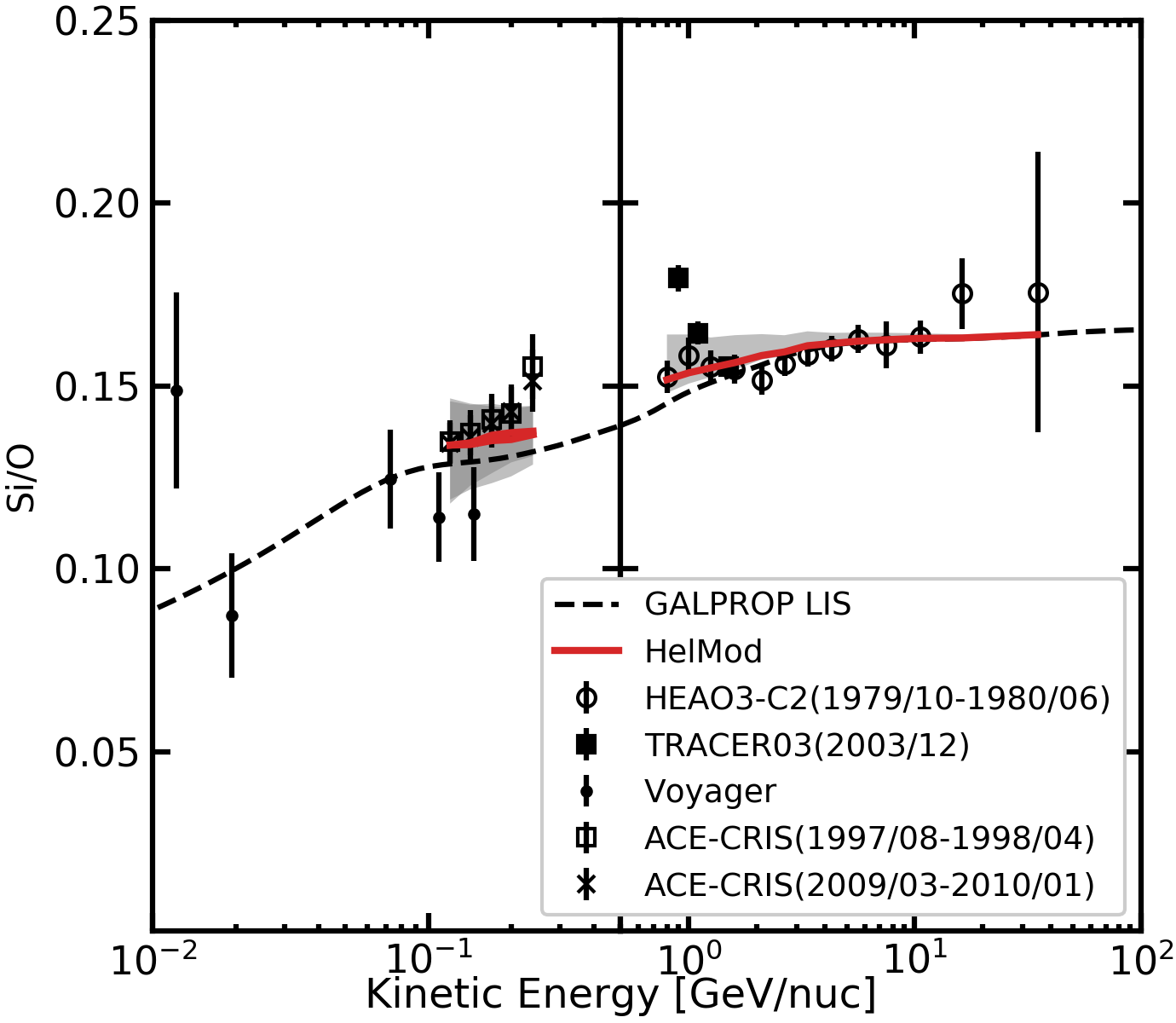}
	\includegraphics[width=0.33\textwidth]{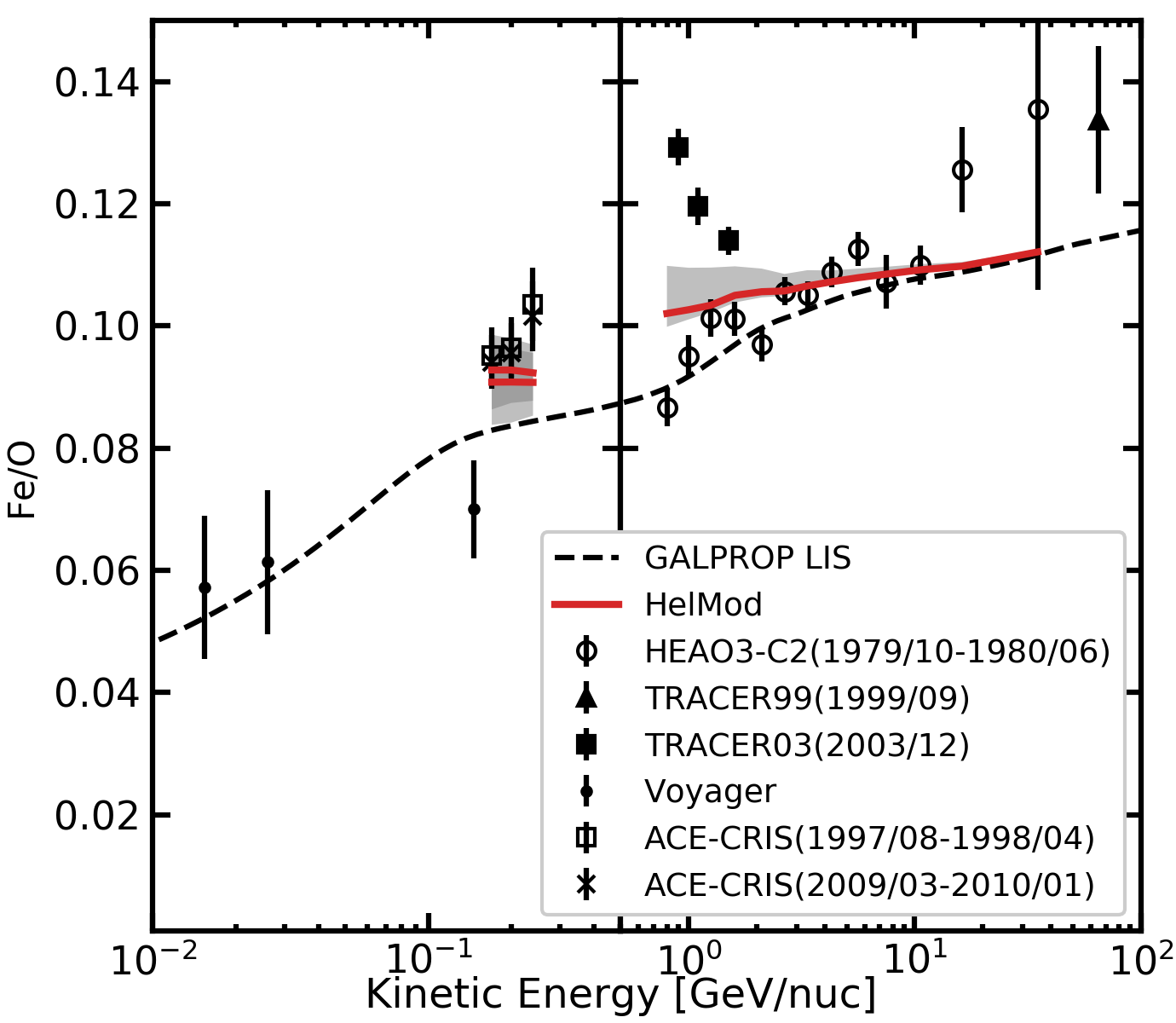}
	\includegraphics[width=0.33\textwidth]{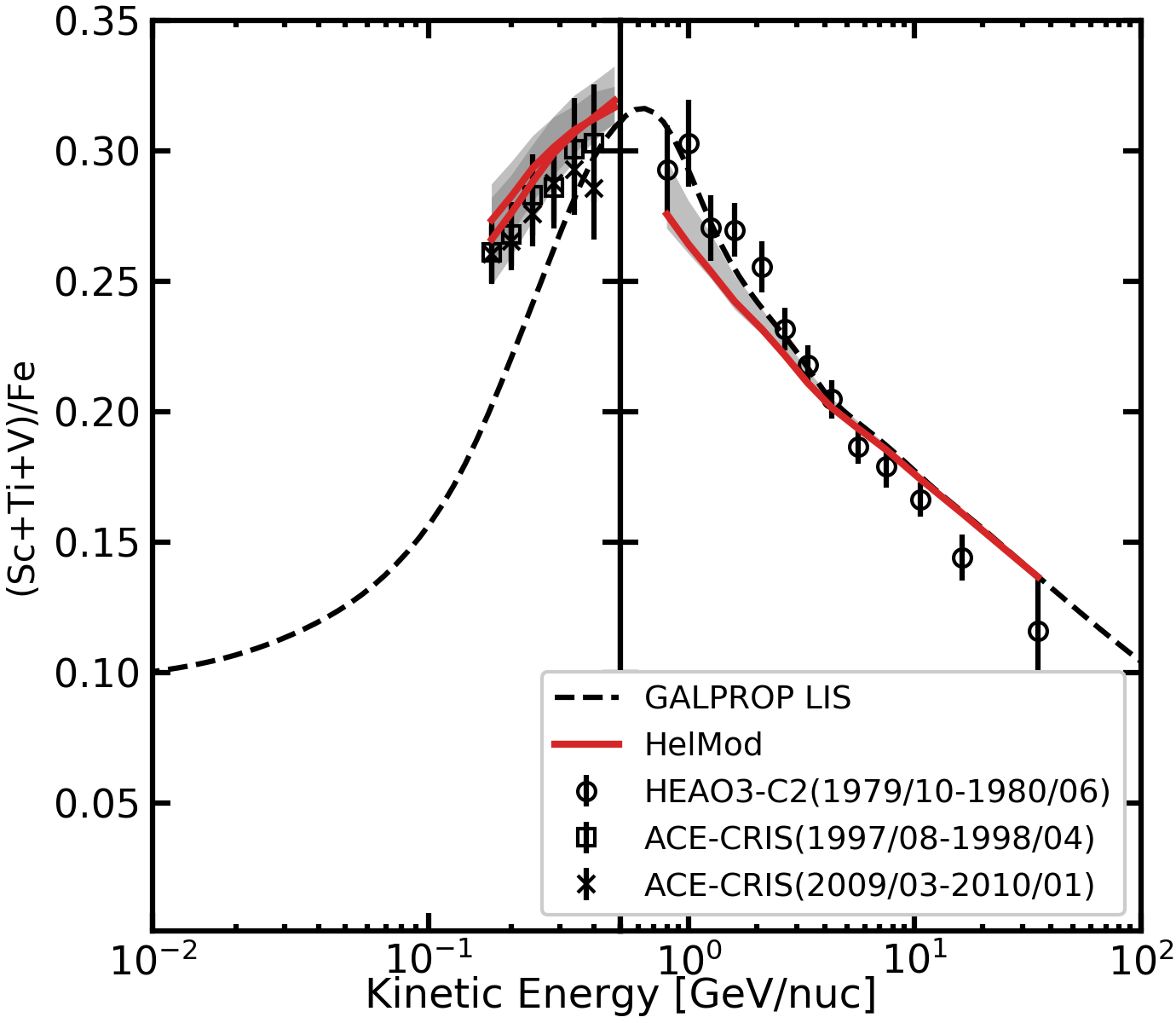}
	\includegraphics[width=0.33\textwidth]{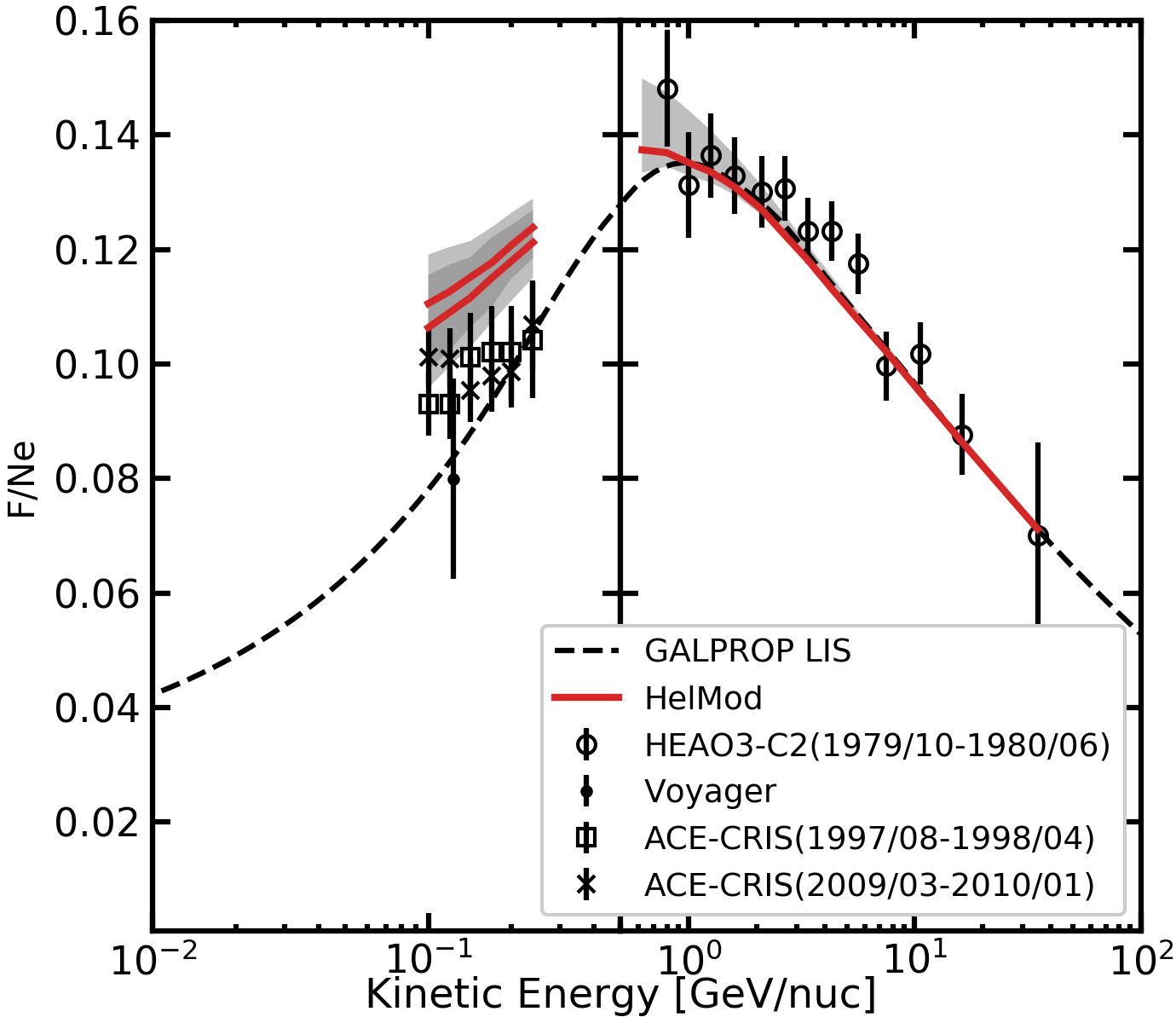}
	\includegraphics[width=0.33\textwidth]{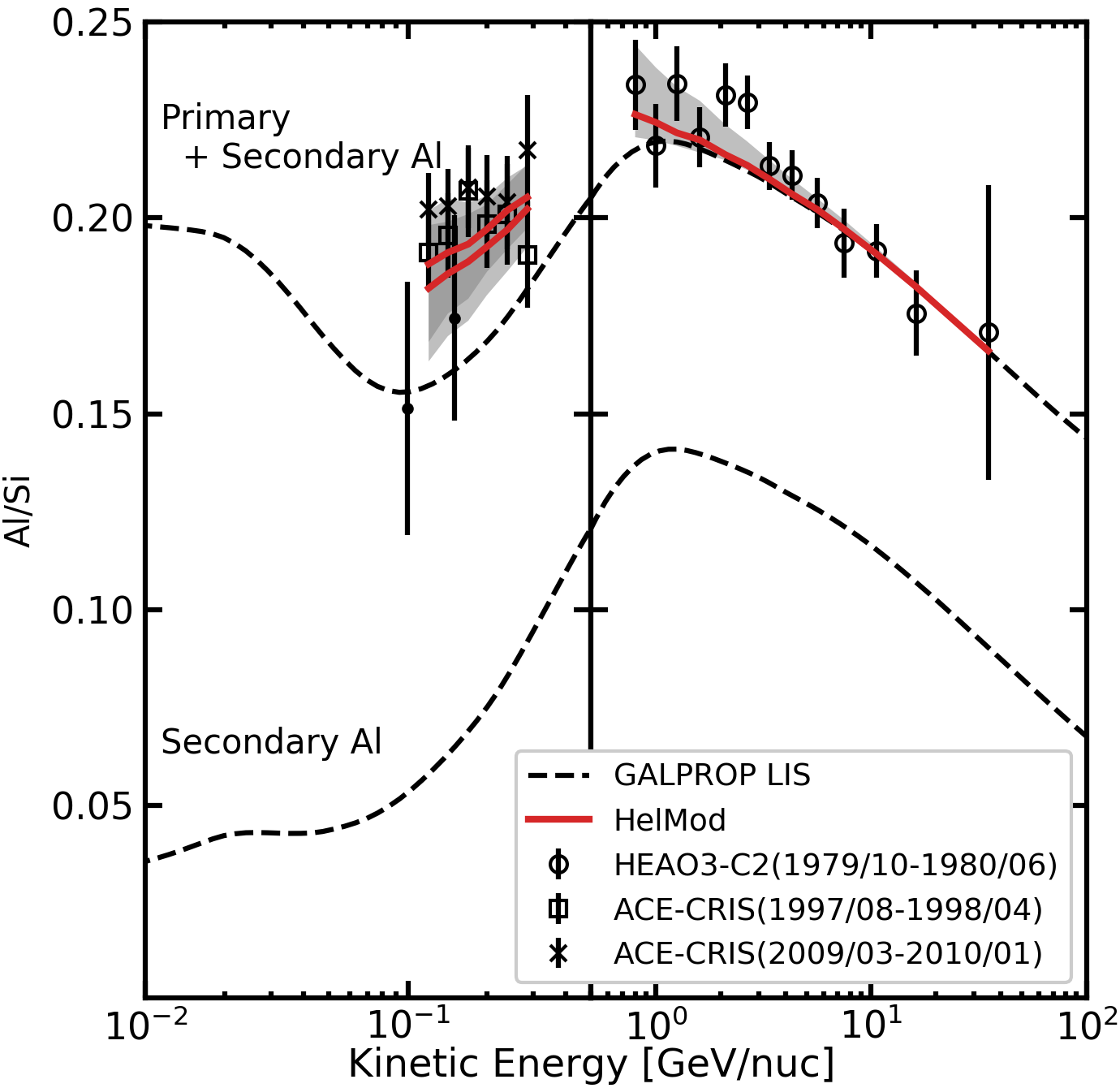}
	\includegraphics[width=0.33\textwidth]{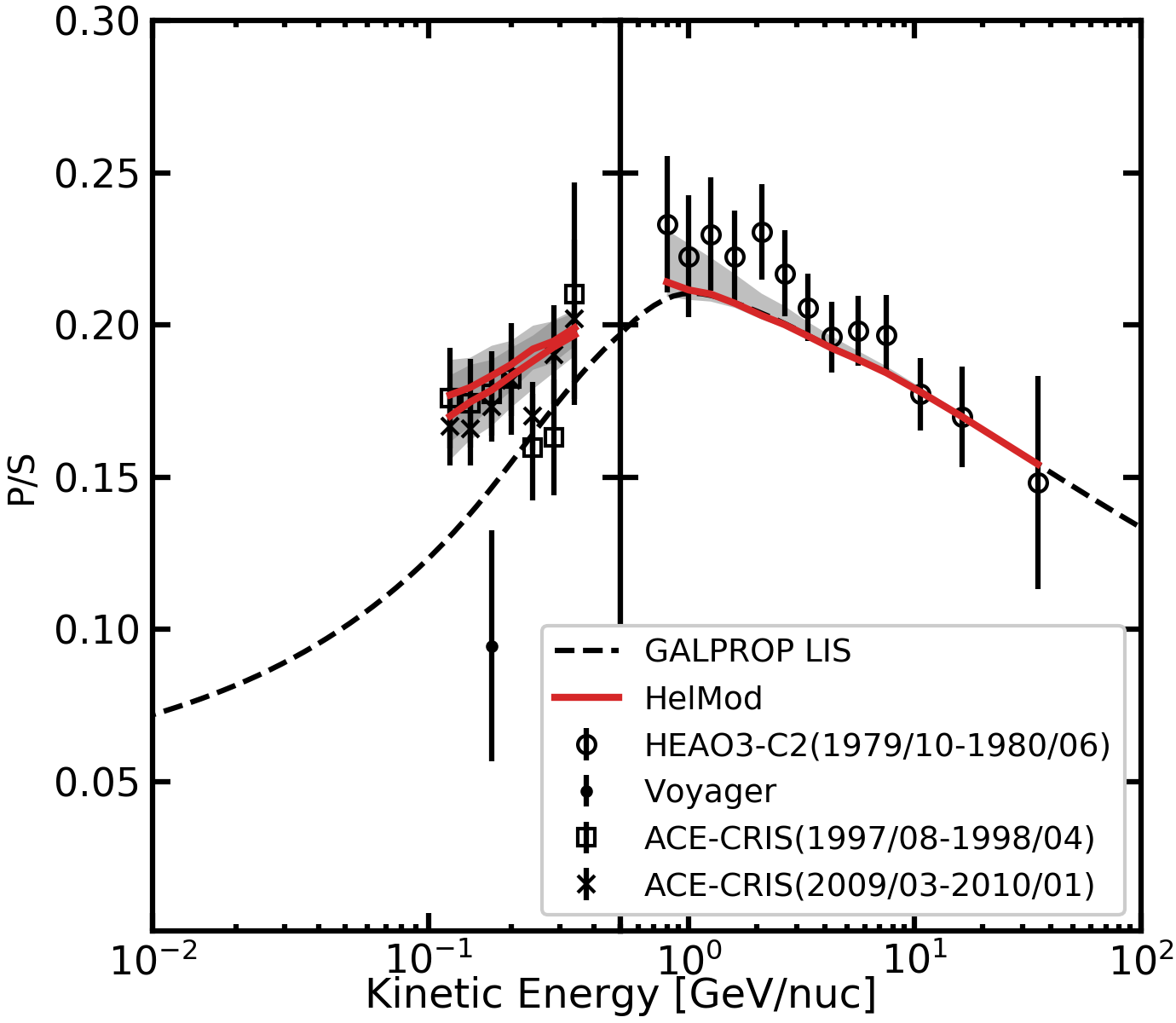}
	\includegraphics[width=0.33\textwidth]{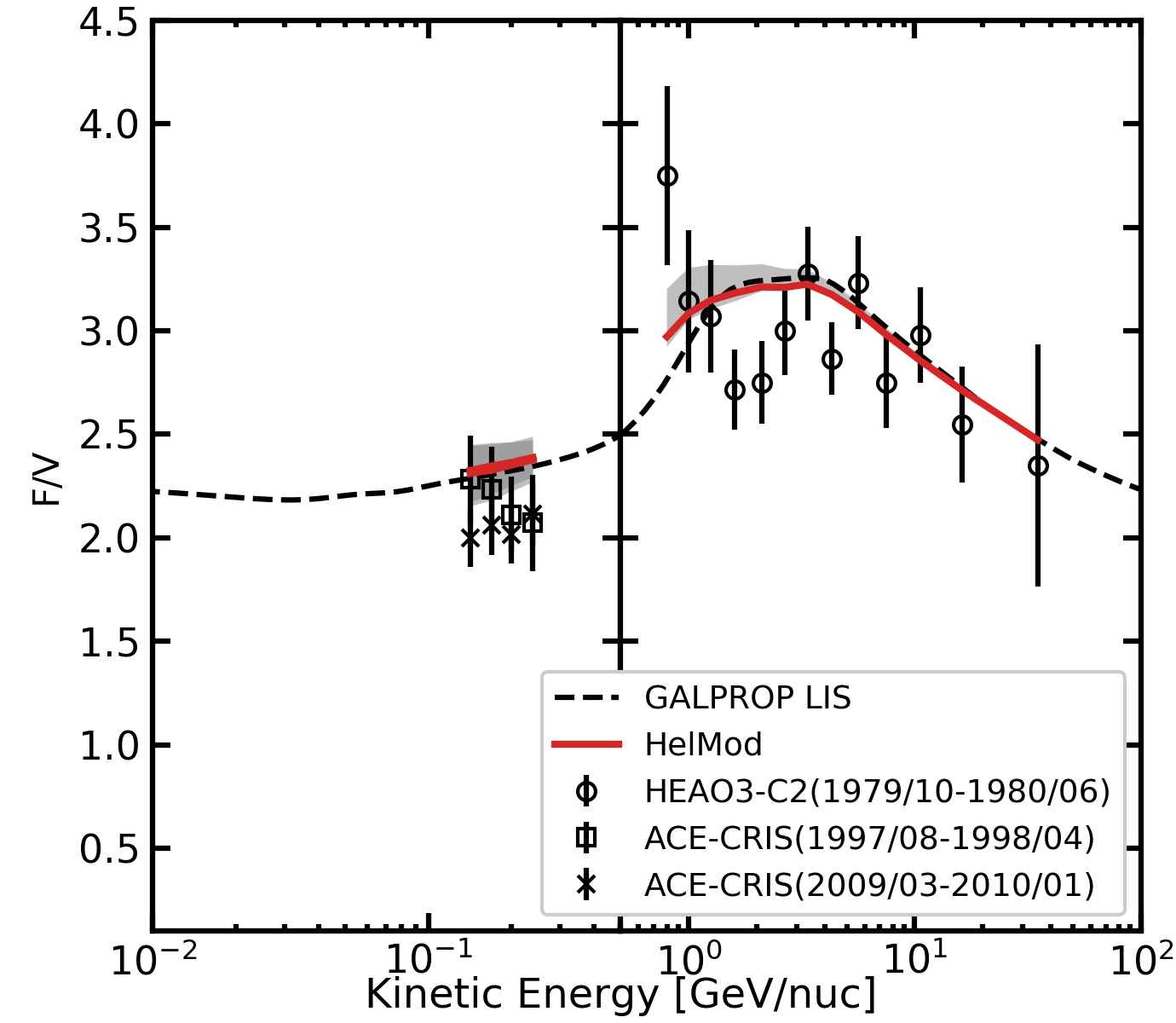}
	\includegraphics[width=0.33\textwidth]{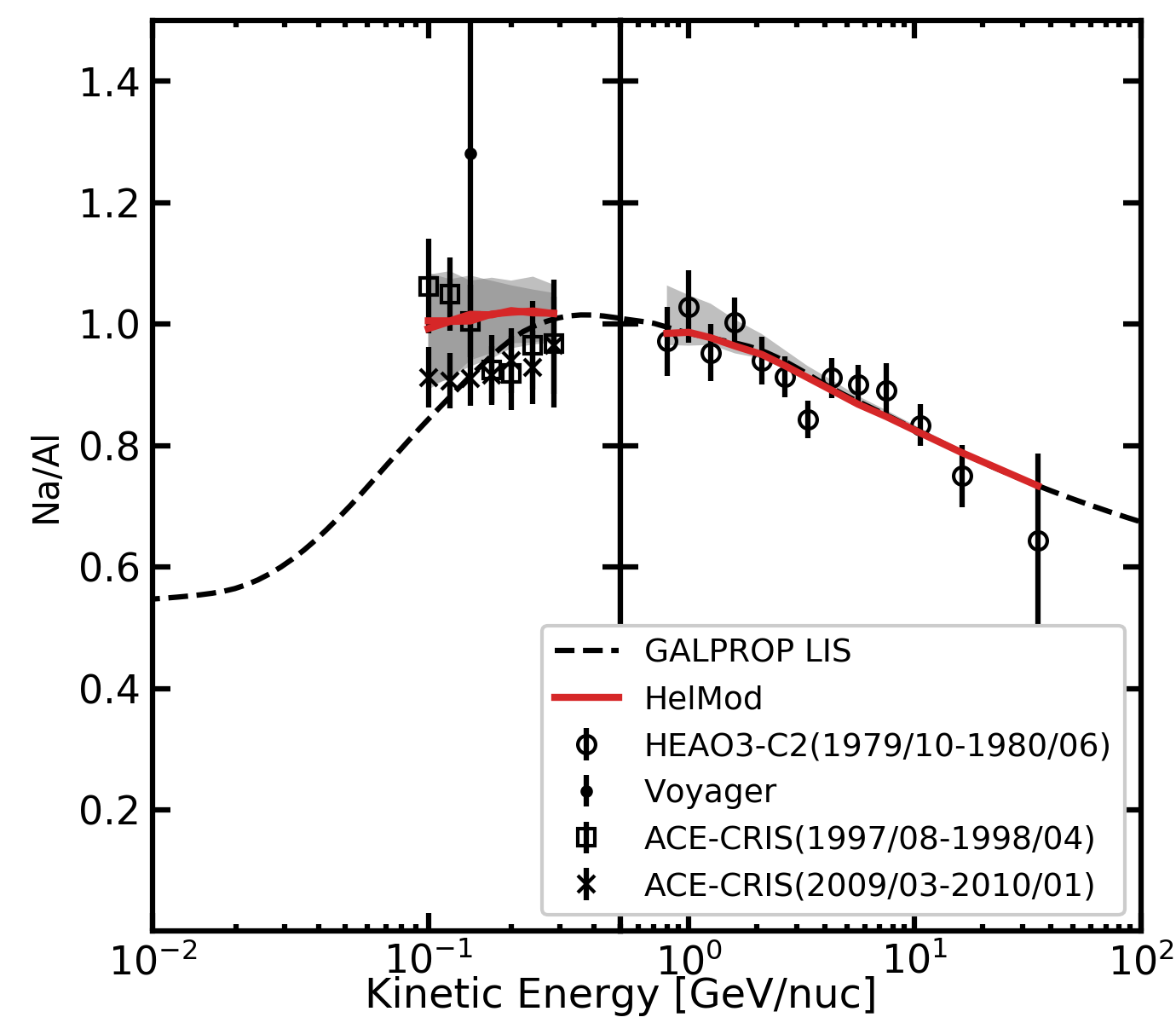}
	\includegraphics[width=0.33\textwidth]{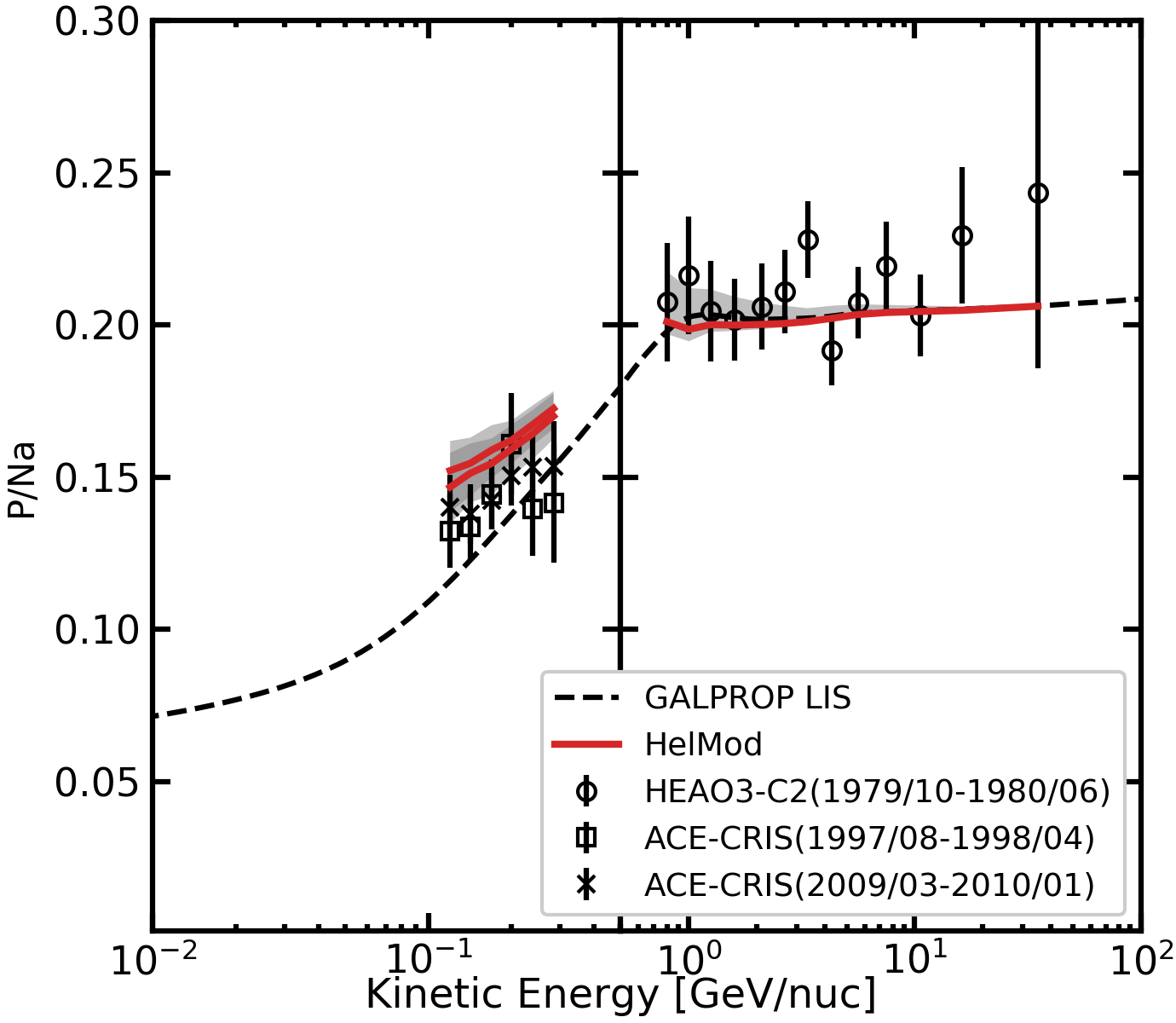}
	\caption{Our model calculations of the ratios of primary/primary species: Si/O, Fe/O, secondary/primary: (Sc+Ti+V)/Fe, F/Ne, Al/Si, P/S, and secondary/secondary: Na/Al, F/V, P/Na, as compared to Voyager 1 \citep{2016ApJ...831...18C}, ACE-CRIS (1997--1998 and 2009--2010), and HEAO-3-C2 \citep{1990A&A...233...96E} data. The line coding as in Figure \ref{fig:He-Ne}.
	}
	\label{fig:ratios}
\end{figure*}

Figure~\ref{fig:ratios} shows the ratios of various nuclei species, such as primary/primary: Si/O, Fe/O, secondary/primary: (Sc+Ti+V)/Fe, F/Ne, Al/Si, P/S, and secondary/secondary: Na/Al, F/V, P/Na as compared to Voyager 1 \citep{2016ApJ...831...18C}, ACE-CRIS (1997--1998 and 2009--2010), and HEAO-3-C2 \citep{1990A&A...233...96E} data. The low-energy part of the ratios is formed by the ionization energy losses that increase sharply with particle charge as $\propto$$Z^2$ and by fragmentation that behaves as $\propto$$Z^{2/3}$. This is very clear in the ratios of the primary species, where no secondary component is involved. One can see that both primary LIS ratios, Si/O and Fe/O, go down at low energies reflecting smaller ionization energy losses of O vs.\ Si or Fe and increase in the fragmentation cross sections for heavier species.

Apart from the ionization energy losses, the fragmentation cross section plays an important role also in {\it flattening} the spectra of heavier species at significantly higher energies where it competes with the escape rate from the Galaxy. Besides that, the large fragmentation cross section of Fe implies that it comes from sources that are closer than the sources of O. The fragmentation becomes less important with energy as particles are leaking from the Galaxy faster. It is interesting to see the difference between the ratios for the species that have about the same fragmentation cross section, Si/O, and for the species, Fe/O, whose fragmentation cross sections differ by about a factor of 3. The former is about flat at high energies while latter is raising at high energies even though all species are primary and the injection spectrum for Fe is slightly steeper than for O (Table~\ref{tbl-inject}). Meanwhile, there are indications that the Fe/O ratio raises even at higher energies, where the leakage from the Galaxy is much faster than the fragmentation timescale. 

Comparing with spectra of other species, one may notice that the observed spectrum of He is harder than the spectrum of protons, while the spectra of C and O have about the same spectral index as He \citep{2018PhRvL.121e1103A}. Given that the most abundant isotopes of C, O, Si all have $A/Z=2$, we see that the increase in $A/Z$ ratio from 1 (protons) to 2 (He, C, O, Si) results in the harder injection spectrum. Heavier elements have even larger $A/Z$ that becomes $\approx2.15$ in the case of $^{56}$Fe, the most abundant isotope of Fe, and may result in even flatter injection spectrum if the described tendency holds. However, the injection index of Fe is $\gamma_3 = 2.19$ vs.\ 2.15 (He), 2.12 (C), 2.13 (O), and 2.19 (Si) in the {\it I}-scenario, and all of them are consistent with each other within the error bars $\pm0.04$. If we look at the injection index $\gamma_2$ (the {\it P}-scenario, $\gamma_3$$=$$\gamma_2$), the picture is even more confusing: $\gamma_2 = 2.40$ (He), 2.43 (C), 2.46 (O), 2.47 (Si), and 2.51 (Fe), i.e.\ the injection spectrum becomes noticeably steeper with $Z$, particularly if we compare He and Fe.

It is interesting to compare our results with two hypotheses, initially based on the observed spectra of H and He, proposing that the observed spectral hardening with $Z$ may be the result of the selective acceleration process. Both of them exploit the fact that a SNR shock weakens with time, thus generating progressively softer spectra. However, physically the two proposed mechanisms are quite different. 

One mechanism proposed by \citet{2011ApJ...729L..13O} and \citet{2016PhRvD..93h3001O} assumes that the early strong SNR shock propagates through the inhomogeneous medium enriched with heavy elements from pre-SNR winds while mixing with more regular ISM at later stages. The effect is then due to the radial layering of different elements. 

The second mechanism does not require the spatial inhomogeneity of element distribution and relies on an analytical study of the dependence of shock injection efficiency upon $A/Z$ ratio \citep{1998PhRvE..58.4911M}. It was predicted to grow and saturate with $A/Z$, but is strongly dependent on the current shock Mach number. Young SNR shocks with higher Mach numbers must preferentially inject elements with higher $A/Z$. After integration over the SNR lifetime, the models by \citet{2012PhRvL.108h1104M} and \citet{2019ApJ...872..108H} recover the hardening of He and other elements with $A/Z \ga 2$. The mechanisms of saturation and ultimate efficiency decline with $A/Z$ that physically should occur in this model are still under investigation. The hybrid simulations \citep{2019ApJ...872..108H} predict the saturation near $A/Z=8-10$ with the most rapid growth between $A/Z=1-2$ (from H to He). 

However, these results are not yet confirmed by independent simulations. Some contradictions with simulations that yield unlimited growth with $A/Z$ have been discussed in the above-mentioned paper. Our calculations show that the injection spectral indices remain the same or even steepen as we move from He to C, O, Si, and Fe. This pattern could be consistent with the injection saturation and decline with $A/Z$. Nevertheless, the two mechanisms described above are not mutually exclusive, which may indicate that the origin of the apparent spectral hardening could be more complicated.

The next plot shows the secondary/primary sub-Fe/Fe = (Sc+Ti+V)/Fe ratio, that is the heavy nuclei analogy of the widely used B/C ratio. This ratio includes species with large fragmentation cross sections and, therefore, is a probe of propagation properties of the {\it local ISM.} The agreement is good in the ``plateau'' energy range given the accuracy of measurements of individual species. Out of three ``sub-Fe'' elements, Ti and V are contributing most. Meanwhile, V is the least accurately measured with two points in the ``plateau'' range being 20\% or $\sim$3$\sigma$ too high relative to the other four (Figure~\ref{fig:Ca-Ni}). This is an indication of large statistical fluctuations or additional systematics. Consequently, minor deviations can be observed in the (Sc+Ti+V)/Fe ratio at lower and higher energies outside of the ``plateau'' that are still consistent with predictions within model uncertainties. These deviations are likely experimental artifacts similar to those observed in the B/C ratio rather than the result of the  differences in the propagation properties of the local ISM. This is also implied by a good agreement with ACE and Voyager 1 data at low energies. Whether such interpretation is correct will become clear after AMS-02 releases the elemental spectra of the iron group. 

The following three plots show different secondary/primary ratios: F/Ne, Al/Si, P/S. In many respects these are medium-nuclei versions of the B/C ratio. The data for all ratios agree well with the predictions indicating that the propagation parameters for the medium nuclei are not changing vs.\ the light species. The upturn in the Al/Si ratio below 100 MeV nucleon$^{-1}$ is partly due to the sharp peak in the Al production cross section in the reactions $p +{} _{14}^{28}$Si $\to {}_{\phm{26,}13}^{26,27}$Al below 100 MeV nucleon$^{-1}$. To illustrate the effect of secondary production, we added a plot of the LIS ratio for the case of pure secondary Al (the primary abundance is set to zero), which also shows a distinct flattening at low energies smoothed by the fast ionization energy losses. The second reason for the upturn is the scattering in the Voyager~1 data points for Al and Si (Figure~\ref{fig:Voy}). The three data points at 20--80 MeV nucleon$^{-1}$ in the Si spectrum appear somewhat too low and so does the model Si spectrum tuned to the data. In the case of Al, it is opposite, and so the Al/Si ratio raises. Higher statistics collected by Voyager 1, 2 spacecraft should resolve this issue.

Finally, we show the mostly secondary/secondary ratios: Na/Al, F/V, P/Na. They are not exactly flat at high energies. The most interesting is the F/V ratio that is analogy to the primary/primary Fe/O ratio. The large fragmentation cross section of Fe nuclei and thus its flatter spectrum compared to lighter O species results in the secondary V spectrum also being flatter than the spectrum of F produced from fragmentation of Si-group nuclei. Meanwhile, the low-energy behavior of the F/V ratio is quite different from the ratio of primaries due to a balance between the competing processes of the ionization energy losses, their fragmentation into the lighter species, and their production from fragmentation of heavier species. In addition, the low-energy parts of the spectra of the Si-group nuclei producing F and of the Fe-group fragmenting onto V are also different due to the differences in fragmentation cross sections and energy losses. 

The decreasing Na/Al ratio reflects the difference in the proportion of the primary component with Na having relatively larger secondary fraction than Al species (Figure~\ref{fig:abund1}). The ratio P/Na is almost flat at high energies that corresponds to about the same proportion of primary component in both species. The slight raise in the ratio reflects the fact that P is produced in fragmentations of heavier species that have the flatter spectrum than the mid-range Si group (cf.\ the behavior of the primary/primary Fe/O and secondary/secondary F/V ratios at high energies, where the heavier species exhibit a flatter spectrum). 

Finally, we wish to share some thoughts about the new breaks in the TeV range observed in the spectra of H and He, Figure \ref{fig:TeV_H_He}. These breaks make the spectra of CR species softer, in contrast to the break at $\sim$370 GV observed in spectra of H through Si, at least, that makes the spectra harder after the break. The spectral indices of primary and secondary species below and above the 370 GV break are well-tuned and are consistent with either {\it I}- or {\it P}-scenario, with the preference given to the {\it P}-scenario that require significantly less free parameters and is consistent with the CR anisotropy measurements. Meanwhile, the breaks in TeV energy range seem to pose a challenge to the {\it P}-scenario, which would require the spectrum of interstellar turbulence and thus the diffusion coefficient to experience corresponding twists. More accurate data from CALET, DAMPE, or ISS-CREAM indicating whether the new breaks are observed at the same rigidity or not and whether they are observed in spectra of other species may provide a clue to their origin.  

\section{Summary}
%%%%%%%%%%%%%%%%%%%%%%%%%%%%%%%%%%%%%%%%%%%%%%%%%%%
%%%%%%%%%%%%%%%%%%%%%%%%%%%%%%%%%%%%%%%%%%%%%%%%%%%

The direct precise measurements of spectra of CR species in the wide energy range are vitally important for CR studies, interpretation of \gray{} and microwave observations, and searches for signatures of new phenomena. The most precise data on CR species that are available in the GV---TV range come from the unique magnetic spectrometer AMS-02 onboard of the ISS. However, moving from the most abundant and light CR nuclei to rare and heavier species takes time to collect statistics and analyze. So far, spectra of only eleven CR nuclei were published by the AMS-02 collaboration, $_1$H--\,$_8$O, $_{10}$Ne, $_{12}$Mg, $_{14}$Si, which show significant deviations from data taken by other instruments in the same energy range. This makes the full-range interpretation of astrophysical measurements difficult and relying on the data from several missions with unknown systematic uncertainties. Folding such uncertainties into the calculations makes the results less reliable and more model-dependent. Meanwhile, waiting for AMS-02 to deliver the accurate spectra of all CR species from $_1$H--\,$_{28}$Ni may take years.

A thorough comparison of the available AMS-02 spectra of CR nuclei and HEAO-3-C2 results reveals a certain energy range, the so-called ``plateau'' from 2.65--10.6 GeV nucleon$^{-1}$, where the appropriately modulated HEAO-3-C2 measurements agree well with more precise AMS-02 data. The analysis of the description of the HEAO-3-C2 counters made possible to unveil the systematics associated with HEAO-3-C2 data that supports our finding. Therefore, the HEAO-3-C2 data in the ``plateau'' energy range can be used as a substitute of AMS-02 data for those species for which AMS-02 measurements are not yet available, meanwhile data outside of this range can be neglected. This is a significant breakthrough that allows reliable CR propagation calculations to be made for all species $_1$H--\,$_{28}$Ni while AMS-02 data are still being analyzed.  

Using the \galprop{}---\helmod{} framework and available data from a number of instruments we derived a self-consistent set of LIS for $_1$H--\,$_{28}$Ni nuclei, and $e^-$, $\bar{p}$ for the first time. The LIS energy range covers 8--9 orders of magnitude in energy from 1 MeV nucleon$^{-1}$ to $\sim$100--500 TeV nucleon$^{-1}$. We provide the final set of propagation parameters as well as the injection spectra and relative abundances for each isotope $_1^1$H--\,$_{28}^{64}$Ni, while $e^-$ and $\bar{p}$ LIS can be found in our previous publications \citep{2017ApJ...840..115B, 2018ApJ...854...94B}. For each element we also provide the analytical parameterization of the LIS as well as their numerical tables that tabulate the LIS in rigidity $R$ and in kinetic energy $E_{\rm kin}$ per nucleon. This is a significant step forward that allows the propagation in the Galaxy and in the heliosphere to be disentangled, while each future measurement can be analyzed within a self-consistent framework.

%%%%%%%%%%%%%%%%%%%%%%%%%%%%%%%%%%%%%%%%%%%%
%%%%%%%%%%%%%%%%%%%%%%%%%%%%%%%%%%%%%%%%%%%%
\acknowledgements
We thank the anonymous referee for valuable comments. Special thanks to Pavol Bobik, Giuliano Boella, Karel Kudela, Marian Putis, and Mario Zannoni for their continuous support of the \helmod{} project and many useful suggestions. This work is supported by ASI (Agenzia Spaziale Italiana) through a contract ASI-INFN I/002/13/0 and by ESA (European Space Agency) through a contract 4000116146/16/NL/HK. Igor V.\ Moskalenko and Troy A.\ Porter acknowledge support from NASA Grant No.~NNX17AB48G. We thank the ACE CRIS instrument team and the ACE Science Center for providing the ACE data. This research has made use of the SSDC Cosmic rays database
%\footnote{https://tools.ssdc.asi.it/CosmicRays/chargedCosmicRays.jsp} 
\citep{2017ICRC...35.1073D} and LPSC Database of Charged Cosmic Rays
%\footnote{https://lpsc.in2p3.fr/cosmic-rays-db/} 
\citep{2014A&A...569A..32M}.

\appendix
\restartappendixnumbering
\section{Plots of the spectra of He--O nuclei}

\begin{figure*}[!hp] 
	\centering
	\includegraphics[width=0.315\textwidth]{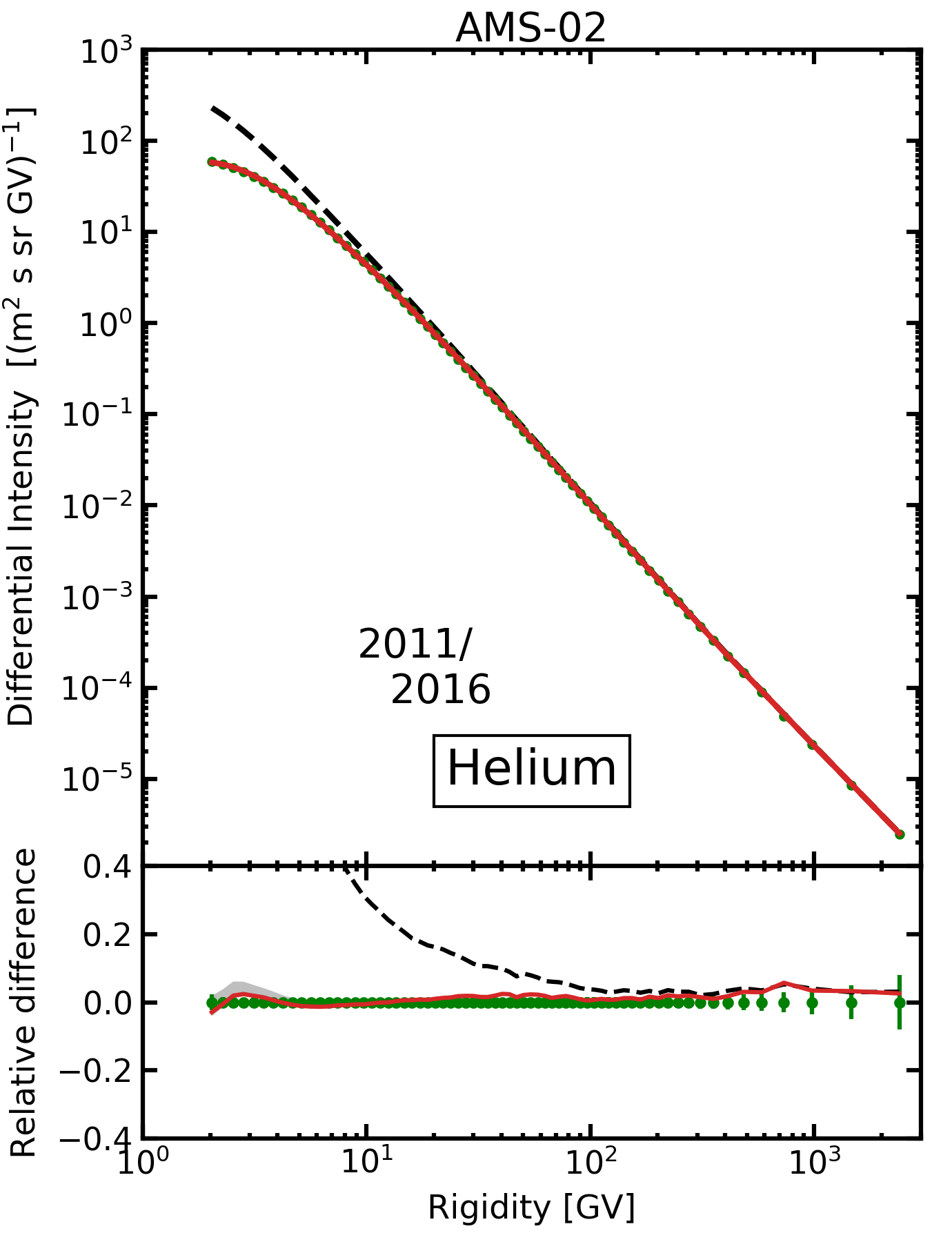}
	\includegraphics[width=0.315\textwidth]{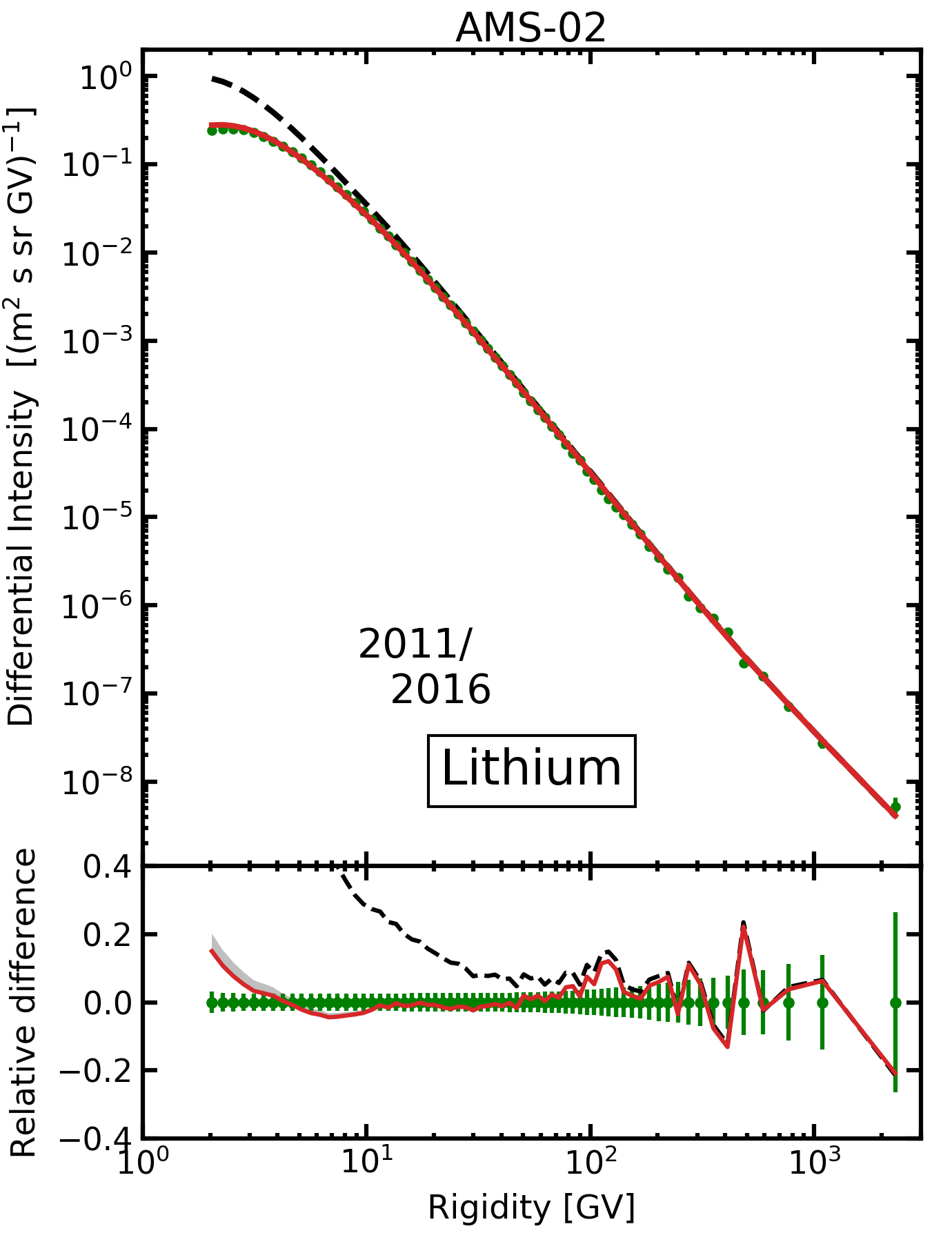}
	\includegraphics[width=0.315\textwidth]{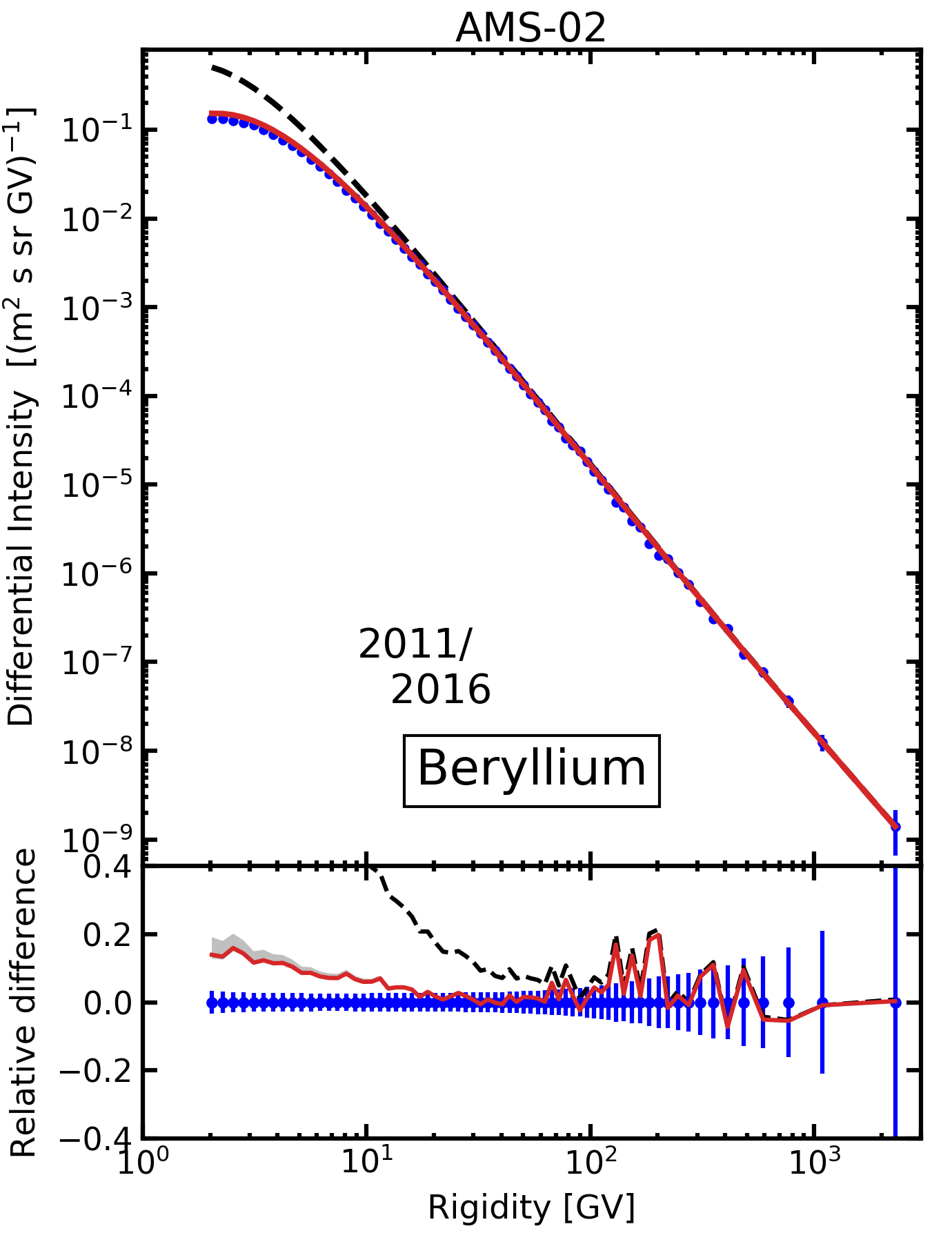}
	\includegraphics[width=0.315\textwidth]{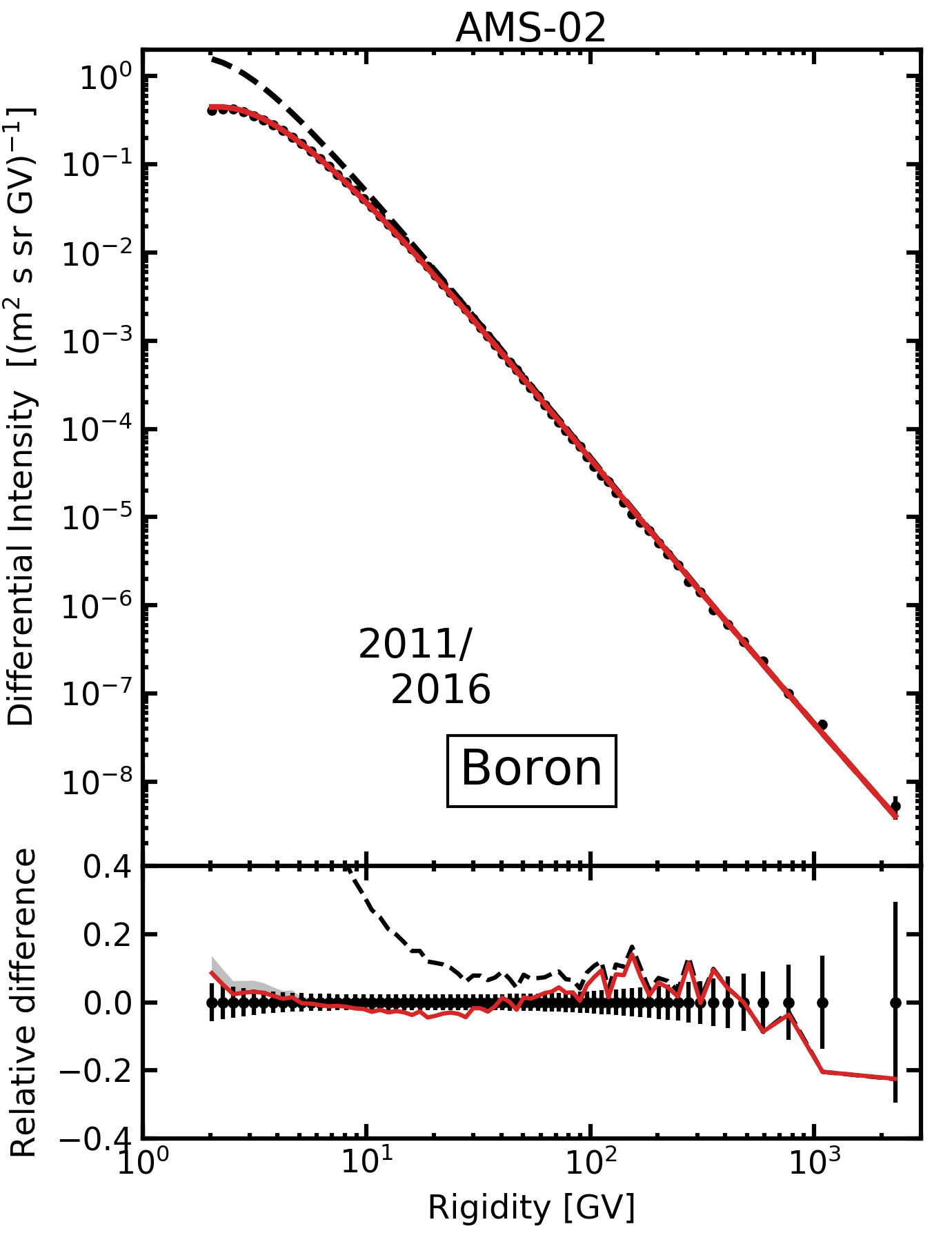}
	\includegraphics[width=0.315\textwidth]{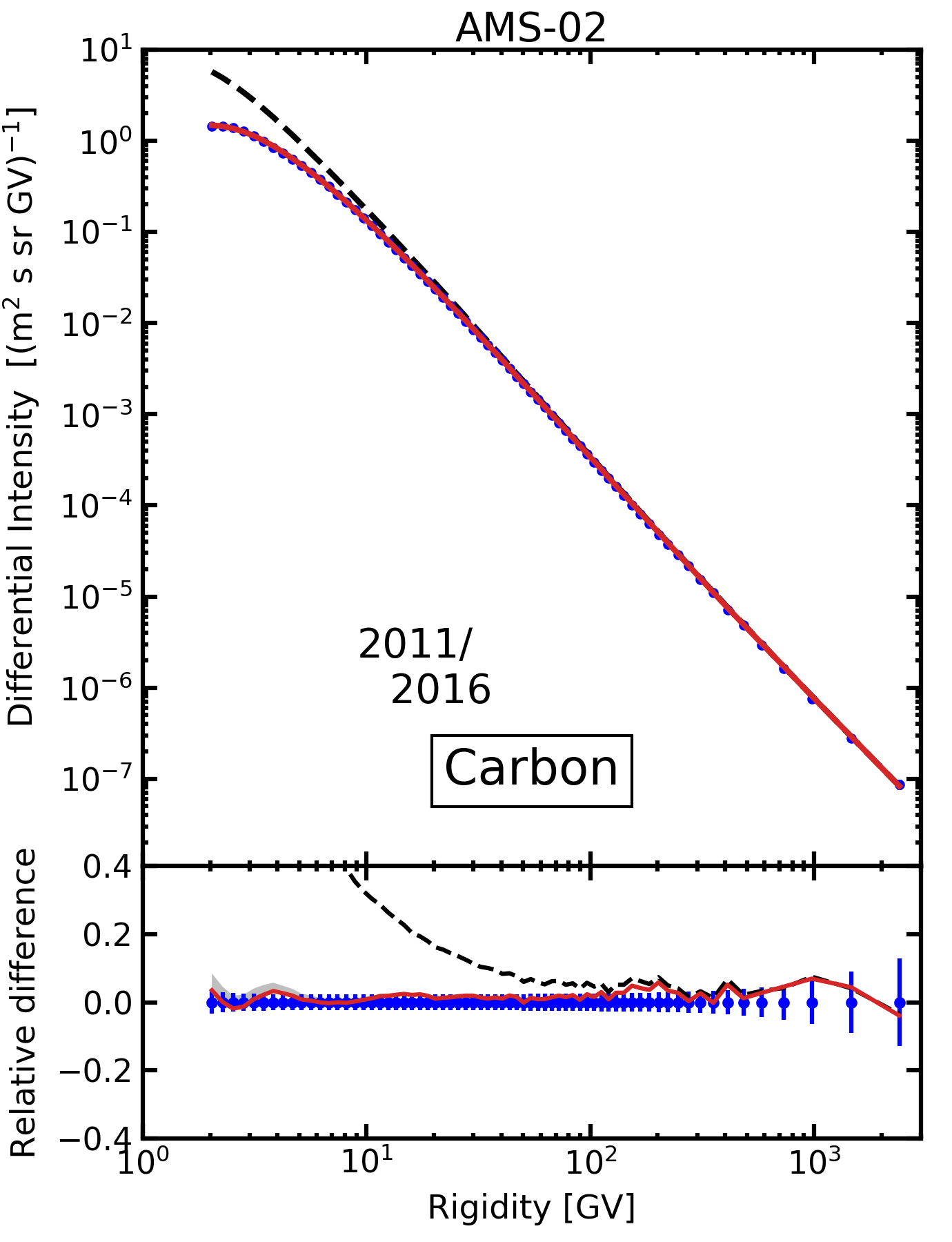}
	\includegraphics[width=0.315\textwidth]{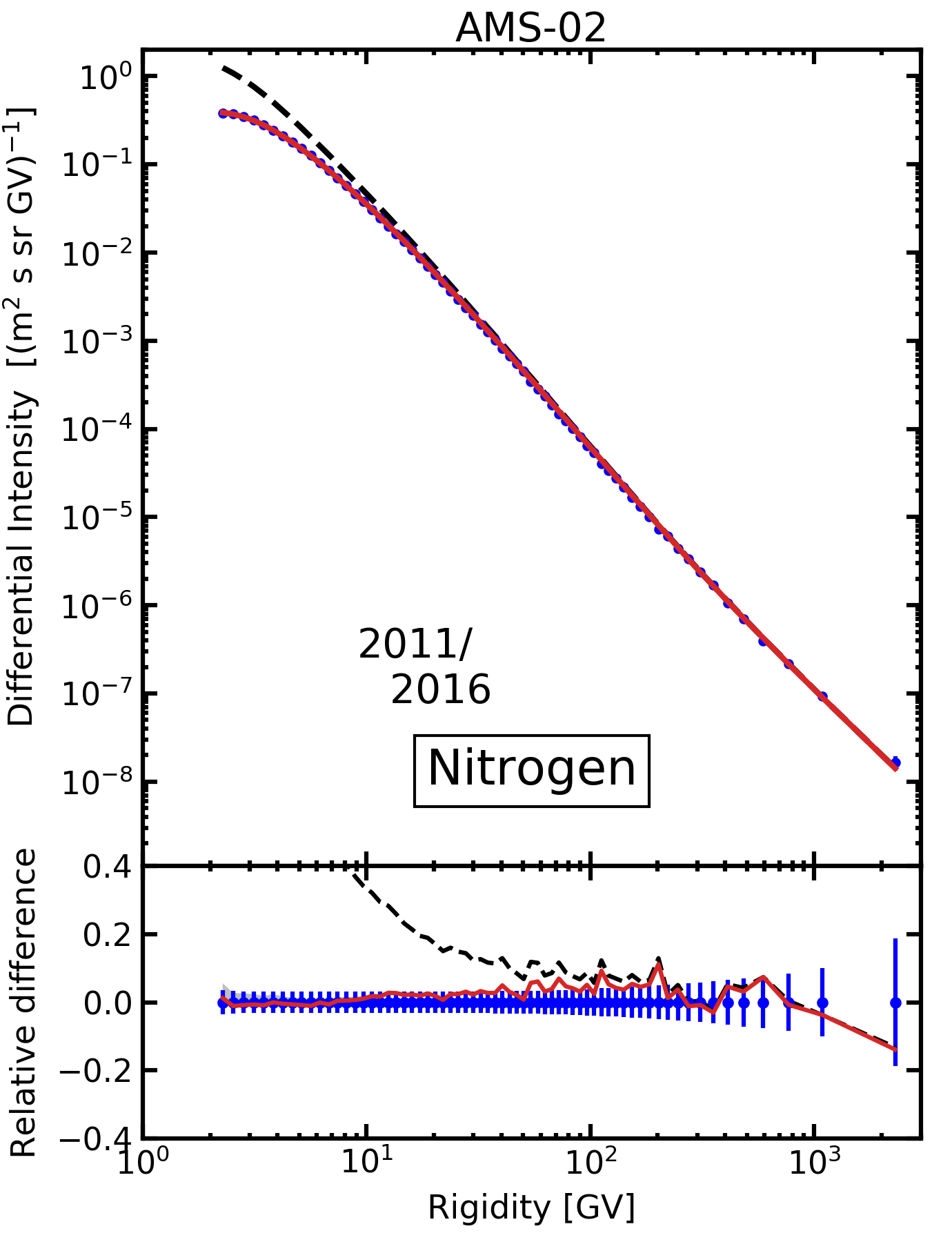}
	\includegraphics[width=0.315\textwidth]{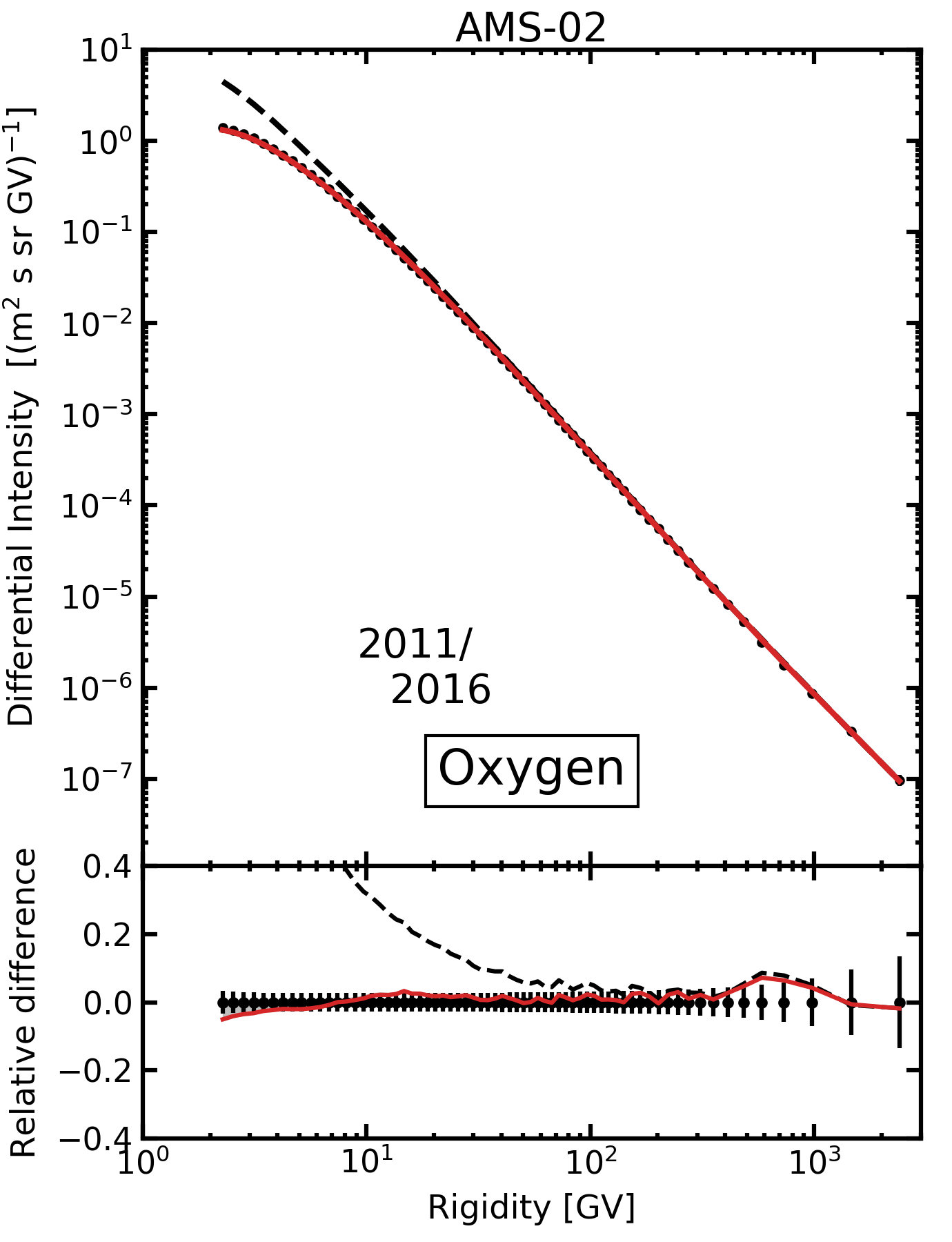}
	\caption{Our updated model calculations compared with AMS-02 data for $_4$He--\,$_8$O nuclei \citep{2015PhRvL.114q1103A, 2015PhRvL.115u1101A, 2016PhRvL.117w1102A, 2017PhRvL.119y1101A, 2018PhRvL.120b1101A, 2018PhRvL.121e1103A}, plotted with the same style that was used in our earlier papers. The line coding as in Figure \ref{fig:He-Ne}. Spectra of protons, Ne, Mg, and Si are shown in Figures \ref{fig:Ne-Si} and \ref{fig:p}.
		}
	\label{fig:AMS_nuclei}
\end{figure*}

For completeness, Figure~\ref{fig:AMS_nuclei} shows the spectra of He--O nuclei published by AMS-02 \citep{2015PhRvL.114q1103A, 2015PhRvL.115u1101A, 2016PhRvL.117w1102A, 2017PhRvL.119y1101A, 2018PhRvL.120b1101A, 2018PhRvL.121e1103A} compared to the updated spectra calculated with the \galprop{}--\helmod{} framework. Spectra of protons, Ne, Mg, and Si are shown in Figures \ref{fig:Ne-Si} and \ref{fig:p}.

\restartappendixnumbering
\section{HEAO-3-C2 counters and measurements of the elemental spectra of CRs}\label{heao3}
%%%%%%%%%%%%%%%%%%%%%%%%%%%%%%%%%%%%%%%%%%%%%%%%%%%
%%%%%%%%%%%%%%%%%%%%%%%%%%%%%%%%%%%%%%%%%%%%%%%%%%%

The French-Danish experiment C2 onboard the NASA HEAO-3 satellite has measured the spectra of CR nuclei for $4\le Z\le 28$ in 14 energy windows from 0.62 to 35 GeV nucleon$^{-1}$ \citep{1990A&A...233...96E}. These spectra were abundantly used in astrophysics to derive parameters of CR propagation and source abundances. Until recently there was no alternative in this energy range, while data at lower (e.g., ACE-CRIS) and at higher energies (e.g., CREAM, \citealt{2008APh....30..133A, 2009ApJ...707..593A} and ATIC-2, \citealt{2009BRASP..73..564P}) connect smoothly without visible breaks, thanks to the energy gap between the ACE-CRIS data and HEAO-3-C2 and the large experimental errors at high energies.

New precise measurements of CR nuclei $_4$Be--\,$_8$O by AMS-02 experiment \citep{2017PhRvL.119y1101A, 2018PhRvL.120b1101A, 2018PhRvL.121e1103A} indicate that there are clear discrepancies with HEAO-3-C2 data at low and high energies \citep{2018ApJ...858...61B, 2020ApJ...889..167B}, while in the middle range between 2.65 and 10.6 GeV nucleon$^{-1}$ the agreement is fair. If there is a particular reason why the whole energy range 0.62 to 35 GeV nucleon$^{-1}$ of HEAO-3-C2 to be clearly split into three intervals with the middle one being the most accurate, it would provide an exiting opportunity to use this limited energy range to build the LIS for all nuclei ($Z\le28$) where the AMS-02 are still unavailable. In this section we look into the published details of the HEAO-3-C2 telescope assembly and discuss possible reasons for the observed split. Because this issue is crucial for future studies, we feel it is necessary to quote some places from the original publications. 

\subsection{The counters}\label{counters}
%%%%%%%%%%%%%%%%%%%%%%%%%%%%%%%%%%%%%%%%%%%%%%%%%%%

The HEAO-3-C2 telescope had three Cherenkov counters for measurements of the particle momentum designed to cover the whole energy range. Each counter was made of a different material with a different refractive index to tune it for a particular energy interval \citep[see Table 1 in][]{1990A&A...233...96E}. The materials used for the counters are: teflon (refractive index 1.33), aerogel block (1.053), and aerogel ``sand'' (1.012). Here is an excerpt from the original paper:

``The determination of the charge and momentum of each incoming particle relies upon the double Cherenkov technique \citep{1970NucIM..81....1C}; the three inner detectors are used primarily for velocity determination and the top and bottom counters for charge determination. The main characteristics of the counters are summarized in Table 1.

Each counter is made of one or two discs of radiating material within a light diffusion box of 60 cm in diameter and viewed by twelve 5-inch photomultiplier tubes. Silica aerogel is used as a Cherenkov material in both C2 and C4 counters. This new material with adjustable refractive index was developed by \citet{1974NucIM.118..177C}, in order to match the particle spectrum observable along the HEAO-3 orbit. The C2 radiator is a mosaic of hexagonal blocks of 5.6 cm thick aerogel; its refractive index is 1.052. Silica aerogel is also used for the C4 counter, but with a lower refractive index (1.012). This material being too brittle to be used in the form of blocks was crushed up into an `aerogel sand', kept in place in the diffusion box by a mylar window. This sand is made of grains about 2 mm in diameter. Two such radiators, each 5.5 cm thick, are placed within the diffusion box \citep{1981ICRC....8...59C}...

In the high `sand energy' region, the signal is distorted by the signal broadening due to the finite resolution of the sand counter. To correct for this effect, we simulate how particles with a given energy are redistributed into measured signal bins, according to both the finite resolution of the counters and the spectral index, then we use these simulation results to correct the number of particles in a given energy window, as explained by \citet{1974ApJ...191..331J}.''

The momentum resolution for each counter is optimal near their thresholds, but quickly degrades as the measured momentum increases, as can be seen in Figure~\ref{fig:heao_counters} (\citealt{1981ICRC....8...59C}, see also Figure A1 in \citealt{1982Ap&SS..84....3B}). However, due to the fluctuations in the background signal, the effective thresholds were set somewhat higher. Here is an excerpt from the original paper \citep{1990A&A...233...96E}:

``...we use teflon spectra up to $\sim$2.4 GeV nucleon$^{-1}$, then aerogel block spectra from $\sim$2.4 to 6.4 GeV nucleon$^{-1}$, and sand spectra above $\sim$6.4 GeV nucleon$^{-1}$. These limits, defining the three energy ranges are chosen slightly higher than the threshold values mentioned in Table 1, as fluctuations in background signal prevent the use of a counter too close to its threshold...

The energy/nucleon limits corresponding to the 14 energy windows are the following:
\begin{itemize}
\item[---] In the teflon range: 0.55, 0.70, 0.91, 1.11, 1.40, 1.82, 2.35 GeV nucleon$^{-1}$;
\item[---] In the aerogel block range: 2.35, 2.96, 3.79, 4.89, 6.42 GeV nucleon$^{-1}$;
\item[---] In the aerogel sand range: 6.42, 8.40, 12.0, 17.8 GeV nucleon$^{-1}$.
\end{itemize}
The last window is the integral momentum ($>$17.8 GeV nucleon$^{-1}$). It is transformed into a differential window, as explained in Appendix 1. The first window can be safely defined only for $Z<10$. At higher charges, the range of the particle becomes comparable to the telescope thickness, so that some particles of lower energy and/or larger incidence angle are absorbed in the telescope matter.'' And further \citep{1982Ap&SS..84....3B}:
``The time-of-flight system is performing well for $Z>6$. For lowest charges the resolution is rather poor...''

The light emission and detection process in the counters was modeled \citep{1981ICRC....8...63L} considering  three sources of the Cherenkov signal. (i) The Cherenkov signal from the primary nucleus is described by the classical expression, and is deemed as well-defined. (ii) As the primary nucleus traverses the instrument it ejects numerous knock-on electrons from the material. Such electrons, if sufficiently fast, will emit Cherenkov radiation in the radiators, and this will contribute significantly to the signal and its fluctuations. They are difficult to evaluate because the electrons can have very high energies ($\sim$$p^2$ MeV), and they are, therefore, not generally confined to the radiator (or material layer) in which they originate. (iii) According to \citet{1981ICRC....8...63L}: ``The light diffusing boxes are lined with white Millipore paper, and the aerogel radiators are held in place by thin mylar foils. Both of these materials emit Cherenkov radiation corresponding to a refractive index $n=1.6$, and this contribution is included in our model.''

\citet{1981ICRC....8...63L} show estimates of the Cherenkov signal contributions from different sources (i)-(iii) (see Figure~1 in their paper). While in teflon and aerogel block counters a sum of the secondary (ii) and foil (iii) contribution does not exceed 10\%, the aerogel sand has a sum of those contributions very flat and approaching 17\% in the whole momentum range.

Figure~\ref{fig:heao_counters1} \citep[reproduced from][]{1981ICRC....8...63L} shows consistency between the final momentum assignments using all three counters. Teflon counter shows overall consistency with aerogel block counter, but exhibits large errors above 4.5 GeV c$^{-1}$ nucleon$^{-1}$. The errors become quite large for aerogel block and sand counters above $\sim$10 GeV c$^{-1}$ nucleon$^{-1}$, which can be seen from the large scattering of the points and shown large typical error bars. Here is an excerpt from the original paper \citep{1981ICRC....8...63L}:

\begin{figure}[tb!]
	\centering
	\includegraphics[width=0.47\textwidth]{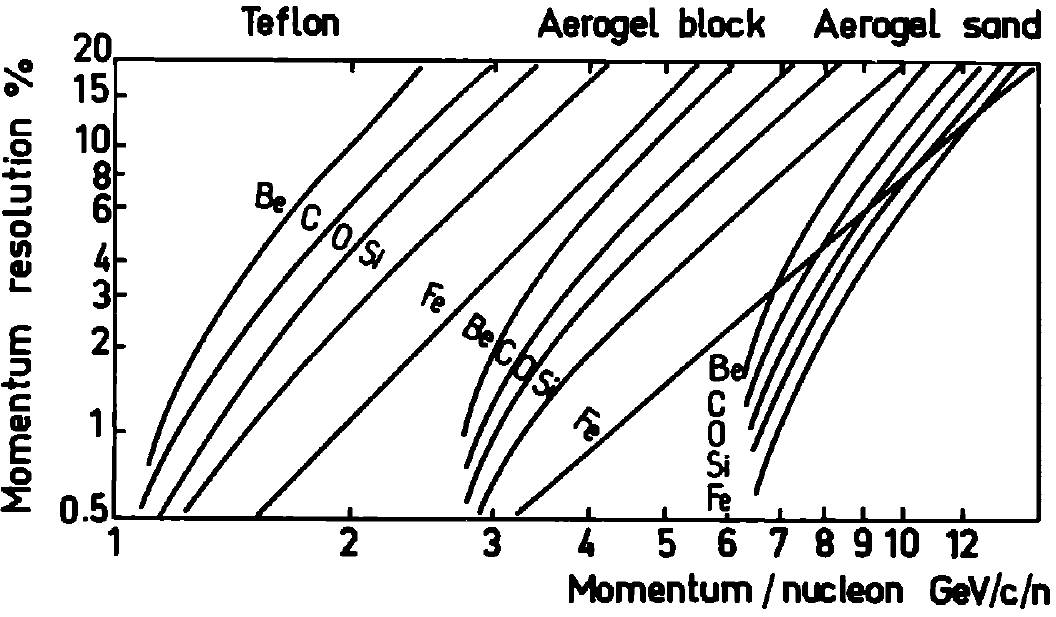}
	\caption{Momentum resolution in the three counters for Be, C, O, Si, and Fe cosmic ray nuclei \citep[reproduced from][]{1981ICRC....8...59C}.}
	\label{fig:heao_counters}
\end{figure}

``All the parameters entering into the above model have been determined by measurements on the ground with exception of the following: a) The effective refractive index for Cherenkov emission in the aerogel sand radiator. The value adopted (for consistency reasons) is 1.012 whereas laboratory optical measurements had predicted $1.015 \pm 0.002$. b)~The rate of occurrence of the `faceplate flashes' (section~6). To explain the observed fluctuations near threshold we had to assume an occurrence frequency of $5\times10^{-4}$ per g cm$^{-2}$ of overlying material for singly charged particles, the effect being proportional to $Z^2$. c) The signal values in all three counters for particles with $\beta$=1. This critical parameter is difficult to determine precisely from the observed signal distributions as both the cosmic ray energy spectrum and the variation of the knock-on contribution to the signal and its fluctuations enter into the calculations in a complex way. Starting from the observed `$\beta$=1 points' for nickel nuclei we have adjusted these for optimum consistency between the three counters. The optimized values, which only differ slightly from the raw ones, are then scaled to other nuclei, assuming $Z^2$ proportionality.''

\subsection{Geomagnetic rigidity cutoff selection}\label{cutoff}
%%%%%%%%%%%%%%%%%%%%%%%%%%%%%%%%%%%%%%%%%%%%%%%%%%%

\begin{figure}[tb!]
	\centering
	\includegraphics[width=0.47\textwidth]{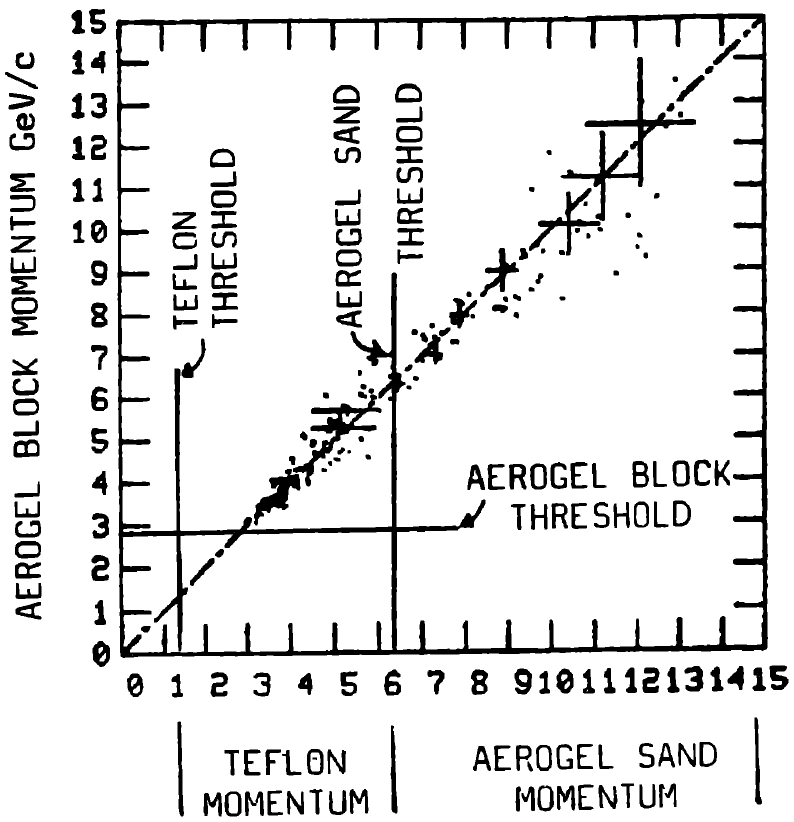}
	\caption{Consistency of final momentum assignment \citep[reproduced from][]{1981ICRC....8...63L}. Examples of the correlation between the momentum determinations from the three counters are shown for nickel nuclei. The error bars shown are one sigma as predicted by the model.
	}
	\label{fig:heao_counters1}
\end{figure}

The accelerator beam calibration was not possible at that time, so the geomagnetic cutoff rigidity that is varying along the HEAO-3 orbit was used instead. Here is how this procedure is described in \citet{1990A&A...233...96E}:

``Now if we want to get the true spectrum of the particles, as they would be observed outside the magnetosphere, we have two possibilities:

(i) either to use all particles above the geomagnetic cut off at the point of measurement (here again we must exclude particles of rigidity close to the cut off). In that case, we get a distorted spectrum, which can be corrected by applying an energy dependent correction factor, for the proportion of time spent at various rigidity cut offs during flight. 

(ii) or to select only the particles registered at rigidity cut offs below that corresponding to the velocity threshold of the counter considered (as an example below 2.3 GV for the teflon counter). In this last approach, only a small proportion of particles are selected (about 10\% in the case of the teflon counter), and therefore we cannot use directly this method to get the energy spectrum of a rare element like P or Cl. We can use it however to get the spectrum of a reference element like oxygen and derive the spectra of other nuclei by using their relative abundances with respect to the reference element, as given in Table 2. In that case of course, the accuracy of every point in each spectrum is limited by the statistical accuracy of the corresponding point of the oxygen spectrum...

In the present paper, we preferred to use a modified version of the second method: by a proper selection of data, we are able to determine absolute flux values for each type of nucleus, in units of particles m$^{-2}$ sr$^{-1}$ s$^{-1}$ (GeV/n)$^{-1}$. More precisely, to determine the spectrum above a threshold rigidity $R_0$, we have selected the data in such a way that:
\begin{itemize}
\item[(i)] the geometry of the instrument is accurately known;
\item[(ii)]  the geomagnetic cut off corresponding to all viewing directions of the telescope is lower than $R_0$; 
\item[(iii)] the time during which these conditions are met is accurately measured. 
\end{itemize}
	
All these conditions are fulfilled by selecting data when the telescope axis lies near the local vertical and when the vertical rigidity cut off is lower than 0.8 $R_0$. The instrument is then single-ended, since it is protected by the solid earth from particles propagating in the reverse direction. The acceptance geometry is known and we have verified that condition (ii) is fulfilled.

Practically, we first select periods of time when the angle between the telescope axis and the local vertical is lower than $25^\circ$. In order to have a well defined restricted geometry, we require in addition that the impact points of the particle in the top and bottom counters lie within a circle of 52 cm in diameter, i.e.\ well inside the radiators of the charge counters, which are 60 cm in diameter. In these conditions, the maximum acceptance angle of the telescope is $28^\circ$ and the geometrical factor: $F=413$ cm$^2$ sr. We then derive the oxygen spectrum between 0.55 and 2.4 GeV nucleon$^{-1}$ from data registered at rigidity cut off values smaller than 1.85 GV and that above 2.4 GeV nucleon$^{-1}$ from data registered at cut off values smaller than 5.0 GV, so as to improve statistics...''

\subsection{Summary of the HEAO-3-C2 telescope performance}\label{summaryHEAO}
%%%%%%%%%%%%%%%%%%%%%%%%%%%%%%%%%%%%%%%%%%%%%%%%%%%

Above we described main reasons that, in our view, are most likely responsible for the observed discrepancies, but many other issues that may affect calibration of the instrument exist. 

It is clear that the whole rigidity range of the instrument is divided between three different counters, which have different responses. The teflon and aerogel block counters are well-studied, while a newly introduced aerogel sand counter is not. Besides small momentum range and large errors in the momentum resolution, its refractive index has to be adjusted after the launch to make it consistent with other counters response. Therefore, it is likely that the discrepancy between AMS-02 and HEAO-3-C2 data observed at high-energies is partly due to the momentum and mass resolutions in the aerogel block and sand counters above $\sim$9 GeV nucleon$^{-1}$, where the resolution of the block counter degrades being far from the threshold, while the sand counter has large intrinsic errors. 

Additionally, the CR spectrum becomes steep at high energies, approaching a power-law index about --2.7. Therefore, the worsening energy resolution of the sand counter has a greater impact on the quality of measurements than that for the other two counters. Interestingly enough, the authors note that ``In comparison with other published data, HEAO-3 data above 10 GeV/n are on the lower side of the distribution, but the spread of the points is relatively large'' \citep[p.104,][]{1990A&A...233...96E}. This further supports our conclusion since in this energy range the solar modulation is moderate. 

The discrepancy at low energies is likely due to the variations in the rigidity cutoff along the orbit of the instrument that was used for the event selection. The HEAO-3 instrument was flown from 20 September 1979---29 May 1981, i.e.\ during the solar maximum conditions (Cycle 21). Meanwhile, during the disturbed periods when the solar wind pressure exceeds $\sim$4 nP, and which occur more often when the solar activity is high, the magnetic field carried by the solar wind affects the geomagnetic field leading to incorrect estimate of the geomagnetic cutoff. In turn, it can result in incorrect estimates of the particle rigidities during a significant fraction of the mission. The effect could be quite large and is more important for lower rigidities. The influence of such disturbances on AMS-02 was investigated, and the appropriate code was developed.\footnote{http://www.geomagsphere.org}

In our analysis, we rely on the mid-range results of the HEAO-3-C2 telescope as measured by the aerogel block counter and the first two energy bins of the sand counter. The same energy range is used for all species.

\bibliography{bibliography}

\afterpage{\clearpage}
\section{Supplementary Material}\label{app:SupMat}
\subsection{Primary fluorine}\label{fluorine} 
%%%%%%%%%%%%%%%%%%%%%%%%%%%%%%%%%%%%%%%%%%%%%%%%%%%

Among the elements $A\ge5$ produced in stars, the abundance of fluorine is anomalously low. This is mainly due to the fact that fluorine is easily destroyed in stars through either $p$- or $\alpha$-captures. The solar system abundances of fluorine relative to oxygen ranges from F/O $= 7.4\times10^{-5}$ in the photosphere to $10.5\times10^{-5}$ in meteorites \citep{2009ARA&A..47..481A}. This can be compared with the CR sources (Table~\ref{tbl-Ab}), F/O $= 29.1\times10^{-5}$ or [F/O]$_{\rm srs}=0.44-0.60$ dex dependently on the (F/O)$_\odot$ used, where [X/Y] = $\log_{10}$(X/Y)--$\log_{10}$(X/Y)$_\odot$, which are apparently 3--4 times more abundant with fluorine.

The main astrophysical sources of fluorine are thought to be supernovae Type II (SN II), Wolf--Rayet (WR) stars, and the asymptotic giant branch (AGB) of intermediate-mass stars \citep[e.g.,][]{2000A&A...355..176M, 2004MNRAS.354..575R, 2019MNRAS.490.4307O}. These sources become important at different stages of chemical evolution of the Galaxy, with the $\nu$-spallation of $_{10}$Ne in SN II dominating at early times in low metallicity environment, while WR and AGB stars dominating at the later stages at solar and supersolar metallicities. The calculations show that reaching the solar level of [F/O] = 0 at present epoch requires all three types of sources to contribute \citep{2004MNRAS.354..575R}, while exceeding it is problematic. Therefore, the excess fluorine in the CR sources [F/O]$_{\rm srs}=0.44-0.60$ dex may provide important constraints on the origin of CRs.

For quite a long time people noticed that some isotopic ratios in CRs, such as $^{12}$C/$^{16}$O, $^{22}$Ne/$^{20}$Ne, and $^{58}$Fe/$^{56}$Fe, significantly exceed values observed in the solar system. This was interpreted as an evidence that some fraction of CRs is produced from a material of the winds of the WR stars \citep{2007SSRv..130..439B}. The proportion of $\sim$20\% of WR material mixed with 80\% material with solar-system composition reproduces the derived CR source abundances quite well. Since WR stars are evolutionary products of OB stars, and OB stars are formed in associations, the material of such winds can be accelerated by a following SN shock passing through the ISM. A recent observation of the radioactive $^{60}$Fe by ACE-CRIS \citep{2016Sci...352..677B} supports the hypothesis of the SN explosion in the solar neighborhood 2--3 Myr ago. This estimate is consistent with other observations, such as the change of the isotopic composition of the deep-sea manganese crust \citep{2004PhRvL..93q1103K} and elevated ratio $^{60}$Fe/Fe in debris on the lunar surface \citep{2016PhRvL.116o1104F}. 

Observation of the excess fluorine in the CR source composition seems to agree well with these observations. A study of the rotating and non-rotating WR stars with the initial composition corresponding to the solar system shows an enrichment of the WR winds with fluorine relative to oxygen by a factor of $\sim$2 \citep{2001SSRv...99...73M}. The calculation was made for a star of $60M_\odot$ during its WN and WC stages, and in all cases the wind was enriched with fluorine by about the same factor. 

However, the current estimate of the $_9$F abundance in CR sources is not very reliable given the accuracy of the available cross sections. The calculated abundance ratio at the CR sources is (F/Ne)$_{\rm srs}$ $\approx 2.5\times10^{-3}$, while in CRs it is (F/Ne)$_{\rm CR} \approx 0.1$ at 10 GeV nucleon$^{-1}$. Therefore, the relative fraction of primary fluorine in CRs is (F/Ne)$_{\rm srs}$/(F/Ne)$_{\rm CR}\approx 0.025$. The fractional contribution of fragmentation of different CR species into the secondary isotopic production was evaluated by \citet{2013ICRC...33..803M} for all isotopes from $^6_3$Li to $^{62}_{28}$Ni with $^{64}_{28}$Ni being the heaviest contributor. The main production channels for $^{19}_{\phn9}$F are fragmentations of $^{20}_{10}$Ne, $^{22}_{10}$Ne, $^{24}_{12}$Mg, and $^{28}_{14}$Si isotopes with less important contributions from fragmentation of $^{21}_{10}$Ne, $^{23}_{11}$Na, $^{25}_{12}$Mg, $^{26}_{12}$Mg, and $^{27}_{13}$Al. Also contributing are decays of intermediate nuclides $^{19}_{10}$Ne[$\beta^+$] and $^{19}_{\phn8}$O[$\beta^-$]. 

Taking into account only the main channels for production of $^{19}_{\phn9}$F and assuming that the propagated abundances of $_{10}$Ne, $_{12}$Mg, and $_{14}$Si are about the same and these elements consist of only one main isotope, we can simply add their contributions: $\sigma_{\rm F}\approx 26\, {\rm mb}\, (_{10}{\rm Ne}) + 10\, {\rm mb}\, (_{12}{\rm Mg}) + 9\, {\rm mb}\, (_{14}{\rm Si}) + 28\, {\rm mb}\, (^{20}_{10}{\rm Ne} \to ^{19}_{10}{\rm Ne}[\beta^+]) + 4\, {\rm mb}\, (^{24}_{12}{\rm Mg} \to ^{19}_{10}{\rm Ne}[\beta^+]) \approx 80$ mb. Here we used accelerator data at $\sim$600 MeV nucleon$^{-1}$ and assumed the same values at higher energies. The effective total production cross section of $^{19}_{\phn9}$F weighted with the propagated abundances of other contributing isotopes at 10 GeV nucleon$^{-1}$ would then be of the order of 100 mb. The ratio (F/Ne)$_{\rm srs}$/(F/Ne)$_{\rm CR}\approx 0.025$ implies that 2.5\% increase in the effective production cross section, i.e.\ $\approx$2.5 mb, would be enough to compensate for the whole primary abundance of $^{19}_{\phn9}$F. The value 2.5 mb is comparable with the typical 10\% cross section error for the main production channels, such as the fragmentation $^{20}_{10}$Ne $\to ^{19}_{\phn9}$F and $^{20}_{10}$Ne $\to ^{19}_{10}$Ne[$\beta^+$] $\to^{19}_{\phn9}$F, and thus the primary abundance of fluorine is within the cross sections uncertainty and consistent with zero.

Though this exercise does not allow us to make a reliable conclusion about the primary fluorine, it shows how important  the accuracy of the production cross sections is for interpretation of CR measurements.

\subsection{LIS for elements from H to Ni}

Here we provide analytical parameterizations of the LIS for each species, Eqs.~(\ref{eq:H})-(\ref{eq:Ni}), with parameters summarized in Table~\ref{Tbl-1-10}. They are complemented by numerical tables calculated for the {\it I}-scenario for all elements from $_1$H to $_{28}$Ni, Tables~\ref{Tbl-ProtonLIS-EKin}-\ref{Tbl-NickelLIS-Rigi}. The {\it I}-scenario has more free parameters than the {\it P}-scenario and, therefore, gives more accurate representation of the data. For each element, we provide two numerical tables that tabulate the LIS in kinetic energy $E_{\rm kin}$ per nucleon and in rigidity $R$. 

\afterpage{\clearpage}
\pagebreak[4]
\restartappendixnumbering
\allowdisplaybreaks

\subsubsection{Analytical parameterizations of the LIS}

\begin{align}
	    % -------------------------------------------------- Z = 1 Proton ---------------------------------------------
	\label{eq:H}
	F_{\rm H}& (R)  = \\
	&\begin{cases}
	\displaystyle  a R^2 + b L(R)^2 + (c + d\tilde{R})G\left[\frac{R}{L(R)}\right] - f - g\tilde{R} - hL(R) i^{G\left[\frac{R}{L(R)}\right]},    &R\le 2.5\, {\rm GV}, \smallskip\\
	\displaystyle  R^{-2.7} \left[ - l - mR+ nL(R) + oG(pR) + \left(q\tilde{R}  - s^{L(R)} r\right)\cos{(t + u\tilde{R})} \right],  &R> 2.5\, {\rm GV},
	\nonumber
	\end{cases}\\[8pt] 
    % -------------------------------------------------- Z = 2 Helium ---------------------------------------------
	\label{eq:He}
	F_{\rm He}& (R)  = \\
	&\begin{cases}
	\displaystyle  a - b R + c\sqrt{\sin(R)} + \left[d R G(f R)^2  - g R\right]\sin(R) -( h+iR) G(f R),    &R\le 2.5\, {\rm GV}, \smallskip\\
	\displaystyle  R^{-2.0} \left[ l + \frac{m}{R} -\frac{n}{R^2} + \frac{o}{p + qR} + r\tanh\left(\frac{s}{t + uR}\right) - v z^{\frac{3}{R}} \right],  &R> 2.5\, {\rm GV},
	\nonumber
	\end{cases}\\[8pt] 
	% -------------------------------------------------- Z = 3 Lithium ---------------------------------------------
	\label{eq:Li}
	F_{\rm Li}& (R)  = \\
	&\begin{cases}
	\displaystyle  R^{1.5}\left\{ - a + bR + cR^2 - dR^{1.5} +T(R)\left[f+ g G\left(G(R)^{rR}\,T(R)\right) \right]\right\},    &R\le 2.6\, {\rm GV}, \smallskip\\
	\displaystyle  R^{-2.7} \left\{ h + i\sqrt{R} - l R + m R\tilde{R}  + n\sin{\tilde{R}} +\frac{1}{R}\left[ o\tilde{R}^{1.5}+p\sqrt{\tilde{R}}G\left(\sqrt{\tilde{R}}\right)  \right] - q G\left(\sqrt{\tilde{R}}\right)   \right\},  &R> 2.6\, {\rm GV},
	\nonumber
	\end{cases}\\[8pt] 
    % -------------------------------------------------- Z = 4 Beryllium ---------------------------------------------
	\label{eq:Be}
	F_{\rm Be}& (R)  = \\
	&\begin{cases}
	\displaystyle \displaystyle aR - bR^2 + cR^4 + R G(R) \left[- d - f R - g G(R)+ h L(i R^2)\sqrt{R} \right],    &R\le 2.6\, {\rm GV}, \smallskip\\
	\displaystyle R^{-2.7} \left\{ l + m G\left(n + oR + pR^{-1}\right) + q\left[r + R + \frac{s}{R - t}\right]^{-0.5} - uR \right\},  &R> 2.6\, {\rm GV},
	\nonumber
	\end{cases}\\[8pt] 
    % -------------------------------------------------- Z = 5 Boron ---------------------------------------------
	\label{eq:B}
	F_{\rm B}& (R)  = \\
	&\begin{cases}
	\displaystyle a + \cos{R} \left[- b + c R  + d G(f R)R^3 \right] - g R - h G(i R^2),    &R\le 2.7\, {\rm GV}, \smallskip\\
	\displaystyle R^{-2.7} \left[ l -\frac{m}{R} + \frac{n}{\sqrt{R}} -\frac{o}{p + q R} + \frac{r}{s + t R \sqrt{R}} - u\log{\tilde{R}} -v R^2 \sqrt{R} \right],  &R> 2.7\, {\rm GV},
	\nonumber
	\end{cases}\\[8pt] 
    % -------------------------------------------------- Z = 6 Carbon ---------------------------------------------
	\label{eq:C}
	F_{\rm C}& (R)  = \\
	&\begin{cases}
	\displaystyle a + R(- b + c R^2) + G(R)\left\{- d + R\left[f + g R G(R)\right]\right\} - h R G(i R),    &R\le 2.55\, {\rm GV}, \smallskip\\
	\displaystyle R^{-2.7} \left[ l + m R -n R^2 -\frac{o}{R} + \frac{p}{q + R} + r\tilde{R}^2 + s\log{\left(tR + \frac{u}{v + R}\right)}  \right],  &R> 2.55\, {\rm GV},
	\nonumber
	\end{cases}\\[8pt] 
	 % -------------------------------------------------- Z = 7 Nitrogen ---------------------------------------------
	\label{eq:N}
	F_{\rm N}& (R)  = \\
	&\begin{cases}
	\displaystyle  - a - b\tilde{R} + cR \tilde{R} - d\sin{(R)}\tilde{R} + f\sin{(R)} - g\tanh{(hR)}   + iG(l\sin^2{R})  ,    &R\le 3.0\, {\rm GV}, \smallskip\\
	\displaystyle  R^{-2.7} \left[ m -\frac{n}{R} + \frac{o}{p + qR} - rR - s\tilde{R} + t\log(p + qR) + \frac{uR}{\log(p + qR)}  \right],  &R> 3.0\, {\rm GV},
	\nonumber
	\end{cases}\\[8pt] 
	% -------------------------------------------------- Z = 8 Oxygen ---------------------------------------------
	\label{eq:O}
	F_{\rm O}& (R)  = \\
	&\begin{cases}
	\displaystyle   a - bR - c\tilde{R} + d\tanh{(R)} + f\tilde{R}\tanh{(R)}\tanh{(\tilde{R})} + g\tanh(\tilde{R}) + h\tanh^4(\tilde{R}) - i\tanh^2(\tilde{R}),    &R\le 2.65\, {\rm GV}, \smallskip\\
	\displaystyle  R^{-2.7} \left[ lR + m\log{(R + \tilde{R})} + \frac{n}{\sqrt{o + pR}} -\frac{q}{(o + pR)^{0.25}} - r - s\sqrt{o + pR} - tR(o + pR)^{0.25} \right],  &R> 2.65\, {\rm GV},
	\nonumber
	\end{cases}\\[8pt] 
	% -------------------------------------------------- Z = 9 Fluorine ---------------------------------------------
	\label{eq:F}
	F_{\rm F}& (R)  = \\
	&\begin{cases}
	\displaystyle a - bR + \sqrt{cR}(d - fR)  - G(R)\left\{ g + R^2\left[h  + i G(R)^2\right]\right\},    &R\le 2.5\, {\rm GV}, \smallskip\\
	\displaystyle  R^{-2.7} \left[ - l - m \tilde{R} + nR -\frac{o}{R} + \frac{pR - q\tilde{R}}{r + R} + \frac{s\tilde{R}}{t + u\tilde{R} + R\tilde{R}}  \right],  &R> 2.5\, {\rm GV},
	\nonumber
	\end{cases}\\[8pt] 
	% -------------------------------------------------- Z = 10 Neon ---------------------------------------------
	\label{eq:Ne}
	F_{\rm Ne}& (R)  = \\
	&\begin{cases}
	\displaystyle  a - bRH(R) + cH(R)+ dH(R)^2 + G\left[f - H(R)- gRH(R)\right]\left[hH(R)^2- i\right] ,    &R\le 2.4\, {\rm GV}, \smallskip\\
	\displaystyle  R^{-2.7} \left[ l -\frac{m}{\sqrt{R}} + nR^{1.5} - oR^2 + \left( r\sqrt{R}  - s - tR\right)\tanh{\left(pR -\frac{q}{\sqrt{R}}\right)}  \right],  &R> 2.4\, {\rm GV},
		\nonumber
		\end{cases}\\[8pt] 
%\end{align}
%\begin{align}
	% -------------------------------------------------- Z = 11 Sodium ---------------------------------------------
	\label{eq:Na}
	F_{\rm Na}& (R)  = \\
	&\begin{cases}
	\displaystyle a - bR -\frac{c}{G(R)} + \left(dR  - f\right)G(R) - [g+ hG(R)]\sin{\left[2G(R)^3\right]},    &R\le 2.4\, {\rm GV}, \smallskip\\
	\displaystyle R^{-2.7} \left[ i  - lR - n^R m + o\sqrt{R} + p\left( q^R \log(\tilde{R}) + r^R\tanh^5(R) \right)  \right],  &R> 2.4\, {\rm GV},
	\nonumber
	\end{cases}\\[8pt] 
	    % -------------------------------------------------- Z = 12 Magnesium ---------------------------------------------
	\label{eq:Mg}
	F_{\rm Mg}& (R)  = \\
	&\begin{cases}
	\displaystyle - a - bR - cR^2\tilde{R}+L\left(d R^2\right) \left[f- g G(R) \right] + h G\left(iR^2\right),    &R\le 2.5\, {\rm GV}, \smallskip\\
	\displaystyle R^{-2.7} \left[ - l - m R^2 -\frac{n}{R} + o\tilde{R} + p R\tilde{R}^2  + \frac{q}{r + R} + \frac{s}{t + uR}   \right],  &R> 2.5\, {\rm GV},
	\nonumber
	\end{cases}\\[8pt] 
	% -------------------------------------------------- Z = 13 Aluminium ---------------------------------------------
	\label{eq:Al}
	F_{\rm Al}& (R)  = \\ %\times R^{2.7}
	&\begin{cases}
	\displaystyle a + b\tanh(\tilde{R})      + c\tanh[\log{G(\tilde{R})}]              - \tilde{R} (d + f R)             + G(\tilde{R}) [g-h \log{G(\tilde{R})}],    &R\le 2.7\, {\rm GV}, \smallskip\\
	\displaystyle R^{-2.7}\left\{ - i +  R \left(l- m\sqrt{R}\right) - n \sqrt{R}+\tilde{R} \left[- o+ \frac{p}{\sqrt{q\tilde{R}+R\log{(r + R)}}} \right] + s \log{(r + R)}   \right\},  &R> 2.7\, {\rm GV},
	\nonumber
	\end{cases}\\[8pt]
	% -------------------------------------------------- Z = 14 Silicon ---------------------------------------------
	\label{eq:Si}
	F_{\rm Si}& (R)  = \\
	&\begin{cases}
	\displaystyle - a - bR^3 + cR^4 - dR^5 + fR^{-3} + g\tanh^2(R + R^2),    &R\le 2.8\, {\rm GV}, \smallskip\\
	\displaystyle R^{-2.7} \left[ (h -iR^{-1})\tilde{R} + l\sqrt{R} -\frac{m}{\sqrt{R}} + nR^{-0.25}  - \frac{o\sqrt{R}}{p + qR} - r  \right],  &R> 2.8\, {\rm GV},
	\nonumber
	\end{cases}\\[8pt] 
	% -------------------------------------------------- Z = 15 Phosphorus ---------------------------------------------
	\label{eq:P}
	F_{\rm P }& (R)  = \\
	&\begin{cases}
	\displaystyle - a + \tanh(R) \left[bR + \sin(R)(cR - dR^2) \right] + fR^3\sin(R) - g\tanh\left(\sin\left[R^2\tanh(R)\right]\right),    &R\le 2.4\, {\rm GV}, \smallskip\\
	\displaystyle R^{-2.7} \left[ - h -\frac{i}{\tilde{R}} - l\tilde{R}^2 + m\tilde{R}^4 +\log(n + R) (o\tilde{R} + p)   \right],  &R> 2.4\, {\rm GV},
	\nonumber
	\end{cases}\\[8pt] 
    % -------------------------------------------------- Z = 16 Sulfur ---------------------------------------------
	\label{eq:S}
	F_{\rm S}& (R)  = \\
	&\begin{cases}
	\displaystyle R^{-2.7} \left[ aR\tanh^3(R) + \left(bR^2 \tanh^2(R) - cR^3 \right)\tanh(d + R)   \right],    &R\le 3.2\, {\rm GV}, \smallskip\\
	\displaystyle R^{-2.7} \left[ f - gR - h\tilde{R} + i\sqrt{R} -\frac{l}{1 + \sqrt{R}}  - \frac{mR^{0.25}}{n + \sqrt{R}}  \right],  &R> 3.2\, {\rm GV},
	\nonumber
	\end{cases}\\[8pt] 
	% -------------------------------------------------- Z = 17 Chlorine ----------------------------------------
	\label{eq:Cl}
	F_{\rm Cl}& (R)  = \\
	&\begin{cases}
	\displaystyle a + b R + \sin{\left(R + \frac{c}{R}\right)} \left( - d + \frac{f}{R} + g R - h R^2 +i\tanh{R}\right),    &R\le 2.7\, {\rm GV}, \smallskip\\
	\displaystyle R^{-2.7} \left[ l + m R - n R^2  + \frac{1}{R}\left(o  -\frac{p}{\tilde{R}}\right) - q  \tilde{R} - r \tanh{\left(\frac{-s\tilde{R}}{R + \tilde{R}}\right)}  \right],  &R> 2.7\, {\rm GV},
	\nonumber
	\end{cases}\\[8pt] 
	% -------------------------------------------------- Z = 18 Argon ---------------------------------------------
	\label{eq:Ar}
	F_{\rm Ar}& (R)  = \\ %\times R^{2.7}
	&\begin{cases}
	\displaystyle a + R \left[ b + G(R)(c+d R) \right] + \sin{R} \left( f + g \sin{R}- h G(R) \right),    &R\le 2.7\, {\rm GV}, \smallskip\\
	\displaystyle R^{-2.7} \left[ i + l \tanh{\left(-m \sqrt{R}\right)} + \sqrt{R} \left( n - o \tilde{R} \right) + G(\tilde{R}) \left(p + q\tilde{R}^2 - rR\right) \right],  &R> 2.7\, {\rm GV},
	\nonumber
	\end{cases}\\[8pt]  
% -------------------------------------------------- Z = 19 Potassium ---------------------------------------------
	\label{eq:K}
	F_{\rm K}& (R)  = \\
	&\begin{cases}
	\displaystyle R^2 \left\{ - a+ bR + c L(R)  - d\tanh^2(R) - fR\tanh^3(R) - gR\log{\left[\tanh{(R)}\right]} \right\},    &R\le 2.4\, {\rm GV}, \smallskip\\
	\displaystyle R^{-2.7} \left[ - h + iR -\frac{l}{R} - mR^2  + nG\left(-\frac{o}{R^2} + \frac{pR + R^2}{q + rR + R^2}\right)  \right],  &R> 2.4\, {\rm GV},
	\nonumber
	\end{cases}\\[8pt] 
% -------------------------------------------------- Z = 20 Calcium ---------------------------------------------
\label{eq:Ca}
F_{\rm Ca}& (R)  = \\ %\times R^{2.7}
&\begin{cases}
\displaystyle a + b\tanh{\left[R G(R)^2\right]} - c R + G(R) \left\{ - d - f R^2 - G(R) \left[g +h R^2 G(R)^2 \right]\right\},    &R\le 2.9\, {\rm GV}, \smallskip\\
\displaystyle R^{-2.7} \left\{ - i + \tilde{R} (l - m \tilde{R}) + \log{( n + oR)} \left( p + \tilde{R} \left[ q \log{( n + oR)} - r R \tilde{R} \right] \right) \right\},  &R> 2.9\, {\rm GV},
\nonumber
\end{cases}\\[8pt] 
% -------------------------------------------------- Z = 21 Scandium ---------------------------------------------
\label{eq:Sc}
F_{\rm Sc}& (R)  = \\
&\begin{cases}
\displaystyle R\left[a + bR - cR^3  + \left(dR^2 + fR^6 - g\right)G(R) + hG(R)^2\right],    &R\le 2.65\, {\rm GV}, \smallskip\\
\displaystyle R^{-2.7} \left[ - i  + lR - m\tilde{R} + n\tilde{R}^2 + o\log(\tilde{R})  + \frac{p - q\tilde{R}}{\sqrt{R}} + \frac{r\tilde{R}}{s + tR}  \right],  &R> 2.65\, {\rm GV},
\nonumber
\end{cases}\\[8pt] 
% -------------------------------------------------- Z = 22 Titanium ---------------------------------------------
\label{eq:Ti}
F_{\rm Ti}& (R)  = \\
&\begin{cases}
\displaystyle a - bR - cR^2  - dG(R) + fG(R)^3 + g\tanh{\left(hG(R) -\frac{i}{R - l} - m\right)},    &R\le 2.8\, {\rm GV}, \smallskip\\
\displaystyle R^{-2.7} \left[ n  -\frac{o}{R} -\frac{p}{\sqrt{R}} + \left(q + r\sqrt{R} + \frac{s}{t\sqrt{R} + R^{1.5}} \right)\sqrt{\frac{\sqrt{R}}{u + R}}    \right],  &R> 2.8\, {\rm GV},
\nonumber
\end{cases}\\[8pt] 
% -------------------------------------------------- Z = 23 Vanadium ---------------------------------------------
\label{eq:V}
F_{\rm V}& (R)  = \\
&\begin{cases}
\displaystyle  a - bR  - cR^2 - dG(R) + f R G(R) + g G(R)^4\, \sqrt{G(R)} + \tanh\left\{\tanh\left(\tanh\left[hG(R)^2\right]\right)\right\},    &R\le 2.1\, {\rm GV}, \smallskip\\
\displaystyle R^{-2.7} \left[ i + lR - m R^2 -\frac{n}{R^2} -\frac{o}{p + R} + \frac{q}{r + R}  + \frac{s - tR}{\sqrt{R}}  \right],  &R> 2.1\, {\rm GV},
\nonumber
\end{cases}\\[8pt] 
% -------------------------------------------------- Z = 24 Chromium ---------------------------------------------
\label{eq:Cr}
F_{\rm Cr}& (R)  = \\
&\begin{cases}
\displaystyle a - bR - c R^2 + G(R)\left[ - d + G(R)(f R - g )   \right] + hG(2R),    &R\le 2.65\, {\rm GV}, \smallskip\\
\displaystyle R^{-2.7} \left[ i + l R -\frac{m}{R} + n\tilde{R} + \frac{o}{p + R} + \sqrt{R}\left(- q+ r R^2  - sR \right) \right],  &R> 2.65\, {\rm GV},
\nonumber
\end{cases}\\[8pt] 
% -------------------------------------------------- Z = 25 Manganese ---------------------------------------------
\label{eq:Mn}
F_{\rm Mn}& (R)  = \\
&\begin{cases}
\displaystyle a - bR + cRG(R) - dG(R) - fG(R)^2 + g\tanh{\left[G(R)^2\right]} + h\tanh^2\left(G(R)^3\tanh{\left[G(R)^2\right]}\right) ,    &R\le 2.6\, {\rm GV}, \smallskip\\
\displaystyle R^{-2.7} \left[  - i  +lR -\frac{m}{R} + \frac{n}{o + pR^2}+ q\tilde{R} + \frac{r}{\tilde{R}} -\frac{s}{R\tilde{R}} + \frac{t\tilde{R}}{R^2}  \right],  &R> 2.6\, {\rm GV},
\nonumber
\end{cases}\\[8pt] 
% -------------------------------------------------- Z = 26 Iron ---------------------------------------------
\label{eq:Fe}
F_{\rm Fe}& (R)  = \\
&\begin{cases}
\displaystyle a - bR -\frac{c}{\sqrt{R}} +d\tanh{R} + \tanh{\left(\tanh{R^2}\right)}(f - g R+ h R^2\tanh{R^2} ),    &R\le 2.8\, {\rm GV}, \smallskip\\
\displaystyle R^{-2.7} \left[  i + lR + m R^{\frac{1}{4}} - n\sqrt{R}  - oR^{\frac{5}{4}} - p \left(\frac{q}{r + R} + \sqrt{R}\right)^{\frac{1}{4}}  \right],  &R> 2.8\, {\rm GV},
\nonumber
\end{cases}\\[8pt] 
% -------------------------------------------------- Z = 27 Cobalt ---------------------------------------------
\label{eq:Co}
F_{\rm Co}& (R)  = \\
&\begin{cases}
\displaystyle  a + b\tilde{R} - cR + d R^2 G(R)^3 - G(R) (f + g\tilde{R} ) - h\sin{(iR^2 \tilde{R})},    &R\le 2.8\, {\rm GV}, \smallskip\\
\displaystyle R^{-2.7} \left[ l + m \tilde{R} + n \sqrt{R} - \frac{o}{\tilde{R}} -\frac{p\tilde{R}}{R} - q\sqrt{\tilde{R}} - r\sqrt{s + R} \right],  &R> 2.8\, {\rm GV},
\nonumber
\end{cases}\\[8pt] 
    % -------------------------------------------------- Z = 28 Nickel ---------------------------------------------
\label{eq:Ni}
F_{\rm Ni}& (R)  = \\
&\begin{cases}
\displaystyle a - bR - c G(R) - dG(R)^2 + f RG(R)^3   - g G\left[G(R)\right]  - h\sin(iR^2),    &R\le 2.5\, {\rm GV}, \smallskip\\
\displaystyle R^{-2.7} \left[ - l + mR + n\tilde{R} + \frac{o}{p + q\tilde{R}\sqrt{R}} + \frac{r - sR - tR^2}{\sqrt{R}}  \right],  &R> 2.5\, {\rm GV},
	\nonumber
	\end{cases}%\\[8pt] 
\end{align}
where $R$ is the particle rigidity in GV, the values of the fitting parameters from $a$ to $z$ are given in Table \ref{Tbl-1-10}, and the functions $\tilde{R}$, $G(x)$, $L(x)$, $T(x)$, and $H(x)$ are defined as: 
	\begin{align}
	\tilde{R}&=\log{R}\\
	G(x)&=e^{-x^2}\\
	L(x)&=\frac{1}{1+e^{-x}}\\
	T(x)&=(3.764\, x^2)^{x G(x)^{3.764x}}\\
	H(x)&=\tanh{(0.9341\,x)}.
	\end{align}

\begin{deluxetable}{ccccccccccc}[!hp]
	\tablecolumns{11}
	\tablewidth{0mm}
%	\def\arraystretch{0.9}
%	\tabletypesize{\scriptsize}
	\tablecaption{Parameters of the analytical fits to the LIS \label{Tbl-1-10}}
	\tablehead{
%\colhead{}  &  \colhead{Hydrogen} & \colhead{Helium} & \colhead{Lithium} & \colhead{Beryllium} & \colhead{Boron} & \colhead{Carbon} & \colhead{Nitrogen} & \colhead{Oxygen} & \colhead{Fluorine} & \colhead{Neon} \\[-10pt] %\baselineskip]
\colhead{Param}		&  \colhead{$_1$H}    & \colhead{$_2$He}    & \colhead{$_3$Li}      & \colhead{$_4$Be}       & \colhead{$_5$B}   & \colhead{$_6$C}    & \colhead{$_7$N}      & \colhead{$_8$O}          & \colhead{$_9$F}      & \colhead{$_{10}$Ne} 
	}
	\startdata
$a$   & 3.2181e+2 & 2.5869e+2 & 5.7479e-1 & 8.0273e-1 & 2.5879e+0 & 2.5619e+1 & 2.5500e+0 & 8.4830e+0 & 1.5000e-1 & 1.7870e-1\\
$b$   & 1.1729e+4 & 8.8956e+1 & 2.9099e+0 & 2.9420e-1 & 2.0789e+0 & 9.9880e+0 & 2.1800e-1 & 1.5990e+0 & 1.6240e-2 & 3.2716e-1\\
$c$   & 2.9232e+3 & 3.6399e+2 & 1.0510e+0 & 6.0570e-3 & 4.9610e-1 & 2.7833e-1 & 1.3780e+0 & 1.5200e-1 & 3.3000e-2 & 1.1690e-1\\
$d$   & 1.0621e+4 & 1.6337e+3 & 3.5180e+0 & 6.4707e-1 & 2.7110e-1 & 2.5530e+1 & 3.3950e+0 & 1.8090e+0 & 3.9750e-2 & 1.0612e+0\\
$f$   & 1.3867e+3 & 1.9527e+0 & 4.3210e-1 & 5.0133e-1 & 5.5680e-1 & 1.6329e+1 & 6.4660e+0 & 1.9910e-1 & 6.2900e-2 & 5.6280e-1\\
$g$   & 1.0399e+4 & 1.0469e+2 & 4.0360e-1 & 2.3520e-1 & 5.9700e-1 & 4.0345e+0 & 2.9065e+0 & 2.1320e+0 & 1.5090e-1 & 4.9458e-1\\
$h$   & 6.7166e-1 & 3.1229e+2 & 1.5845e+0 & 1.0306e+0 & 5.0461e-1 & 8.4840e+0 & 2.3890e+0 & 1.7750e+0 & 1.2330e-2 & 1.0275e+0\\
$i$   & 1.0528e+4 & 4.4168e+2 & 5.4231e-2 & 7.3387e+0 & 2.5469e+0 & 7.2138e-1 & 1.6350e+0 & 8.6098e+0 & 5.6880e-2 & 2.4220e-1\\
\enddata
\end{deluxetable}
\addtocounter{table}{-1}
\begin{deluxetable}{ccccccccccc}[!hp]
	\tablecolumns{11}
	\tablewidth{0mm}
%	\def\arraystretch{0.9}
%	\tabletypesize{\scriptsize}
	\tablecaption{Parameters of the analytical fits to the LIS (Cont'd)}
	\tablehead{
%\colhead{}  &  \colhead{Sodium} & \colhead{Magnesium} & \colhead{Aluminum} & \colhead{Silicon} & \colhead{Phosphorus} & \colhead{Sulfur} & \colhead{Chlorine} & \colhead{Argon} & \colhead{Potassium} & \colhead{Calcium} \\
\colhead{Param} &  \colhead{$_{11}$Na}    & \colhead{$_{12}$Mg}    & \colhead{$_{13}$Al}      & \colhead{$_{14}$Si}          & \colhead{$_{15}$P}      & \colhead{$_{16}$S}          & \colhead{$_{17}$Cl}      & \colhead{$_{18}$Ar}          & \colhead{$_{19}$K}      & \colhead{$_{20}$Ca} 
	}
	\startdata
	$a$   & 2.8986e-1 & 1.7520e+0 & 1.2799e-1 & 5.2200e-3 & 1.0494e-5 & 5.3240e-2 & 5.3920e-4 & 2.5438e-5 & 2.3390e-2 & 2.0630e-1 \\
	$b$   & 6.8600e-2 & 3.1990e-1 & 9.9660e-2 & 1.9870e-1 & 4.3270e-3 & 3.4588e-1 & 5.8450e-3 & 9.4440e-3 & 1.9640e-2 & 3.7100e-2 \\
	$c$  & 4.5620e-7 & 2.4630e-2 & 4.4010e-2 & 8.8650e-2 & 3.7809e-2 & 6.0390e-2 & 6.9720e-3 & 4.0297e-2 & 4.9840e-2 & 4.9200e-2 \\
	$d$   & 5.0130e-2 & 4.1040e+0 & 3.8345e-3 & 1.1147e-2 & 2.2944e-2 & 3.7590e-4 & 1.2950e-2 & 1.4057e-2 & 6.0257e-3 & 1.1430e-1 \\
	$f$   & 2.7050e-1 & 3.4076e+0 & 2.8640e-2 & 1.3364e-6 & 4.2950e-3 & 2.3330e+1 & 3.9470e-4 & 3.4100e-2 & 2.5495e-2 & 4.5500e-2 \\
	$g$   & 1.3359e-2 & 4.7905e-1 & 3.2260e-2 & 1.2769e+0 & 1.4953e-4 & 6.0000e-6 & 6.5200e-3 & 1.1060e-4 & 1.4850e-2 & 9.1570e-2 \\
	$h$   & 7.1440e-3 & 3.0143e-1 & 5.6330e-3 & 4.1250e+0 & 0.3398e+0 & 1.4470e+0 & 2.0216e-3 & 8.4675e-2 & 2.6180e-2 & 1.0694e-1 \\
	$i$   & 1.2880e+0 & 3.5570e-1 & 5.2299e+0 & 1.9305e+1 & 0.2714e+0 & 1.1580e-2 & 2.2600e-2 & 1.2850e+0 & 3.9300e-7 & 3.3032e+0 \\
\enddata
\end{deluxetable}
\addtocounter{table}{-1}
\begin{deluxetable}{ccccccccc}[!hp]
	\tablecolumns{9}
	\tablewidth{0mm}
%	\def\arraystretch{0.9}
%	\tabletypesize{\scriptsize}
	\tablecaption{Parameters of the analytical fits to the LIS (Cont'd)}
	\tablehead{
%\colhead{}  &  \colhead{Scandium} & \colhead{Titanium} & \colhead{Vanadium} & \colhead{Chromium} & \colhead{Manganese} & \colhead{Iron} & \colhead{Cobalt} & \colhead{Nickel}  \\
\colhead{Param} &  \colhead{$_{21}$Sc}    & \colhead{$_{22}$Ti} & \colhead{$_{23}$V} & \colhead{$_{24}$Cr}  & \colhead{$_{25}$Mn}      & \colhead{$_{26}$Fe}          & \colhead{$_{27}$Co}      & \colhead{$_{28}$Ni}         
	}
	\startdata
	$a$   & 4.8040e-3 & 1.2400e-1 & 5.8719e-2 & 1.4049e-1 & 9.1143e-2 & 2.2973e-2 & 5.4787e-3 & 7.4200e-2 \\
	$b$   & 2.9380e-3 & 6.4262e-3 & 3.9100e-3 & 2.9690e-2 & 2.2760e-2 & 3.4760e-1 & 4.8046e-4 & 1.4580e-2 \\
	$c$   & 3.8183e-4 & 7.0644e-3 & 3.0460e-3 & 1.5200e-3 & 1.8000e-2 & 3.5540e-3 & 1.5370e-3 & 5.0780e-3 \\
	$d$   & 2.1519e-2 & 1.4862e-1 & 7.0142e-2 & 1.3255e-1 & 8.7100e-2 & 1.4447e-1 & 2.0000e-3 & 6.1640e-2 \\
	$f$   & 1.7680e-3 & 2.4763e-2 & 2.0860e-3 & 2.7540e-2 & 4.2450e-2 & 1.3607e+0 & 5.1480e-3 & 1.3707e-2 \\
	$g$   & 6.8726e-3 & 2.0640e-4 & 1.0095e-2 & 3.4667e-2 & 2.7860e-2 & 1.2525e-1 & 3.8710e-4 & 2.0290e-2 \\
	$h$   & 2.0013e-3 & 7.3320e+0 & 1.4046e-3 & 2.6850e-2 & 4.2140e-2 & 5.4440e-2 & 4.6009e-6 & 1.3026e-3 \\
	$i$   & 4.1600e-1 & 2.7142e-3 & 2.2000e-1 & 1.4870e-1 & 1.2500e+0 & 7.2300e+1 & 1.0067e+0 & 4.8310e-1 \\
\enddata
\end{deluxetable}

\afterpage{\clearpage}
\pagebreak[4]

\subsubsection{Numerical tables}

Numerical tables of the LIS for all elements from $_1$H to $_{28}$Ni calculated for the {\it I}-scenario. For each element we provide two tables, tabulated in kinetic energy $E_{\rm kin}$ per nucleon and in rigidity $R$. 

\begin{deluxetable*}{cccccccccc}[!hp]
\tabletypesize{\footnotesize}
    \tablecolumns{10}
    \tablewidth{0mm}
    \tablecaption{$Z=1$ -- Proton LIS\label{Tbl-ProtonLIS-EKin}}
    \tablehead{
        \colhead{$E_{\rm kin}$} & \colhead{Differential} &
        \colhead{$E_{\rm kin}$} & \colhead{Differential} &
        \colhead{$E_{\rm kin}$} & \colhead{Differential} &
        \colhead{$E_{\rm kin}$} & \colhead{Differential} &
        \colhead{$E_{\rm kin}$} & \colhead{Differential} \\ [-2ex] 
        \colhead{GeV/n} & \colhead{intensity} &
        \colhead{GeV/n} & \colhead{intensity} &
        \colhead{GeV/n} & \colhead{intensity} &
        \colhead{GeV/n} & \colhead{intensity} &
        \colhead{GeV/n} & \colhead{intensity}  
    }
\startdata
1.000e-03 & 1.538e+04 & 6.309e-02 & 2.173e+04 & 3.981e+00 & 2.253e+02 & 2.512e+02 & 3.377e-03 & 1.585e+04 & 5.986e-08\\
1.101e-03 & 1.652e+04 & 6.948e-02 & 2.116e+04 & 4.384e+00 & 1.823e+02 & 2.766e+02 & 2.585e-03 & 1.745e+04 & 4.617e-08\\
1.213e-03 & 1.688e+04 & 7.651e-02 & 2.057e+04 & 4.827e+00 & 1.470e+02 & 3.046e+02 & 1.980e-03 & 1.922e+04 & 3.559e-08\\
1.335e-03 & 1.720e+04 & 8.425e-02 & 1.995e+04 & 5.315e+00 & 1.182e+02 & 3.354e+02 & 1.519e-03 & 2.116e+04 & 2.740e-08\\
1.470e-03 & 1.752e+04 & 9.277e-02 & 1.930e+04 & 5.853e+00 & 9.467e+01 & 3.693e+02 & 1.167e-03 & 2.330e+04 & 2.109e-08\\
1.619e-03 & 1.785e+04 & 1.022e-01 & 1.863e+04 & 6.446e+00 & 7.553e+01 & 4.067e+02 & 8.981e-04 & 2.566e+04 & 1.621e-08\\
1.783e-03 & 1.818e+04 & 1.125e-01 & 1.795e+04 & 7.098e+00 & 6.001e+01 & 4.478e+02 & 6.922e-04 & 2.825e+04 & 1.246e-08\\
1.963e-03 & 1.852e+04 & 1.239e-01 & 1.724e+04 & 7.816e+00 & 4.747e+01 & 4.931e+02 & 5.343e-04 & 3.111e+04 & 9.565e-09\\
\enddata
    \tablecomments{Differential Intensity units: (m$^2$ s sr GeV/n)$^{-1}$.}
\end{deluxetable*}

\begin{deluxetable*}{cccccccccc}[!hp]
\tabletypesize{\footnotesize}
    \tablecolumns{10}
    \tablewidth{0mm}
    \tablecaption{$Z=1$ -- Proton LIS\label{Tbl-ProtonLIS-Rigi}}
    \tablehead{
        \colhead{Rigidity} & \colhead{Differential} &
        \colhead{Rigidity} & \colhead{Differential} &
        \colhead{Rigidity} & \colhead{Differential} &
        \colhead{Rigidity} & \colhead{Differential} &
        \colhead{Rigidity} & \colhead{Differential} \\ [-2ex] 
        \colhead{GV} & \colhead{intensity} &
        \colhead{GV} & \colhead{intensity} &
        \colhead{GV} & \colhead{intensity} &
        \colhead{GV} & \colhead{intensity} &
        \colhead{GV} & \colhead{intensity}  
    }
\startdata
4.333e-02 & 7.087e+02 & 3.498e-01 & 7.584e+03 & 4.829e+00 & 2.250e+02 & 2.521e+02 & 3.394e-03 & 1.585e+04 & 5.994e-08\\
4.547e-02 & 7.989e+02 & 3.677e-01 & 7.715e+03 & 5.238e+00 & 1.825e+02 & 2.775e+02 & 2.597e-03 & 1.745e+04 & 4.624e-08\\
4.772e-02 & 8.566e+02 & 3.866e-01 & 7.827e+03 & 5.688e+00 & 1.474e+02 & 3.055e+02 & 1.989e-03 & 1.922e+04 & 3.564e-08\\
5.008e-02 & 9.158e+02 & 4.064e-01 & 7.920e+03 & 6.183e+00 & 1.187e+02 & 3.363e+02 & 1.526e-03 & 2.116e+04 & 2.744e-08\\
5.255e-02 & 9.790e+02 & 4.274e-01 & 7.991e+03 & 6.726e+00 & 9.514e+01 & 3.702e+02 & 1.172e-03 & 2.330e+04 & 2.111e-08\\
5.515e-02 & 1.047e+03 & 4.496e-01 & 8.039e+03 & 7.324e+00 & 7.596e+01 & 4.076e+02 & 9.019e-04 & 2.566e+04 & 1.623e-08\\
5.787e-02 & 1.119e+03 & 4.730e-01 & 8.064e+03 & 7.981e+00 & 6.040e+01 & 4.488e+02 & 6.950e-04 & 2.826e+04 & 1.247e-08\\
6.073e-02 & 1.196e+03 & 4.978e-01 & 8.065e+03 & 8.704e+00 & 4.781e+01 & 4.941e+02 & 5.365e-04 & 3.111e+04 & 9.576e-09\\
\enddata
    \tablecomments{Differential Intensity units: (m$^2$ s sr GV)$^{-1}$.}
\end{deluxetable*}

\begin{deluxetable*}{cccccccccc}[!hp]
\tabletypesize{\footnotesize}
    \tablecolumns{10}
    \tablewidth{0mm}
    \tablecaption{$Z=2$ -- Helium LIS\label{Tbl-HeliumLIS-EKin}}
    \tablehead{
        \colhead{$E_{\rm kin}$} & \colhead{Differential} &
        \colhead{$E_{\rm kin}$} & \colhead{Differential} &
        \colhead{$E_{\rm kin}$} & \colhead{Differential} &
        \colhead{$E_{\rm kin}$} & \colhead{Differential} &
        \colhead{$E_{\rm kin}$} & \colhead{Differential} \\ [-2ex] 
        \colhead{GeV/n} & \colhead{intensity} &
        \colhead{GeV/n} & \colhead{intensity} &
        \colhead{GeV/n} & \colhead{intensity} &
        \colhead{GeV/n} & \colhead{intensity} &
        \colhead{GeV/n} & \colhead{intensity}  
    }
\startdata
1.000e-03 & 1.183e+03 & 6.309e-02 & 1.974e+03 & 3.981e+00 & 1.362e+01 & 2.512e+02 & 2.704e-04 & 1.585e+04 & 6.966e-09\\
1.101e-03 & 1.265e+03 & 6.948e-02 & 1.924e+03 & 4.384e+00 & 1.095e+01 & 2.766e+02 & 2.107e-04 & 1.745e+04 & 5.385e-09\\
1.213e-03 & 1.302e+03 & 7.651e-02 & 1.870e+03 & 4.827e+00 & 8.775e+00 & 3.046e+02 & 1.644e-04 & 1.922e+04 & 4.157e-09\\
1.335e-03 & 1.337e+03 & 8.425e-02 & 1.814e+03 & 5.315e+00 & 7.011e+00 & 3.354e+02 & 1.284e-04 & 2.116e+04 & 3.204e-09\\
1.470e-03 & 1.373e+03 & 9.277e-02 & 1.755e+03 & 5.853e+00 & 5.584e+00 & 3.693e+02 & 1.004e-04 & 2.330e+04 & 2.467e-09\\
1.619e-03 & 1.410e+03 & 1.022e-01 & 1.694e+03 & 6.446e+00 & 4.435e+00 & 4.067e+02 & 7.848e-05 & 2.566e+04 & 1.897e-09\\
1.783e-03 & 1.447e+03 & 1.125e-01 & 1.632e+03 & 7.098e+00 & 3.511e+00 & 4.478e+02 & 6.141e-05 & 2.825e+04 & 1.458e-09\\
1.963e-03 & 1.486e+03 & 1.239e-01 & 1.570e+03 & 7.816e+00 & 2.771e+00 & 4.931e+02 & 4.807e-05 & 3.111e+04 & 1.119e-09\\
\enddata
    \tablecomments{Differential Intensity units: (m$^2$ s sr GeV/n)$^{-1}$.}
\end{deluxetable*}

\begin{deluxetable*}{cccccccccc}[!hp]
\tabletypesize{\footnotesize}
    \tablecolumns{10}
    \tablewidth{0mm}
    \tablecaption{$Z=2$ -- Helium LIS\label{Tbl-HeliumLIS-Rigi}}
    \tablehead{
        \colhead{Rigidity} & \colhead{Differential} &
        \colhead{Rigidity} & \colhead{Differential} &
        \colhead{Rigidity} & \colhead{Differential} &
        \colhead{Rigidity} & \colhead{Differential} &
        \colhead{Rigidity} & \colhead{Differential} \\ [-2ex] 
        \colhead{GV} & \colhead{intensity} &
        \colhead{GV} & \colhead{intensity} &
        \colhead{GV} & \colhead{intensity} &
        \colhead{GV} & \colhead{intensity} &
        \colhead{GV} & \colhead{intensity}  
    }
\startdata
8.635e-02 & 2.744e+01 & 6.972e-01 & 3.609e+02 & 9.646e+00 & 6.237e+00 & 5.042e+02 & 1.331e-04 & 3.169e+04 & 3.472e-09\\
9.061e-02 & 3.079e+01 & 7.328e-01 & 3.696e+02 & 1.047e+01 & 5.038e+00 & 5.550e+02 & 1.038e-04 & 3.490e+04 & 2.684e-09\\
9.509e-02 & 3.324e+01 & 7.704e-01 & 3.779e+02 & 1.137e+01 & 4.055e+00 & 6.110e+02 & 8.102e-05 & 3.843e+04 & 2.072e-09\\
9.979e-02 & 3.582e+01 & 8.101e-01 & 3.869e+02 & 1.235e+01 & 3.252e+00 & 6.726e+02 & 6.331e-05 & 4.232e+04 & 1.597e-09\\
1.047e-01 & 3.860e+01 & 8.519e-01 & 3.940e+02 & 1.344e+01 & 2.600e+00 & 7.405e+02 & 4.950e-05 & 4.660e+04 & 1.230e-09\\
1.099e-01 & 4.160e+01 & 8.961e-01 & 3.981e+02 & 1.464e+01 & 2.072e+00 & 8.152e+02 & 3.873e-05 & 5.132e+04 & 9.460e-10\\
1.153e-01 & 4.482e+01 & 9.428e-01 & 4.003e+02 & 1.595e+01 & 1.645e+00 & 8.975e+02 & 3.032e-05 & 5.651e+04 & 7.269e-10\\
1.210e-01 & 4.829e+01 & 9.922e-01 & 4.004e+02 & 1.740e+01 & 1.302e+00 & 9.881e+02 & 2.375e-05 & 6.223e+04 & 5.580e-10\\
\enddata
    \tablecomments{Differential Intensity units: (m$^2$ s sr GV)$^{-1}$.}
\end{deluxetable*}

\begin{deluxetable*}{cccccccccc}[!hp]
\tabletypesize{\footnotesize}
    \tablecolumns{10}
    \tablewidth{0mm}
    \tablecaption{$Z=3$ -- Lithium LIS\label{Tbl-LithiumLIS-EKin}}
    \tablehead{
        \colhead{$E_{\rm kin}$} & \colhead{Differential} &
        \colhead{$E_{\rm kin}$} & \colhead{Differential} &
        \colhead{$E_{\rm kin}$} & \colhead{Differential} &
        \colhead{$E_{\rm kin}$} & \colhead{Differential} &
        \colhead{$E_{\rm kin}$} & \colhead{Differential} \\ [-2ex] 
        \colhead{GeV/n} & \colhead{intensity} &
        \colhead{GeV/n} & \colhead{intensity} &
        \colhead{GeV/n} & \colhead{intensity} &
        \colhead{GeV/n} & \colhead{intensity} &
        \colhead{GeV/n} & \colhead{intensity}  
    }
\startdata
1.000e-03 & 1.268e-01 & 6.309e-02 & 3.318e+00 & 3.981e+00 & 6.860e-02 & 2.512e+02 & 4.059e-07 & 1.585e+04 & 9.815e-12\\
1.101e-03 & 1.400e-01 & 6.948e-02 & 3.396e+00 & 4.384e+00 & 5.452e-02 & 2.766e+02 & 3.104e-07 & 1.745e+04 & 7.783e-12\\
1.213e-03 & 1.505e-01 & 7.651e-02 & 3.440e+00 & 4.827e+00 & 4.306e-02 & 3.046e+02 & 2.377e-07 & 1.922e+04 & 6.174e-12\\
1.335e-03 & 1.617e-01 & 8.425e-02 & 3.463e+00 & 5.315e+00 & 3.379e-02 & 3.354e+02 & 1.824e-07 & 2.116e+04 & 4.900e-12\\
1.470e-03 & 1.737e-01 & 9.277e-02 & 3.480e+00 & 5.853e+00 & 2.636e-02 & 3.693e+02 & 1.401e-07 & 2.330e+04 & 3.890e-12\\
1.619e-03 & 1.867e-01 & 1.022e-01 & 3.499e+00 & 6.446e+00 & 2.046e-02 & 4.067e+02 & 1.078e-07 & 2.566e+04 & 3.090e-12\\
1.783e-03 & 2.007e-01 & 1.125e-01 & 3.517e+00 & 7.098e+00 & 1.580e-02 & 4.478e+02 & 8.302e-08 & 2.825e+04 & 2.455e-12\\
1.963e-03 & 2.159e-01 & 1.239e-01 & 3.535e+00 & 7.816e+00 & 1.215e-02 & 4.931e+02 & 6.402e-08 & 3.111e+04 & 1.951e-12\\
\enddata
    \tablecomments{Differential Intensity units: (m$^2$ s sr GeV/n)$^{-1}$.}
\end{deluxetable*}

\begin{deluxetable*}{cccccccccc}[!hp]
\tabletypesize{\footnotesize}
    \tablecolumns{10}
    \tablewidth{0mm}
    \tablecaption{$Z=3$ -- Lithium LIS\label{Tbl-LithiumLIS-Rigi}}
    \tablehead{
        \colhead{Rigidity} & \colhead{Differential} &
        \colhead{Rigidity} & \colhead{Differential} &
        \colhead{Rigidity} & \colhead{Differential} &
        \colhead{Rigidity} & \colhead{Differential} &
        \colhead{Rigidity} & \colhead{Differential} \\ [-2ex] 
        \colhead{GV} & \colhead{intensity} &
        \colhead{GV} & \colhead{intensity} &
        \colhead{GV} & \colhead{intensity} &
        \colhead{GV} & \colhead{intensity} &
        \colhead{GV} & \colhead{intensity}  
    }
\startdata
1.007e-01 & 3.246e-03 & 8.134e-01 & 6.036e-01 & 1.125e+01 & 2.530e-02 & 5.882e+02 & 1.556e-07 & 3.698e+04 & 4.105e-12\\
1.057e-01 & 3.718e-03 & 8.550e-01 & 6.354e-01 & 1.221e+01 & 2.017e-02 & 6.475e+02 & 1.193e-07 & 4.072e+04 & 3.259e-12\\
1.109e-01 & 4.201e-03 & 8.988e-01 & 6.640e-01 & 1.326e+01 & 1.598e-02 & 7.128e+02 & 9.158e-08 & 4.484e+04 & 2.588e-12\\
1.164e-01 & 4.746e-03 & 9.451e-01 & 6.919e-01 & 1.441e+01 & 1.256e-02 & 7.847e+02 & 7.044e-08 & 4.937e+04 & 2.056e-12\\
1.222e-01 & 5.363e-03 & 9.939e-01 & 7.218e-01 & 1.568e+01 & 9.815e-03 & 8.639e+02 & 5.425e-08 & 5.437e+04 & 1.634e-12\\
1.282e-01 & 6.061e-03 & 1.045e+00 & 7.555e-01 & 1.708e+01 & 7.625e-03 & 9.511e+02 & 4.184e-08 & 5.987e+04 & 1.299e-12\\
1.345e-01 & 6.852e-03 & 1.100e+00 & 7.910e-01 & 1.861e+01 & 5.893e-03 & 1.047e+03 & 3.230e-08 & 6.593e+04 & 1.033e-12\\
1.412e-01 & 7.748e-03 & 1.158e+00 & 8.273e-01 & 2.029e+01 & 4.535e-03 & 1.153e+03 & 2.497e-08 & 7.260e+04 & 8.221e-13\\
\enddata
    \tablecomments{Differential Intensity units: (m$^2$ s sr GV)$^{-1}$.}
\end{deluxetable*}

\begin{deluxetable*}{cccccccccc}[!hp]
\tabletypesize{\footnotesize}
    \tablecolumns{10}
    \tablewidth{0mm}
    \tablecaption{$Z=4$ -- Beryllium LIS\label{Tbl-BerylliumLIS-EKin}}
    \tablehead{
        \colhead{$E_{\rm kin}$} & \colhead{Differential} &
        \colhead{$E_{\rm kin}$} & \colhead{Differential} &
        \colhead{$E_{\rm kin}$} & \colhead{Differential} &
        \colhead{$E_{\rm kin}$} & \colhead{Differential} &
        \colhead{$E_{\rm kin}$} & \colhead{Differential} \\ [-2ex] 
        \colhead{GeV/n} & \colhead{intensity} &
        \colhead{GeV/n} & \colhead{intensity} &
        \colhead{GeV/n} & \colhead{intensity} &
        \colhead{GeV/n} & \colhead{intensity} &
        \colhead{GeV/n} & \colhead{intensity}  
    }
\startdata
1.000e-03 & 4.246e-02 & 6.309e-02 & 1.441e+00 & 3.981e+00 & 4.095e-02 & 2.512e+02 & 2.320e-07 & 1.585e+04 & 1.343e-12\\
1.101e-03 & 4.687e-02 & 6.948e-02 & 1.496e+00 & 4.384e+00 & 3.223e-02 & 2.766e+02 & 1.744e-07 & 1.745e+04 & 1.014e-12\\
1.213e-03 & 5.079e-02 & 7.651e-02 & 1.543e+00 & 4.827e+00 & 2.531e-02 & 3.046e+02 & 1.313e-07 & 1.922e+04 & 7.659e-13\\
1.335e-03 & 5.504e-02 & 8.425e-02 & 1.583e+00 & 5.315e+00 & 1.981e-02 & 3.354e+02 & 9.889e-08 & 2.116e+04 & 5.784e-13\\
1.470e-03 & 5.966e-02 & 9.277e-02 & 1.617e+00 & 5.853e+00 & 1.545e-02 & 3.693e+02 & 7.454e-08 & 2.330e+04 & 4.367e-13\\
1.619e-03 & 6.469e-02 & 1.022e-01 & 1.645e+00 & 6.446e+00 & 1.200e-02 & 4.067e+02 & 5.622e-08 & 2.566e+04 & 3.297e-13\\
1.783e-03 & 7.015e-02 & 1.125e-01 & 1.668e+00 & 7.098e+00 & 9.293e-03 & 4.478e+02 & 4.242e-08 & 2.825e+04 & 2.490e-13\\
1.963e-03 & 7.610e-02 & 1.239e-01 & 1.688e+00 & 7.816e+00 & 7.169e-03 & 4.931e+02 & 3.203e-08 & 3.111e+04 & 1.880e-13\\
\enddata
    \tablecomments{Differential Intensity units: (m$^2$ s sr GeV/n)$^{-1}$.}
\end{deluxetable*}

\afterpage{\clearpage} %imos

\begin{deluxetable*}{cccccccccc}[!hp]
\tabletypesize{\footnotesize}
    \tablecolumns{10}
    \tablewidth{0mm}
    \tablecaption{$Z=4$ -- Beryllium LIS\label{Tbl-BerylliumLIS-Rigi}}
    \tablehead{
        \colhead{Rigidity} & \colhead{Differential} &
        \colhead{Rigidity} & \colhead{Differential} &
        \colhead{Rigidity} & \colhead{Differential} &
        \colhead{Rigidity} & \colhead{Differential} &
        \colhead{Rigidity} & \colhead{Differential} \\ [-2ex] 
        \colhead{GV} & \colhead{intensity} &
        \colhead{GV} & \colhead{intensity} &
        \colhead{GV} & \colhead{intensity} &
        \colhead{GV} & \colhead{intensity} &
        \colhead{GV} & \colhead{intensity}  
    }
\startdata
7.555e-02 & 6.222e-04 & 6.101e-01 & 2.296e-01 & 8.440e+00 & 2.862e-02 & 4.412e+02 & 1.768e-07 & 2.773e+04 & 1.001e-12\\
7.929e-02 & 8.547e-04 & 6.412e-01 & 2.510e-01 & 9.157e+00 & 2.283e-02 & 4.856e+02 & 1.327e-07 & 3.054e+04 & 7.562e-13\\
8.320e-02 & 1.072e-03 & 6.741e-01 & 2.731e-01 & 9.945e+00 & 1.813e-02 & 5.346e+02 & 9.965e-08 & 3.363e+04 & 5.710e-13\\
8.731e-02 & 1.290e-03 & 7.088e-01 & 2.959e-01 & 1.081e+01 & 1.432e-02 & 5.885e+02 & 7.492e-08 & 3.703e+04 & 4.311e-13\\
9.163e-02 & 1.513e-03 & 7.454e-01 & 3.186e-01 & 1.176e+01 & 1.122e-02 & 6.479e+02 & 5.637e-08 & 4.078e+04 & 3.255e-13\\
9.616e-02 & 1.747e-03 & 7.841e-01 & 3.405e-01 & 1.281e+01 & 8.760e-03 & 7.133e+02 & 4.245e-08 & 4.490e+04 & 2.457e-13\\
1.009e-01 & 1.998e-03 & 8.250e-01 & 3.614e-01 & 1.396e+01 & 6.815e-03 & 7.853e+02 & 3.199e-08 & 4.945e+04 & 1.855e-13\\
1.059e-01 & 2.276e-03 & 8.682e-01 & 3.819e-01 & 1.522e+01 & 5.284e-03 & 8.646e+02 & 2.412e-08 & 5.445e+04 & 1.401e-13\\
\enddata
    \tablecomments{Differential Intensity units: (m$^2$ s sr GV)$^{-1}$.}
\end{deluxetable*}

\begin{deluxetable*}{cccccccccc}[!hp]
\tabletypesize{\footnotesize}
    \tablecolumns{10}
    \tablewidth{0mm}
    \tablecaption{$Z=5$ -- Boron LIS\label{Tbl-BoronLIS-EKin}}
    \tablehead{
        \colhead{$E_{\rm kin}$} & \colhead{Differential} &
        \colhead{$E_{\rm kin}$} & \colhead{Differential} &
        \colhead{$E_{\rm kin}$} & \colhead{Differential} &
        \colhead{$E_{\rm kin}$} & \colhead{Differential} &
        \colhead{$E_{\rm kin}$} & \colhead{Differential} \\ [-2ex] 
        \colhead{GeV/n} & \colhead{intensity} &
        \colhead{GeV/n} & \colhead{intensity} &
        \colhead{GeV/n} & \colhead{intensity} &
        \colhead{GeV/n} & \colhead{intensity} &
        \colhead{GeV/n} & \colhead{intensity}  
    }
\startdata
1.000e-03 & 1.769e-01 & 6.309e-02 & 5.363e+00 & 3.981e+00 & 9.708e-02 & 2.512e+02 & 5.837e-07 & 1.585e+04 & 3.499e-12\\
1.101e-03 & 1.946e-01 & 6.948e-02 & 5.508e+00 & 4.384e+00 & 7.660e-02 & 2.766e+02 & 4.396e-07 & 1.745e+04 & 2.643e-12\\
1.213e-03 & 2.108e-01 & 7.651e-02 & 5.621e+00 & 4.827e+00 & 6.021e-02 & 3.046e+02 & 3.314e-07 & 1.922e+04 & 1.996e-12\\
1.335e-03 & 2.282e-01 & 8.425e-02 & 5.706e+00 & 5.315e+00 & 4.714e-02 & 3.354e+02 & 2.499e-07 & 2.116e+04 & 1.508e-12\\
1.470e-03 & 2.472e-01 & 9.277e-02 & 5.768e+00 & 5.853e+00 & 3.675e-02 & 3.693e+02 & 1.887e-07 & 2.330e+04 & 1.139e-12\\
1.619e-03 & 2.679e-01 & 1.022e-01 & 5.811e+00 & 6.446e+00 & 2.855e-02 & 4.067e+02 & 1.425e-07 & 2.566e+04 & 8.600e-13\\
1.783e-03 & 2.903e-01 & 1.125e-01 & 5.845e+00 & 7.098e+00 & 2.210e-02 & 4.478e+02 & 1.077e-07 & 2.825e+04 & 6.495e-13\\
1.963e-03 & 3.147e-01 & 1.239e-01 & 5.873e+00 & 7.816e+00 & 1.706e-02 & 4.931e+02 & 8.137e-08 & 3.111e+04 & 4.905e-13\\
\enddata
    \tablecomments{Differential Intensity units: (m$^2$ s sr GeV/n)$^{-1}$.}
\end{deluxetable*}

\begin{deluxetable*}{cccccccccc}[!hp]
\tabletypesize{\footnotesize}
    \tablecolumns{10}
    \tablewidth{0mm}
    \tablecaption{$Z=5$ -- Boron LIS\label{Tbl-BoronLIS-Rigi}}
    \tablehead{
        \colhead{Rigidity} & \colhead{Differential} &
        \colhead{Rigidity} & \colhead{Differential} &
        \colhead{Rigidity} & \colhead{Differential} &
        \colhead{Rigidity} & \colhead{Differential} &
        \colhead{Rigidity} & \colhead{Differential} \\ [-2ex] 
        \colhead{GV} & \colhead{intensity} &
        \colhead{GV} & \colhead{intensity} &
        \colhead{GV} & \colhead{intensity} &
        \colhead{GV} & \colhead{intensity} &
        \colhead{GV} & \colhead{intensity}  
    }
\startdata
9.498e-02 & 4.177e-03 & 7.669e-01 & 9.235e-01 & 1.061e+01 & 4.093e-02 & 5.546e+02 & 2.535e-07 & 3.486e+04 & 1.523e-12\\
9.967e-02 & 4.796e-03 & 8.061e-01 & 9.882e-01 & 1.151e+01 & 3.240e-02 & 6.105e+02 & 1.910e-07 & 3.839e+04 & 1.151e-12\\
1.046e-01 & 5.450e-03 & 8.475e-01 & 1.051e+00 & 1.250e+01 & 2.553e-02 & 6.721e+02 & 1.440e-07 & 4.228e+04 & 8.691e-13\\
1.098e-01 & 6.194e-03 & 8.911e-01 & 1.111e+00 & 1.359e+01 & 2.003e-02 & 7.399e+02 & 1.086e-07 & 4.655e+04 & 6.564e-13\\
1.152e-01 & 7.041e-03 & 9.371e-01 & 1.171e+00 & 1.479e+01 & 1.564e-02 & 8.145e+02 & 8.201e-08 & 5.126e+04 & 4.958e-13\\
1.209e-01 & 8.006e-03 & 9.857e-01 & 1.229e+00 & 1.610e+01 & 1.217e-02 & 8.967e+02 & 6.195e-08 & 5.645e+04 & 3.744e-13\\
1.269e-01 & 9.105e-03 & 1.037e+00 & 1.286e+00 & 1.755e+01 & 9.438e-03 & 9.873e+02 & 4.681e-08 & 6.216e+04 & 2.828e-13\\
1.331e-01 & 1.036e-02 & 1.091e+00 & 1.342e+00 & 1.913e+01 & 7.295e-03 & 1.087e+03 & 3.539e-08 & 6.845e+04 & 2.135e-13\\
\enddata
    \tablecomments{Differential Intensity units: (m$^2$ s sr GV)$^{-1}$.}
\end{deluxetable*}

\begin{deluxetable*}{cccccccccc}[!hp]
\tabletypesize{\footnotesize}
    \tablecolumns{10}
    \tablewidth{0mm}
    \tablecaption{$Z=6$ -- Carbon LIS\label{Tbl-CarbonLIS-EKin}}
    \tablehead{
        \colhead{$E_{\rm kin}$} & \colhead{Differential} &
        \colhead{$E_{\rm kin}$} & \colhead{Differential} &
        \colhead{$E_{\rm kin}$} & \colhead{Differential} &
        \colhead{$E_{\rm kin}$} & \colhead{Differential} &
        \colhead{$E_{\rm kin}$} & \colhead{Differential} \\ [-2ex] 
        \colhead{GeV/n} & \colhead{intensity} &
        \colhead{GeV/n} & \colhead{intensity} &
        \colhead{GeV/n} & \colhead{intensity} &
        \colhead{GeV/n} & \colhead{intensity} &
        \colhead{GeV/n} & \colhead{intensity}  
    }
\startdata
1.000e-03 & 4.765e+00 & 6.309e-02 & 3.576e+01 & 3.981e+00 & 3.838e-01 & 2.512e+02 & 8.785e-06 & 1.585e+04 & 2.474e-10\\
1.101e-03 & 5.133e+00 & 6.948e-02 & 3.604e+01 & 4.384e+00 & 3.103e-01 & 2.766e+02 & 6.847e-06 & 1.745e+04 & 1.939e-10\\
1.213e-03 & 5.456e+00 & 7.651e-02 & 3.617e+01 & 4.827e+00 & 2.501e-01 & 3.046e+02 & 5.342e-06 & 1.922e+04 & 1.520e-10\\
1.335e-03 & 5.798e+00 & 8.425e-02 & 3.615e+01 & 5.315e+00 & 2.009e-01 & 3.354e+02 & 4.171e-06 & 2.116e+04 & 1.191e-10\\
1.470e-03 & 6.160e+00 & 9.277e-02 & 3.597e+01 & 5.853e+00 & 1.608e-01 & 3.693e+02 & 3.259e-06 & 2.330e+04 & 9.339e-11\\
1.619e-03 & 6.544e+00 & 1.022e-01 & 3.562e+01 & 6.446e+00 & 1.283e-01 & 4.067e+02 & 2.548e-06 & 2.566e+04 & 7.319e-11\\
1.783e-03 & 6.951e+00 & 1.125e-01 & 3.510e+01 & 7.098e+00 & 1.020e-01 & 4.478e+02 & 1.994e-06 & 2.825e+04 & 5.736e-11\\
1.963e-03 & 7.381e+00 & 1.239e-01 & 3.441e+01 & 7.816e+00 & 8.094e-02 & 4.931e+02 & 1.560e-06 & 3.111e+04 & 4.496e-11\\
\enddata
    \tablecomments{Differential Intensity units: (m$^2$ s sr GeV/n)$^{-1}$.}
\end{deluxetable*}

\begin{deluxetable*}{cccccccccc}[!hp]
\tabletypesize{\footnotesize}
    \tablecolumns{10}
    \tablewidth{0mm}
    \tablecaption{$Z=6$ -- Carbon LIS\label{Tbl-CarbonLIS-Rigi}}
    \tablehead{
        \colhead{Rigidity} & \colhead{Differential} &
        \colhead{Rigidity} & \colhead{Differential} &
        \colhead{Rigidity} & \colhead{Differential} &
        \colhead{Rigidity} & \colhead{Differential} &
        \colhead{Rigidity} & \colhead{Differential} \\ [-2ex] 
        \colhead{GV} & \colhead{intensity} &
        \colhead{GV} & \colhead{intensity} &
        \colhead{GV} & \colhead{intensity} &
        \colhead{GV} & \colhead{intensity} &
        \colhead{GV} & \colhead{intensity}  
    }
\startdata
9.354e-02 & 1.335e-01 & 7.553e-01 & 6.713e+00 & 1.045e+01 & 1.557e-01 & 5.462e+02 & 3.587e-06 & 3.434e+04 & 1.012e-10\\
9.816e-02 & 1.488e-01 & 7.939e-01 & 7.012e+00 & 1.134e+01 & 1.261e-01 & 6.012e+02 & 2.798e-06 & 3.781e+04 & 7.929e-11\\
1.030e-01 & 1.659e-01 & 8.346e-01 & 7.290e+00 & 1.231e+01 & 1.017e-01 & 6.619e+02 & 2.184e-06 & 4.163e+04 & 6.215e-11\\
1.081e-01 & 1.850e-01 & 8.776e-01 & 7.541e+00 & 1.338e+01 & 8.175e-02 & 7.287e+02 & 1.706e-06 & 4.585e+04 & 4.871e-11\\
1.134e-01 & 2.061e-01 & 9.229e-01 & 7.758e+00 & 1.456e+01 & 6.546e-02 & 8.022e+02 & 1.334e-06 & 5.049e+04 & 3.818e-11\\
1.191e-01 & 2.296e-01 & 9.708e-01 & 7.937e+00 & 1.586e+01 & 5.225e-02 & 8.831e+02 & 1.043e-06 & 5.559e+04 & 2.992e-11\\
1.249e-01 & 2.558e-01 & 1.021e+00 & 8.075e+00 & 1.728e+01 & 4.158e-02 & 9.723e+02 & 8.164e-07 & 6.122e+04 & 2.345e-11\\
1.311e-01 & 2.849e-01 & 1.075e+00 & 8.168e+00 & 1.885e+01 & 3.299e-02 & 1.070e+03 & 6.392e-07 & 6.741e+04 & 1.838e-11\\
\enddata
    \tablecomments{Differential Intensity units: (m$^2$ s sr GV)$^{-1}$.}
\end{deluxetable*}

\begin{deluxetable*}{cccccccccc}[!hp]
\tabletypesize{\footnotesize}
    \tablecolumns{10}
    \tablewidth{0mm}
    \tablecaption{$Z=7$ -- Nitrogen LIS\label{Tbl-NitrogenLIS-EKin}}
    \tablehead{
        \colhead{$E_{\rm kin}$} & \colhead{Differential} &
        \colhead{$E_{\rm kin}$} & \colhead{Differential} &
        \colhead{$E_{\rm kin}$} & \colhead{Differential} &
        \colhead{$E_{\rm kin}$} & \colhead{Differential} &
        \colhead{$E_{\rm kin}$} & \colhead{Differential} \\ [-2ex] 
        \colhead{GeV/n} & \colhead{intensity} &
        \colhead{GeV/n} & \colhead{intensity} &
        \colhead{GeV/n} & \colhead{intensity} &
        \colhead{GeV/n} & \colhead{intensity} &
        \colhead{GeV/n} & \colhead{intensity}  
    }
\startdata
1.000e-03 & 4.180e-01 & 6.309e-02 & 6.085e+00 & 3.981e+00 & 9.794e-02 & 2.512e+02 & 1.229e-06 & 1.585e+04 & 5.477e-11\\
1.101e-03 & 4.525e-01 & 6.948e-02 & 6.258e+00 & 4.384e+00 & 7.853e-02 & 2.766e+02 & 9.571e-07 & 1.745e+04 & 4.372e-11\\
1.213e-03 & 4.843e-01 & 7.651e-02 & 6.405e+00 & 4.827e+00 & 6.273e-02 & 3.046e+02 & 7.470e-07 & 1.922e+04 & 3.490e-11\\
1.335e-03 & 5.183e-01 & 8.425e-02 & 6.525e+00 & 5.315e+00 & 4.991e-02 & 3.354e+02 & 5.841e-07 & 2.116e+04 & 2.786e-11\\
1.470e-03 & 5.548e-01 & 9.277e-02 & 6.615e+00 & 5.853e+00 & 3.956e-02 & 3.693e+02 & 4.574e-07 & 2.330e+04 & 2.225e-11\\
1.619e-03 & 5.938e-01 & 1.022e-01 & 6.673e+00 & 6.446e+00 & 3.125e-02 & 4.067e+02 & 3.587e-07 & 2.566e+04 & 1.777e-11\\
1.783e-03 & 6.357e-01 & 1.125e-01 & 6.695e+00 & 7.098e+00 & 2.458e-02 & 4.478e+02 & 2.816e-07 & 2.825e+04 & 1.419e-11\\
1.963e-03 & 6.806e-01 & 1.239e-01 & 6.677e+00 & 7.816e+00 & 1.927e-02 & 4.931e+02 & 2.214e-07 & 3.111e+04 & 1.134e-11\\
\enddata
    \tablecomments{Differential Intensity units: (m$^2$ s sr GeV/n)$^{-1}$.}
\end{deluxetable*}

\begin{deluxetable*}{cccccccccc}[!hp]
\tabletypesize{\footnotesize}
    \tablecolumns{10}
    \tablewidth{0mm}
    \tablecaption{$Z=7$ -- Nitrogen LIS\label{Tbl-NitrogenLIS-Rigi}}
    \tablehead{
        \colhead{Rigidity} & \colhead{Differential} &
        \colhead{Rigidity} & \colhead{Differential} &
        \colhead{Rigidity} & \colhead{Differential} &
        \colhead{Rigidity} & \colhead{Differential} &
        \colhead{Rigidity} & \colhead{Differential} \\ [-2ex] 
        \colhead{GV} & \colhead{intensity} &
        \colhead{GV} & \colhead{intensity} &
        \colhead{GV} & \colhead{intensity} &
        \colhead{GV} & \colhead{intensity} &
        \colhead{GV} & \colhead{intensity}  
    }
\startdata
9.252e-02 & 1.082e-02 & 7.470e-01 & 1.084e+00 & 1.034e+01 & 4.238e-02 & 5.402e+02 & 5.349e-07 & 3.396e+04 & 2.344e-11\\
9.709e-02 & 1.216e-02 & 7.852e-01 & 1.160e+00 & 1.121e+01 & 3.405e-02 & 5.946e+02 & 4.167e-07 & 3.739e+04 & 1.870e-11\\
1.019e-01 & 1.365e-02 & 8.254e-01 & 1.235e+00 & 1.218e+01 & 2.724e-02 & 6.546e+02 & 3.252e-07 & 4.118e+04 & 1.493e-11\\
1.069e-01 & 1.532e-02 & 8.679e-01 & 1.308e+00 & 1.324e+01 & 2.170e-02 & 7.206e+02 & 2.542e-07 & 4.534e+04 & 1.192e-11\\
1.122e-01 & 1.719e-02 & 9.128e-01 & 1.378e+00 & 1.440e+01 & 1.722e-02 & 7.934e+02 & 1.990e-07 & 4.993e+04 & 9.514e-12\\
1.177e-01 & 1.929e-02 & 9.601e-01 & 1.444e+00 & 1.568e+01 & 1.361e-02 & 8.734e+02 & 1.560e-07 & 5.498e+04 & 7.597e-12\\
1.236e-01 & 2.164e-02 & 1.010e+00 & 1.503e+00 & 1.709e+01 & 1.071e-02 & 9.616e+02 & 1.224e-07 & 6.055e+04 & 6.067e-12\\
1.297e-01 & 2.429e-02 & 1.063e+00 & 1.554e+00 & 1.864e+01 & 8.394e-03 & 1.059e+03 & 9.619e-08 & 6.667e+04 & 4.845e-12\\
\enddata
    \tablecomments{Differential Intensity units: (m$^2$ s sr GV)$^{-1}$.}
\end{deluxetable*}

\begin{deluxetable*}{cccccccccc}[!hp]
\tabletypesize{\footnotesize}
    \tablecolumns{10}
    \tablewidth{0mm}
    \tablecaption{$Z=8$ -- Oxygen LIS\label{Tbl-OxygenLIS-EKin}}
    \tablehead{
        \colhead{$E_{\rm kin}$} & \colhead{Differential} &
        \colhead{$E_{\rm kin}$} & \colhead{Differential} &
        \colhead{$E_{\rm kin}$} & \colhead{Differential} &
        \colhead{$E_{\rm kin}$} & \colhead{Differential} &
        \colhead{$E_{\rm kin}$} & \colhead{Differential} \\ [-2ex] 
        \colhead{GeV/n} & \colhead{intensity} &
        \colhead{GeV/n} & \colhead{intensity} &
        \colhead{GeV/n} & \colhead{intensity} &
        \colhead{GeV/n} & \colhead{intensity} &
        \colhead{GeV/n} & \colhead{intensity}  
    }
\startdata
1.000e-03 & 5.328e+00 & 6.309e-02 & 3.892e+01 & 3.981e+00 & 3.735e-01 & 2.512e+02 & 9.794e-06 & 1.585e+04 & 2.846e-10\\
1.101e-03 & 5.724e+00 & 6.948e-02 & 3.897e+01 & 4.384e+00 & 3.043e-01 & 2.766e+02 & 7.642e-06 & 1.745e+04 & 2.231e-10\\
1.213e-03 & 6.088e+00 & 7.651e-02 & 3.883e+01 & 4.827e+00 & 2.471e-01 & 3.046e+02 & 5.969e-06 & 1.922e+04 & 1.748e-10\\
1.335e-03 & 6.473e+00 & 8.425e-02 & 3.850e+01 & 5.315e+00 & 1.999e-01 & 3.354e+02 & 4.667e-06 & 2.116e+04 & 1.370e-10\\
1.470e-03 & 6.881e+00 & 9.277e-02 & 3.798e+01 & 5.853e+00 & 1.612e-01 & 3.693e+02 & 3.651e-06 & 2.330e+04 & 1.074e-10\\
1.619e-03 & 7.314e+00 & 1.022e-01 & 3.728e+01 & 6.446e+00 & 1.295e-01 & 4.067e+02 & 2.858e-06 & 2.566e+04 & 8.414e-11\\
1.783e-03 & 7.772e+00 & 1.125e-01 & 3.639e+01 & 7.098e+00 & 1.037e-01 & 4.478e+02 & 2.239e-06 & 2.825e+04 & 6.593e-11\\
1.963e-03 & 8.257e+00 & 1.239e-01 & 3.533e+01 & 7.816e+00 & 8.274e-02 & 4.931e+02 & 1.755e-06 & 3.111e+04 & 5.166e-11\\
\enddata
    \tablecomments{Differential Intensity units: (m$^2$ s sr GeV/n)$^{-1}$.}
\end{deluxetable*}

\begin{deluxetable*}{cccccccccc}[!hp]
\tabletypesize{\footnotesize}
    \tablecolumns{10}
    \tablewidth{0mm}
    \tablecaption{$Z=8$ -- Oxygen LIS\label{Tbl-OxygenLIS-Rigi}}
    \tablehead{
        \colhead{Rigidity} & \colhead{Differential} &
        \colhead{Rigidity} & \colhead{Differential} &
        \colhead{Rigidity} & \colhead{Differential} &
        \colhead{Rigidity} & \colhead{Differential} &
        \colhead{Rigidity} & \colhead{Differential} \\ [-2ex] 
        \colhead{GV} & \colhead{intensity} &
        \colhead{GV} & \colhead{intensity} &
        \colhead{GV} & \colhead{intensity} &
        \colhead{GV} & \colhead{intensity} &
        \colhead{GV} & \colhead{intensity}  
    }
\startdata
9.714e-02 & 1.627e-01 & 7.844e-01 & 7.497e+00 & 1.085e+01 & 1.375e-01 & 5.672e+02 & 3.618e-06 & 3.566e+04 & 1.057e-10\\
1.019e-01 & 1.815e-01 & 8.244e-01 & 7.739e+00 & 1.177e+01 & 1.121e-01 & 6.244e+02 & 2.827e-06 & 3.926e+04 & 8.284e-11\\
1.070e-01 & 2.025e-01 & 8.667e-01 & 7.944e+00 & 1.279e+01 & 9.106e-02 & 6.873e+02 & 2.211e-06 & 4.324e+04 & 6.492e-11\\
1.123e-01 & 2.258e-01 & 9.113e-01 & 8.108e+00 & 1.390e+01 & 7.365e-02 & 7.567e+02 & 1.730e-06 & 4.761e+04 & 5.087e-11\\
1.178e-01 & 2.517e-01 & 9.584e-01 & 8.227e+00 & 1.512e+01 & 5.935e-02 & 8.330e+02 & 1.355e-06 & 5.243e+04 & 3.987e-11\\
1.236e-01 & 2.805e-01 & 1.008e+00 & 8.301e+00 & 1.647e+01 & 4.765e-02 & 9.171e+02 & 1.061e-06 & 5.773e+04 & 3.124e-11\\
1.297e-01 & 3.126e-01 & 1.061e+00 & 8.326e+00 & 1.794e+01 & 3.812e-02 & 1.010e+03 & 8.319e-07 & 6.357e+04 & 2.448e-11\\
1.362e-01 & 3.482e-01 & 1.116e+00 & 8.303e+00 & 1.957e+01 & 3.040e-02 & 1.112e+03 & 6.525e-07 & 7.001e+04 & 1.918e-11\\
\enddata
    \tablecomments{Differential Intensity units: (m$^2$ s sr GV)$^{-1}$.}
\end{deluxetable*}

\begin{deluxetable*}{cccccccccc}[!hp]
\tabletypesize{\footnotesize}
    \tablecolumns{10}
    \tablewidth{0mm}
    \tablecaption{$Z=9$ -- Fluorine LIS\label{Tbl-FluorineLIS-EKin}}
    \tablehead{
        \colhead{$E_{\rm kin}$} & \colhead{Differential} &
        \colhead{$E_{\rm kin}$} & \colhead{Differential} &
        \colhead{$E_{\rm kin}$} & \colhead{Differential} &
        \colhead{$E_{\rm kin}$} & \colhead{Differential} &
        \colhead{$E_{\rm kin}$} & \colhead{Differential} \\ [-2ex] 
        \colhead{GeV/n} & \colhead{intensity} &
        \colhead{GeV/n} & \colhead{intensity} &
        \colhead{GeV/n} & \colhead{intensity} &
        \colhead{GeV/n} & \colhead{intensity} &
        \colhead{GeV/n} & \colhead{intensity}  
    }
\startdata
1.000e-03 & 8.092e-03 & 6.309e-02 & 2.302e-01 & 3.981e+00 & 6.870e-03 & 2.512e+02 & 5.529e-08 & 1.585e+04 & 5.341e-13\\
1.101e-03 & 8.836e-03 & 6.948e-02 & 2.434e-01 & 4.384e+00 & 5.497e-03 & 2.766e+02 & 4.203e-08 & 1.745e+04 & 4.085e-13\\
1.213e-03 & 9.564e-03 & 7.651e-02 & 2.564e-01 & 4.827e+00 & 4.383e-03 & 3.046e+02 & 3.199e-08 & 1.922e+04 & 3.125e-13\\
1.335e-03 & 1.035e-02 & 8.425e-02 & 2.690e-01 & 5.315e+00 & 3.481e-03 & 3.354e+02 & 2.438e-08 & 2.116e+04 & 2.390e-13\\
1.470e-03 & 1.121e-02 & 9.277e-02 & 2.811e-01 & 5.853e+00 & 2.753e-03 & 3.693e+02 & 1.860e-08 & 2.330e+04 & 1.828e-13\\
1.619e-03 & 1.214e-02 & 1.022e-01 & 2.924e-01 & 6.446e+00 & 2.168e-03 & 4.067e+02 & 1.419e-08 & 2.566e+04 & 1.399e-13\\
1.783e-03 & 1.315e-02 & 1.125e-01 & 3.026e-01 & 7.098e+00 & 1.701e-03 & 4.478e+02 & 1.084e-08 & 2.825e+04 & 1.070e-13\\
1.963e-03 & 1.425e-02 & 1.239e-01 & 3.117e-01 & 7.816e+00 & 1.328e-03 & 4.931e+02 & 8.282e-09 & 3.111e+04 & 8.186e-14\\
\enddata
    \tablecomments{Differential Intensity units: (m$^2$ s sr GeV/n)$^{-1}$.}
\end{deluxetable*}

\begin{deluxetable*}{cccccccccc}[!hp]
\tabletypesize{\footnotesize}
    \tablecolumns{10}
    \tablewidth{0mm}
    \tablecaption{$Z=9$ -- Fluorine LIS\label{Tbl-FluorineLIS-Rigi}}
    \tablehead{
        \colhead{Rigidity} & \colhead{Differential} &
        \colhead{Rigidity} & \colhead{Differential} &
        \colhead{Rigidity} & \colhead{Differential} &
        \colhead{Rigidity} & \colhead{Differential} &
        \colhead{Rigidity} & \colhead{Differential} \\ [-2ex] 
        \colhead{GV} & \colhead{intensity} &
        \colhead{GV} & \colhead{intensity} &
        \colhead{GV} & \colhead{intensity} &
        \colhead{GV} & \colhead{intensity} &
        \colhead{GV} & \colhead{intensity}  
    }
\startdata
9.114e-02 & 1.775e-04 & 7.359e-01 & 3.822e-02 & 1.018e+01 & 3.195e-03 & 5.322e+02 & 2.619e-08 & 3.346e+04 & 2.530e-13\\
9.565e-02 & 2.033e-04 & 7.735e-01 & 4.220e-02 & 1.105e+01 & 2.563e-03 & 5.858e+02 & 1.991e-08 & 3.684e+04 & 1.935e-13\\
1.004e-01 & 2.309e-04 & 8.132e-01 & 4.641e-02 & 1.200e+01 & 2.049e-03 & 6.449e+02 & 1.515e-08 & 4.057e+04 & 1.480e-13\\
1.053e-01 & 2.623e-04 & 8.551e-01 & 5.082e-02 & 1.304e+01 & 1.630e-03 & 7.100e+02 & 1.155e-08 & 4.467e+04 & 1.132e-13\\
1.105e-01 & 2.980e-04 & 8.992e-01 & 5.538e-02 & 1.419e+01 & 1.292e-03 & 7.816e+02 & 8.809e-09 & 4.919e+04 & 8.660e-14\\
1.160e-01 & 3.387e-04 & 9.459e-01 & 6.003e-02 & 1.545e+01 & 1.019e-03 & 8.605e+02 & 6.723e-09 & 5.417e+04 & 6.625e-14\\
1.217e-01 & 3.850e-04 & 9.952e-01 & 6.473e-02 & 1.684e+01 & 8.002e-04 & 9.474e+02 & 5.134e-09 & 5.965e+04 & 5.068e-14\\
1.277e-01 & 4.377e-04 & 1.047e+00 & 6.941e-02 & 1.836e+01 & 6.256e-04 & 1.043e+03 & 3.923e-09 & 6.569e+04 & 3.877e-14\\
\enddata
    \tablecomments{Differential Intensity units: (m$^2$ s sr GV)$^{-1}$.}
\end{deluxetable*}

\begin{deluxetable*}{cccccccccc}[!hp]
\tabletypesize{\footnotesize}
    \tablecolumns{10}
    \tablewidth{0mm}
    \tablecaption{$Z=10$ -- Neon LIS\label{Tbl-NeonLIS-EKin}}
    \tablehead{
        \colhead{$E_{\rm kin}$} & \colhead{Differential} &
        \colhead{$E_{\rm kin}$} & \colhead{Differential} &
        \colhead{$E_{\rm kin}$} & \colhead{Differential} &
        \colhead{$E_{\rm kin}$} & \colhead{Differential} &
        \colhead{$E_{\rm kin}$} & \colhead{Differential} \\ [-2ex] 
        \colhead{GeV/n} & \colhead{intensity} &
        \colhead{GeV/n} & \colhead{intensity} &
        \colhead{GeV/n} & \colhead{intensity} &
        \colhead{GeV/n} & \colhead{intensity} &
        \colhead{GeV/n} & \colhead{intensity}  
    }
\startdata
1.000e-03 & 2.609e-01 & 6.309e-02 & 3.420e+00 & 3.981e+00 & 5.945e-02 & 2.512e+02 & 1.441e-06 & 1.585e+04 & 7.915e-11\\
1.101e-03 & 2.821e-01 & 6.948e-02 & 3.507e+00 & 4.384e+00 & 4.841e-02 & 2.766e+02 & 1.138e-06 & 1.745e+04 & 6.300e-11\\
1.213e-03 & 3.025e-01 & 7.651e-02 & 3.582e+00 & 4.827e+00 & 3.928e-02 & 3.046e+02 & 9.010e-07 & 1.922e+04 & 5.015e-11\\
1.335e-03 & 3.244e-01 & 8.425e-02 & 3.644e+00 & 5.315e+00 & 3.176e-02 & 3.354e+02 & 7.143e-07 & 2.116e+04 & 3.991e-11\\
1.470e-03 & 3.479e-01 & 9.277e-02 & 3.690e+00 & 5.853e+00 & 2.557e-02 & 3.693e+02 & 5.669e-07 & 2.330e+04 & 3.177e-11\\
1.619e-03 & 3.730e-01 & 1.022e-01 & 3.718e+00 & 6.446e+00 & 2.051e-02 & 4.067e+02 & 4.504e-07 & 2.566e+04 & 2.528e-11\\
1.783e-03 & 3.999e-01 & 1.125e-01 & 3.727e+00 & 7.098e+00 & 1.640e-02 & 4.478e+02 & 3.581e-07 & 2.825e+04 & 2.012e-11\\
1.963e-03 & 4.287e-01 & 1.239e-01 & 3.716e+00 & 7.816e+00 & 1.305e-02 & 4.931e+02 & 2.849e-07 & 3.111e+04 & 1.601e-11\\
\enddata
    \tablecomments{Differential Intensity units: (m$^2$ s sr GeV/n)$^{-1}$.}
\end{deluxetable*}

\begin{deluxetable*}{cccccccccc}[!hp]
\tabletypesize{\footnotesize}
    \tablecolumns{10}
    \tablewidth{0mm}
    \tablecaption{$Z=10$ -- Neon LIS\label{Tbl-NeonLIS-Rigi}}
    \tablehead{
        \colhead{Rigidity} & \colhead{Differential} &
        \colhead{Rigidity} & \colhead{Differential} &
        \colhead{Rigidity} & \colhead{Differential} &
        \colhead{Rigidity} & \colhead{Differential} &
        \colhead{Rigidity} & \colhead{Differential} \\ [-2ex] 
        \colhead{GV} & \colhead{intensity} &
        \colhead{GV} & \colhead{intensity} &
        \colhead{GV} & \colhead{intensity} &
        \colhead{GV} & \colhead{intensity} &
        \colhead{GV} & \colhead{intensity}  
    }
\startdata
9.498e-02 & 6.991e-03 & 7.669e-01 & 6.329e-01 & 1.061e+01 & 2.427e-02 & 5.546e+02 & 5.939e-07 & 3.486e+04 & 3.266e-11\\
9.967e-02 & 7.881e-03 & 8.061e-01 & 6.736e-01 & 1.151e+01 & 1.979e-02 & 6.105e+02 & 4.698e-07 & 3.839e+04 & 2.599e-11\\
1.046e-01 & 8.866e-03 & 8.475e-01 & 7.136e-01 & 1.250e+01 & 1.607e-02 & 6.721e+02 & 3.722e-07 & 4.228e+04 & 2.069e-11\\
1.098e-01 & 9.974e-03 & 8.911e-01 & 7.519e-01 & 1.359e+01 & 1.300e-02 & 7.399e+02 & 2.953e-07 & 4.655e+04 & 1.647e-11\\
1.152e-01 & 1.122e-02 & 9.371e-01 & 7.879e-01 & 1.479e+01 & 1.047e-02 & 8.145e+02 & 2.345e-07 & 5.126e+04 & 1.310e-11\\
1.209e-01 & 1.262e-02 & 9.857e-01 & 8.209e-01 & 1.610e+01 & 8.395e-03 & 8.967e+02 & 1.864e-07 & 5.645e+04 & 1.043e-11\\
1.269e-01 & 1.419e-02 & 1.037e+00 & 8.500e-01 & 1.755e+01 & 6.706e-03 & 9.873e+02 & 1.482e-07 & 6.216e+04 & 8.299e-12\\
1.331e-01 & 1.595e-02 & 1.091e+00 & 8.747e-01 & 1.913e+01 & 5.337e-03 & 1.087e+03 & 1.179e-07 & 6.845e+04 & 6.604e-12\\
\enddata
    \tablecomments{Differential Intensity units: (m$^2$ s sr GV)$^{-1}$.}
\end{deluxetable*}

\begin{deluxetable*}{cccccccccc}[!hp]
\tabletypesize{\footnotesize}
    \tablecolumns{10}
    \tablewidth{0mm}
    \tablecaption{$Z=11$ -- Sodium LIS\label{Tbl-SodiumLIS-EKin}}
    \tablehead{
        \colhead{$E_{\rm kin}$} & \colhead{Differential} &
        \colhead{$E_{\rm kin}$} & \colhead{Differential} &
        \colhead{$E_{\rm kin}$} & \colhead{Differential} &
        \colhead{$E_{\rm kin}$} & \colhead{Differential} &
        \colhead{$E_{\rm kin}$} & \colhead{Differential} \\ [-2ex] 
        \colhead{GeV/n} & \colhead{intensity} &
        \colhead{GeV/n} & \colhead{intensity} &
        \colhead{GeV/n} & \colhead{intensity} &
        \colhead{GeV/n} & \colhead{intensity} &
        \colhead{GeV/n} & \colhead{intensity}  
    }
\startdata
1.000e-03 & 3.756e-02 & 6.309e-02 & 5.707e-01 & 3.981e+00 & 1.117e-02 & 2.512e+02 & 1.311e-07 & 1.585e+04 & 2.136e-12\\
1.101e-03 & 4.068e-02 & 6.948e-02 & 5.867e-01 & 4.384e+00 & 8.981e-03 & 2.766e+02 & 1.005e-07 & 1.745e+04 & 1.661e-12\\
1.213e-03 & 4.374e-02 & 7.651e-02 & 6.007e-01 & 4.827e+00 & 7.194e-03 & 3.046e+02 & 7.711e-08 & 1.922e+04 & 1.291e-12\\
1.335e-03 & 4.702e-02 & 8.425e-02 & 6.128e-01 & 5.315e+00 & 5.739e-03 & 3.354e+02 & 5.924e-08 & 2.116e+04 & 1.004e-12\\
1.470e-03 & 5.055e-02 & 9.277e-02 & 6.225e-01 & 5.853e+00 & 4.559e-03 & 3.693e+02 & 4.555e-08 & 2.330e+04 & 7.812e-13\\
1.619e-03 & 5.434e-02 & 1.022e-01 & 6.297e-01 & 6.446e+00 & 3.608e-03 & 4.067e+02 & 3.505e-08 & 2.566e+04 & 6.077e-13\\
1.783e-03 & 5.841e-02 & 1.125e-01 & 6.343e-01 & 7.098e+00 & 2.844e-03 & 4.478e+02 & 2.699e-08 & 2.825e+04 & 4.728e-13\\
1.963e-03 & 6.279e-02 & 1.239e-01 & 6.364e-01 & 7.816e+00 & 2.232e-03 & 4.931e+02 & 2.080e-08 & 3.111e+04 & 3.679e-13\\
\enddata
    \tablecomments{Differential Intensity units: (m$^2$ s sr GeV/n)$^{-1}$.}
\end{deluxetable*}

\begin{deluxetable*}{cccccccccc}[!hp]
\tabletypesize{\footnotesize}
    \tablecolumns{10}
    \tablewidth{0mm}
    \tablecaption{$Z=11$ -- Sodium LIS\label{Tbl-SodiumLIS-Rigi}}
    \tablehead{
        \colhead{Rigidity} & \colhead{Differential} &
        \colhead{Rigidity} & \colhead{Differential} &
        \colhead{Rigidity} & \colhead{Differential} &
        \colhead{Rigidity} & \colhead{Differential} &
        \colhead{Rigidity} & \colhead{Differential} \\ [-2ex] 
        \colhead{GV} & \colhead{intensity} &
        \colhead{GV} & \colhead{intensity} &
        \colhead{GV} & \colhead{intensity} &
        \colhead{GV} & \colhead{intensity} &
        \colhead{GV} & \colhead{intensity}  
    }
\startdata
9.027e-02 & 8.317e-04 & 7.289e-01 & 9.567e-02 & 1.008e+01 & 5.246e-03 & 5.271e+02 & 6.269e-08 & 3.313e+04 & 1.022e-12\\
9.473e-02 & 9.453e-04 & 7.661e-01 & 1.027e-01 & 1.094e+01 & 4.229e-03 & 5.802e+02 & 4.806e-08 & 3.649e+04 & 7.943e-13\\
9.941e-02 & 1.066e-03 & 8.054e-01 & 1.098e-01 & 1.188e+01 & 3.395e-03 & 6.387e+02 & 3.688e-08 & 4.018e+04 & 6.176e-13\\
1.043e-01 & 1.203e-03 & 8.469e-01 & 1.169e-01 & 1.292e+01 & 2.714e-03 & 7.032e+02 & 2.833e-08 & 4.424e+04 & 4.803e-13\\
1.095e-01 & 1.357e-03 & 8.906e-01 & 1.238e-01 & 1.405e+01 & 2.160e-03 & 7.741e+02 & 2.178e-08 & 4.872e+04 & 3.736e-13\\
1.149e-01 & 1.530e-03 & 9.369e-01 & 1.306e-01 & 1.530e+01 & 1.712e-03 & 8.523e+02 & 1.676e-08 & 5.365e+04 & 2.906e-13\\
1.206e-01 & 1.726e-03 & 9.857e-01 & 1.370e-01 & 1.668e+01 & 1.351e-03 & 9.383e+02 & 1.291e-08 & 5.908e+04 & 2.261e-13\\
1.265e-01 & 1.947e-03 & 1.037e+00 & 1.431e-01 & 1.819e+01 & 1.062e-03 & 1.033e+03 & 9.948e-09 & 6.506e+04 & 1.760e-13\\
\enddata
    \tablecomments{Differential Intensity units: (m$^2$ s sr GV)$^{-1}$.}
\end{deluxetable*}

\begin{deluxetable*}{cccccccccc}[!hp]
\tabletypesize{\footnotesize}
    \tablecolumns{10}
    \tablewidth{0mm}
    \tablecaption{$Z=12$ -- Magnesium LIS\label{Tbl-MagnesiumLIS-EKin}}
    \tablehead{
        \colhead{$E_{\rm kin}$} & \colhead{Differential} &
        \colhead{$E_{\rm kin}$} & \colhead{Differential} &
        \colhead{$E_{\rm kin}$} & \colhead{Differential} &
        \colhead{$E_{\rm kin}$} & \colhead{Differential} &
        \colhead{$E_{\rm kin}$} & \colhead{Differential} \\ [-2ex] 
        \colhead{GeV/n} & \colhead{intensity} &
        \colhead{GeV/n} & \colhead{intensity} &
        \colhead{GeV/n} & \colhead{intensity} &
        \colhead{GeV/n} & \colhead{intensity} &
        \colhead{GeV/n} & \colhead{intensity}  
    }
\startdata
1.000e-03 & 4.505e-01 & 6.309e-02 & 6.316e+00 & 3.981e+00 & 7.408e-02 & 2.512e+02 & 1.705e-06 & 1.585e+04 & 4.597e-11\\
1.101e-03 & 4.883e-01 & 6.948e-02 & 6.402e+00 & 4.384e+00 & 6.023e-02 & 2.766e+02 & 1.329e-06 & 1.745e+04 & 3.596e-11\\
1.213e-03 & 5.256e-01 & 7.651e-02 & 6.444e+00 & 4.827e+00 & 4.878e-02 & 3.046e+02 & 1.037e-06 & 1.922e+04 & 2.812e-11\\
1.335e-03 & 5.658e-01 & 8.425e-02 & 6.440e+00 & 5.315e+00 & 3.934e-02 & 3.354e+02 & 8.102e-07 & 2.116e+04 & 2.199e-11\\
1.470e-03 & 6.090e-01 & 9.277e-02 & 6.388e+00 & 5.853e+00 & 3.160e-02 & 3.693e+02 & 6.332e-07 & 2.330e+04 & 1.720e-11\\
1.619e-03 & 6.555e-01 & 1.022e-01 & 6.289e+00 & 6.446e+00 & 2.530e-02 & 4.067e+02 & 4.953e-07 & 2.566e+04 & 1.345e-11\\
1.783e-03 & 7.055e-01 & 1.125e-01 & 6.150e+00 & 7.098e+00 & 2.018e-02 & 4.478e+02 & 3.876e-07 & 2.825e+04 & 1.052e-11\\
1.963e-03 & 7.593e-01 & 1.239e-01 & 5.979e+00 & 7.816e+00 & 1.604e-02 & 4.931e+02 & 3.034e-07 & 3.111e+04 & 8.224e-12\\
\enddata
    \tablecomments{Differential Intensity units: (m$^2$ s sr GeV/n)$^{-1}$.}
\end{deluxetable*}

\begin{deluxetable*}{cccccccccc}[!hp]
\tabletypesize{\footnotesize}
    \tablecolumns{10}
    \tablewidth{0mm}
    \tablecaption{$Z=12$ -- Magnesium LIS\label{Tbl-MagnesiumLIS-Rigi}}
    \tablehead{
        \colhead{Rigidity} & \colhead{Differential} &
        \colhead{Rigidity} & \colhead{Differential} &
        \colhead{Rigidity} & \colhead{Differential} &
        \colhead{Rigidity} & \colhead{Differential} &
        \colhead{Rigidity} & \colhead{Differential} \\ [-2ex] 
        \colhead{GV} & \colhead{intensity} &
        \colhead{GV} & \colhead{intensity} &
        \colhead{GV} & \colhead{intensity} &
        \colhead{GV} & \colhead{intensity} &
        \colhead{GV} & \colhead{intensity}  
    }
\startdata
9.354e-02 & 1.211e-02 & 7.553e-01 & 1.167e+00 & 1.045e+01 & 3.067e-02 & 5.462e+02 & 7.117e-07 & 3.434e+04 & 1.922e-11\\
9.816e-02 & 1.368e-02 & 7.939e-01 & 1.224e+00 & 1.134e+01 & 2.497e-02 & 6.012e+02 & 5.552e-07 & 3.781e+04 & 1.503e-11\\
1.030e-01 & 1.546e-02 & 8.346e-01 & 1.273e+00 & 1.231e+01 & 2.024e-02 & 6.619e+02 & 4.335e-07 & 4.163e+04 & 1.176e-11\\
1.081e-01 & 1.746e-02 & 8.776e-01 & 1.313e+00 & 1.338e+01 & 1.633e-02 & 7.287e+02 & 3.388e-07 & 4.585e+04 & 9.194e-12\\
1.134e-01 & 1.971e-02 & 9.229e-01 & 1.344e+00 & 1.456e+01 & 1.312e-02 & 8.022e+02 & 2.649e-07 & 5.049e+04 & 7.190e-12\\
1.191e-01 & 2.226e-02 & 9.708e-01 & 1.367e+00 & 1.586e+01 & 1.051e-02 & 8.831e+02 & 2.073e-07 & 5.559e+04 & 5.622e-12\\
1.249e-01 & 2.513e-02 & 1.021e+00 & 1.381e+00 & 1.728e+01 & 8.379e-03 & 9.723e+02 & 1.622e-07 & 6.122e+04 & 4.396e-12\\
1.311e-01 & 2.837e-02 & 1.075e+00 & 1.387e+00 & 1.885e+01 & 6.659e-03 & 1.070e+03 & 1.271e-07 & 6.741e+04 & 3.437e-12\\
\enddata
    \tablecomments{Differential Intensity units: (m$^2$ s sr GV)$^{-1}$.}
\end{deluxetable*}%

\end{document}